\documentclass[seceq]{ptptex}

\usepackage{graphicx,amssymb,amsfonts,amsmath,dcolumn,bm}

\title{
Re-examining Bogoliubov's theory of an interacting Bose gas
}

\author{
A.M. \textsc{Ettouhami}\footnote{E-mail: ettouhami@gmail.com}
}

\inst{
29 Lloyd Gibson Cres., Whitby, Ontario, L1R 2H6, Canada
}

\abst{
As is well-known, in the conventional formulation of Bogoliubov's theory of an interacting Bose gas,
the Hamiltonian $\hat{H}$ is written as a decoupled sum of contributions from different momenta of the form 
$\hat{H} = \sum_{\bf k\neq 0}\hat{H}_{\bf k}$. Then, each of the single-mode Hamiltonians $\hat{H}_{\bf k}$
is diagonalized separately, and the resulting ground state wavefunction of the total Hamiltonian $\hat{H}$
is written as a simple product of the ground state wavefunctions of each of the single-mode Hamiltonians 
$\hat{H}_{\bf k}$. While this way of diagonalizing the total Hamiltonian
$\hat{H}$ may seem to be valid from the perspective of the standard,
number non-conserving Bogoliubov's method, where the $\bf k=0$ state
is removed from the Hilbert space and hence the individual Hilbert spaces where the Hamiltonians
$\{\hat{H}_{\bf k}\}$ are diagonalized are disjoint from one another, we argue that from a number-conserving perspective 
this diagonalization method may not be adequate since the true Hilbert spaces
where the Hamiltonians $\{\hat{H}_{\bf k}\}$ should be diagonalized all have the  
${\bf k}=0$ state in common, and hence the ground state wavefunction of the total Hamiltonian $\hat{H}$
may {\em not} be written as a simple product of the ground state wavefunctions of the $\hat{H}_{\bf k}$'s. 
In this paper, we give a thorough review of Bogoliubov's method, and discuss a variational 
and number-conserving formulation of this theory in which the ${\bf k}=0$ state is restored
to the Hilbert space of the interacting gas, and where, instead of diagonalizing the Hamiltonians
$\hat{H}_{\bf k}$ separately, we diagonalize the total Hamiltonian $\hat{H}$ as a whole. 
When this is done, we find that the ground state energy is lowered below the Bogoliubov result,
and the depletion of bosons is significantly reduced with respect to the one obtained in the number non-conserving treatment.
We also find that the spectrum of the usual $\alpha_{\bf k}$ excitations of Bogoliubov's 
method changes from a gapless one, as predicted by the standard, number non-conserving formulation of this
theory, to one which exhibits a finite gap in the $k\to 0$ limit.
We discuss the presence of a gap in the spectrum of the $\alpha_{\bf k}$'s in light of Goldstone's theorem, 
and show that there is no contradiction with the latter.
}

\PTPindex{322, 320}  

\begin{document}

\maketitle


\section{Introduction} 
\label{Sec:Introduction}

There has been a lot of interest in the properties of interacting Bose systems during the past fifteen years 
following the experimental observation of Bose-Einstein condensation (BEC) in ultracold vapors of 
trapped rubidium\cite{Anderson1995} and sodium gases.\cite{Davis1995} On the theoretical front, 
much of the effort to understand the properties of these systems has focused on using approaches 
of the Bogoliubov type at the lowest temperatures, 
\cite{Bogoliubov1947,Lee1957,Bruckner1957,Beliaev1958,LeeHuangYang1957,Hugenholtz1959,Sawada1959,Gavoret1964,Hohenberg1965,Cheung1971,Wong1974}
and finite temperature generalizations of Bogoliubov's theory
\cite{Popov1965,Singh1967,Szepfalusy1974,Griffin1993,Griffin1996,Bijlsma1997,Shi1998,Andersen2004} 
at higher temperatures. At a fundamental level, theories of the Bogoliubov type are based on two major approximations. The first approximation is the so-called Bogoliubov prescription (BP), which consists in replacing the creation and annihilation operators of the condensate with a $c$-number. The Hamiltonian which is obtained from that procedure does not commute with the operator $\hat{N}$ describing 
the total number of particles in the system, which implies that the total number of particles is not conserved.
The second approximation, sometimes called\cite{Zagrebnov2001,Suto2008} the Bogoliubov truncation (BT), 
consists in truncating the Hamiltonian of the system, keeping only enough terms to allow the truncated Hamiltonian to be
exactly diagonalizable with the help of a ``canonical transformation."
The question then arises as to whether the physics of Bose condensed systems
is correctly described by Bogoliubov type theories, given these two major approximations that are made, and given also a number of other, lesser and mainly technical approximations having to do with the choice of self-energies in the field-theoretic formulation of these theories. Over the years, many authors have tried to shed light on these kinds of questions using different degrees of mathematical rigor
(see References \citen{Suto2008}-\citen{Yukalov2006} for an overview and a list of relevant references). Yet, and despite considerable activity, a closer look reveals that several aspects of Bogoliubov type theories are still not fully
understood.\cite{Suto2008,Yukalov2006,Girardeau1959,Gardiner1997,Girardeau1998,Castin1998,Ginibre1968,Lieb2005}

\addvspace{1.5mm}

In this paper, we will examine Bogoliubov's theory at zero temperature in quite some detail, shedding more light on the validity and
accuracy of the Bogoliubov prescription at a microscopic level,
and raising a few issues with this theory which have been overlooked in the past, and which 
in the author's view still need to be clarified. 
Chief among these new issues that we want to address is the question of trying to understand how
the decoupled way in which the Hamiltonian is diagonalized in Bogoliubov's theory affects
the nature of the Bogoliubov ground state.
Indeed, and as is well-known, in the standard formulation of Bogoliubov's theory,
after the creation and annihilation operators of the condensate,
$a_0^\dagger$ and $a_0$ respectively, are replaced
by the c-number $\simeq\sqrt{N}$ ($N$ being the total number of bosons in the system), 
the Hamiltonian $\hat{H}$ can be written as a decoupled sum of contributions from different momenta of the form 
$\hat{H} = \sum_{\bf k\neq 0}\hat{H}_{\bf k}$, where each Hamiltonian $\hat{H}_{\bf k}$
describes the interaction of bosons in the condensed $\bf k = 0$ state 
with bosons in the momentum modes $\pm\bf k$. 
Then, each of the single-mode Hamiltonians $\hat{H}_{\bf k}$
is diagonalized {\em separately} and the ground state (GS) wavefunction of $\hat{H}$
is obtained as the product of the GS wavefunctions of the $\hat{H}_{\bf k}$'s.
Here we shall argue that, while this way of diagonalizing the total Hamiltonian
$\hat{H}$ may seem to be valid from the perspective of the conventional,
number non-conserving Bogoliubov's method, where the $\bf k=0$ state
is removed from the Hilbert space and hence the individual Hilbert spaces where the Hamiltonians
$\{\hat{H}_{\bf k}\}$ are diagonalized are disjoint from one another, from a number-conserving perspective 
this diagonalization method is not appropriate, since the true Hilbert spaces
where the Hamiltonians $\{\hat{H}_{\bf k}\}$ should be diagonalized all have the  
${\bf k}=0$ state in common, and hence the ground state wavefunction of the Hamiltonian $\hat{H}$
may {\em not} be written as a simple product of the ground state wavefunctions of the $\hat{H}_{\bf k}$'s. 
We then shall discuss a variational, number-conserving 
generalization of Bogoliubov's theory in which the ${\bf k}=0$ state is restored
into the Hilbert space of the interacting gas, and where, instead of diagonalizing the Hamiltonians
$\hat{H}_{\bf k}$ separately, we diagonalize the total Hamiltonian $\hat{H}$ as a whole. 
When this is done, we find that the ground state energy is lowered below the Bogoliubov result,
and the depletion of bosons is significantly reduced with respect to the number non-conserving treatment.
Moreover, the spectrum of excitations of the system changes from a gapless one, 
as predicted by the standard, number non-conserving
Bogoliubov method, to one which exhibits a finite gap in the $k\to 0$ limit.

\addvspace{1.5mm}

The rest of this article consists of two main parts. The first part is almost entirely devoted
to assessing the accuracy of Bogoliubov's prescription, and consists of Secs. 
\ref{Sec:StandardBogoliubov} through \ref{Sec:ExactDiagOneMode}. In the second part of the paper, consisting
mainly of Sec. \ref{Sec:DiagFullH}, we go beyond the conventional
formulation of Bogoliubov's theory by trying to enforce the conservation of the number of bosons
in the system, making sure we keep an accurate count of the number of bosons
in the ${\bf k}=0$ state. As mentioned above, when this is done, the results of Bogoliubov's theory
are changed both in quantitative and qualitative ways, and these changes 
are discussed in quite some detail in Sec. \ref{Sec:Discussion}.

\addvspace{1.5mm}

We now want to give the reader a more detailed overview 
of the contents of each one of the remaining eight Sections of this
article. In Sec. \ref{Sec:StandardBogoliubov},
we shall start by reviewing the standard Bogoliubov treatment of an interacting Bose gas. 
As we mentioned above, after the BP is performed the total number of bosons is not conserved, which leads
to a number of unphysical features. As an example, the average number of bosons $N_{\bf k}$ in the
single-particle state of momentum ${\bf k}$ is found to diverge like $1/k$ as $k\to 0$, 
which does not of course make much sense for a system with a fixed number of bosons $N$. 
In Sec. \ref{Sec:Issues} we discuss this and other similar unphysical predictions of Bogoliubov's method, 
and their consequence on the evaluation of expectation values of physical 
observables in the Bogoliubov ground state. 
To be able to formulate Bogoliubov's theory within a number-conserving framework
and so get rid of the aforementioned unphysical features,
in Sec. \ref{Sec:DerivationH_Bog} we proceed to derive a number-conserving version
of Bogoliubov's Hamiltonian, which we show can be written as a sum 
of decoupled Hamiltonians $\hat{H}_{\bf k}$ for each momentum mode ${\bf k}$,
$\hat{H} = \sum_{\bf k\neq 0}\hat{H}_{\bf k}$. Then, in order to substantiate the claim made
above according to which each one of the single-mode Hamiltonians $\hat{H}_{\bf k}$
corresponding to a given momentum mode ${\bf k}$ is diagonalized independently in Bogoliubov's theory,   
in Sec. \ref{Sec:VariationalMethod} we present a detailed analysis of a variational, number-conserving 
approach to the single-mode Hamiltonian $\hat{H}_{\bf k}$, and show explicitly that
the results obtained from the diagonalization of this single-mode Hamiltonian
perfectly coincide with the results of Bogoliubov's method in the thermodynamic $N\to \infty$ limit.
At this point, and in order to assess the accuracy of the above diagonalization of the single-mode Hamiltonian $\hat{H}_{\bf k}$, 
in Sec. \ref{Sec:ExactDiagOneMode} we perform an {\em exact} numerical diagonalization
of this Hamiltonian. Our numerical results corroborate the variational approach of Sec. \ref{Sec:VariationalMethod}, 
which is found to be surprizingly accurate.
The main conclusion of the analysis carried out in Secs. \ref{Sec:VariationalMethod}
and \ref{Sec:ExactDiagOneMode} is that the treatment of the truncated many-body Hamiltonian 
carried out in Bogoliubov's theory amounts to diagonalizing each one of the Hamiltonians 
$\hat{H}_{\bf k}$ (representing the kinetic energy plus the interaction energy 
of bosons having a momentum $+{\bf k}$ or $-{\bf k}$ with the condensate) 
independently from the other momentum contributions $\hat{H}_{\bf k'(\neq k)}$, and is in fact
exceedingly good at predicting the ground state energy
of each one of these single-mode Hamiltonian $\hat{H}_{\bf k}$ taken {\em separately}. 
While this would be perfectly legitimate
{\em if} the various Hilbert spaces used to diagonalize the Hamiltonians $\hat{H}_{\bf k}$
were disjoint, in our case all these Hilbert spaces have the single-particle state with ${\bf k}={\bf 0}$ in common,
and so the above decoupled diagonalization procedure, strictly speaking, is mathematically not valid.

\addvspace{1.5mm}

In the second part of the paper, and 
in order to assess the quantitative accuracy of diagonalzing the $\hat{H}_{\bf k}$'s in the decoupled
way described above, we generalize the variational approach of Sec. \ref{Sec:VariationalMethod}
for the single-mode Hamiltonian $\hat{H}_{\bf k}$
to the {\em total} Hamiltonian $\hat{H}=\sum_{\bf k\neq 0}\hat{H}_{\bf k}$
in Sec. \ref{Sec:DiagFullH}, and we do that
in such a way as to keep an accurate count of the number of bosons in the ${\bf k}=0$ state,  
and with the requirement that the total number of bosons in {\em all} momentum modes be conserved. 
This leads, quite surprizingly, to an excitation spectrum of bosons which presents
a gap as $k\to 0$, by contrast to the usual Bogoliubov method in which the spectrum of excitations is gapless.
In Section \ref{Sec:Discussion} we give a discussion of our results, in view of their apparent violation of
Goldstone's theorem, and in Section \ref{Sec:Conclusion}
we present our conclusions.

\section{Review of the standard, number non-conserving formulation of Bogoliubov's theory}
\label{Sec:StandardBogoliubov}

We shall begin by reviewing the standard, number non-conserving formulation of Bogoliubov's theory.
To this end, let us consider the Hamiltonian of a gas of $N$ spinless bosons, interacting
through a two-body potential $v({\bf r})$:
\begin{align}
\hat{H} = \int d{\bf r} \;\hat\Psi^\dagger({\bf r})\Big(-\frac{\hbar^2{\bm\nabla}^2}{2m}\Big)
\hat\Psi({\bf r})
+ \frac{1}{2}\int d{\bf r}d{\bf r}'\, \hat\Psi^\dagger({\bf r})
\hat\Psi^\dagger({\bf r}')v({\bf r}-{\bf r}')
\hat\Psi({\bf r}') \hat\Psi({\bf r}),
\label{Eq:defH}
\end{align}
where $\hat\Psi({\bf r})$ is a boson field operator in second quantized language,
which has the Fourier decomposition $\hat\Psi({\bf r})=\sum_{\bf k}a_{\bf k}e^{i{\bf k}\cdot{\bf r}}/\sqrt{V}$
(periodic boundary conditions will be assumed throughout this paper),
$a_{\bf k}$ being a boson annihilation operator, and
$V$ being the volume of the system. In Fourier space, the Hamiltonian $\hat{H}$
has the following expression:
\begin{equation}
\hat{H} = \sum_{\bf k\neq 0}\varepsilon_{\bf k} a_{\bf k}^\dagger a_{\bf k}
+\frac{1}{2V}\sum_{\bf k,k'}\sum_{\bf q} v({\bf q}) a_{\bf k + q}^\dagger a_{\bf k' - q}^\dagger
a_{\bf k'}a_{\bf k},
\label{Eq:HFourierSpace}
\end{equation}
where $v({\bf k})$ is the Fourier transform of $v({\bf r})$ and
$\varepsilon_{\bf k} = \hbar^2k^2/2m$ is the kinetic energy of bosons with wavevector ${\bf k}$.

\addvspace{1.5mm}

According to Bogoliubov's prescription,\cite{Bogoliubov1947,FW} one replaces
the creation and annihilation operators $a_0^\dagger$ and $a_0$ 
of the ${\bf k}=0$ state by $\sqrt{N_0}$, where $N_0$ is the 
number of bosons in the condensate, upon which the above Hamiltonian takes the form 
$\hat{H} = H_0 + \hat{H}_2 + \hat{H}_3 + \hat{H}_4$, with ($n_0=N_0/V$ is the density of condensed bosons):
\begin{subequations}
\begin{align}
H_0 & = \frac{1}{2}V n_0^2v({\bf 0}),
\label{Eq:defH0}
\\
\hat{H}_2 & = \sum_{\bf k\neq 0} \Big\{ 
\xi_{\bf k} a_{\bf k}^\dagger a_{\bf k} 
+  \frac{1}{2}n_0 v({\bf k}) \big[ a_{\bf k}^\dagger a_{-\bf k}^\dagger
+ a_{\bf k} a_{-\bf k}\big]\Big\},
\label{Eq:defH2}
\\
\hat{H}_3 & = \sqrt{\frac{n_0}{V}}\sum_{\bf q, k} v({\bf q}) \big[
a_{\bf k + q}^\dagger a_{\bf k} a_{\bf q} 
+ a_{\bf k}^\dagger a_{\bf q} a_{\bf k + q}
\big],
\label{Eq:defH3}
\\
\hat{H}_4 & = \frac{1}{2V}\sum_{\bf q, k,k'} v({\bf q}) a_{\bf k + q}^\dagger a_{\bf k' -q}^\dagger
a_{\bf k'} a_{\bf k}.
\label{Eq:defH4}
\end{align}
\label{Eq:defH_i}
\end{subequations}
In Eq. (\ref{Eq:defH2}), $\xi_{\bf k}$ is the shifted boson energy:
\begin{equation}
\xi_{\bf k} = \varepsilon_{\bf k} + n_0\big[v({\bf k}) + v({\bf 0})\big],
\end{equation}
and it is understood that no creation and annihilation operators of the condensate ($a_0^\dagger$ and $a_0$)
appear on the {\em rhs} of Eqs. (\ref{Eq:defH3}) and (\ref{Eq:defH4}).

\addvspace{1.5mm}
   
As it can be seen, the new expression of the Hamiltonian after the Bogoliubov prescription is performed
does not conserve the number of bosons. To deal with this unphysical artifact of Bogoliubov's prescription, 
Bogoliubov suggests that one should work with the ``grand canonical Hamiltonian" 
$\tilde{H} = \hat{H} -\mu \hat{N}$, $\mu$ being the chemical potential and
the operator $\hat{N}=\sum_{\bf k} a^\dagger_{\bf k}a_{\bf k}$ being the total number of bosons.
This amounts to replacing $H_0$ and $\hat{H}_2$ in Eqs. (\ref{Eq:defH_i})
by  $\tilde{H}_0 = H_0 - \mu N_0$ and $\tilde{H}_2 = \hat{H}_2 - \mu \hat{N}_1$, respectively,
where $\hat{N}_1 = \sum_{\bf k\neq 0} a_{\bf k}^\dagger a_{\bf k}$ is the operator counting the 
total number of bosons outside the condensate. The ground state energy is then found by
diagonalizing $\tilde{H}_2$, which is done by introducing a new set of creation and annihilation operators, 
$\alpha^\dagger_{\bf k}$ and $\alpha_{\bf k}$, such that \cite{FW}:
\begin{equation}
\alpha_{\bf k} = u_{\bf k}a_{\bf k} + v_{\bf k}a^\dagger_{-\bf k},
\quad
\alpha_{\bf k}^\dagger = u_{\bf k}a_{\bf k}^\dagger + v_{\bf k}a_{-\bf k}.
\label{Eq:defalphas0}
\end{equation}
These expressions can easily be inverted, with the result:
\begin{equation}
a_{\bf k} = u_{\bf k}\alpha_{\bf k} - v_{\bf k} \alpha^\dagger_{-\bf k},
\quad
a^\dagger_{\bf k} = u_{\bf k}\alpha^\dagger_{\bf k} - v_{\bf k} \alpha_{-\bf k}.
\label{Eq:defalphas}
\end{equation}
In Eqs. (\ref{Eq:defalphas0}) and (\ref{Eq:defalphas}), $u_{\bf k}$ and $v_{\bf k}$ are assumed to be real, 
spherically symmetric functions of the wavevector ${\bf k}$. For the newly defined operators 
$\alpha_{\bf k}$ and $\alpha_{\bf k}^\dagger$ to describe boson
excitations, it is required that they should obey bosonic commutation relations:
\begin{align}
[\alpha_{\bf k},\alpha_{\bf k'}]=[\alpha_{\bf k}^\dagger,\alpha_{\bf k'}^\dagger]=0, 
\qquad 
[\alpha_{\bf k},\alpha_{\bf k'}^\dagger]=\delta_{\bf k,k'},
\end{align}
which results in the condition \cite{FW} 
\begin{equation}
u_{\bf k}^2 - v_{\bf k}^2 = 1.
\end{equation}
If we now use Eq. (\ref{Eq:defalphas}) to rewrite the Hamiltonian 
$\tilde{H}_2$ in terms of the operators $\alpha_{\bf k}$ and $\alpha^\dagger_{\bf k}$, and if 
in addition we require that the resulting expression not contain products 
of the form $\alpha_{\bf k}\alpha_{-\bf k}$ or $\alpha^\dagger_{\bf k}\alpha^\dagger_{-\bf k}$, 
we obtain the following expressions of the so-called ``coherence factors" 
$u_{\bf k}$ and $v_{\bf k}$:
\begin{equation}
u_{\bf k}^2 = \frac{1}{2}\Big(\frac{\xi_{\bf k} - \mu}{{E}_{\bf k}} + 1\Big), 
\quad 
v_{\bf k}^2 = \frac{1}{2}\Big(\frac{\xi_{\bf k} -\mu}{{E}_{\bf k}} - 1\Big),
\label{Eq:uv}
\end{equation}
where ${E}_{\bf k}$ is the energy spectrum:
\begin{equation}
{E}_{\bf k} = \sqrt{(\xi_{\bf k} - \mu)^2 - n_0^2v({\bf k})^2}.
\label{Eq:defEk}
\end{equation}
The Hamiltonian $\tilde{H}_2$ then takes the quadratic form:
\begin{equation}
\tilde{H}_2 = \tilde{E}_2 +\sum_{\bf k\neq 0} {E}_{\bf k}\alpha_{\bf k}^\dagger\alpha_{\bf k},
\label{Eq:diagtildeH2}
\end{equation}
where we denote by $\tilde{E}_2$ the following quantity:
\begin{equation} 
\tilde{E}_2 = \frac{1}{2}\sum_{\bf k\neq 0} \big[ {E}_{\bf k} - 
\varepsilon_{\bf k} - n_0v({\bf k})\big].
\label{Eq:H2_2}
\end{equation}
Requiring the spectrum $E_{\bf k}$ of Eq. (\ref{Eq:defEk}) to be gapless\cite{Hugenholtz1959} 
as $k\to 0$ results in the following expression of the chemical potential:
\begin{equation}
\mu = n_0 v({\bf 0}),
\label{Eq:mu_0}
\end{equation}
upon which ${E}_{\bf k}$ takes the celebrated Bogoliubov form:
\begin{equation}
{E}_{\bf k} = \sqrt{\varepsilon_{\bf k}\big[\varepsilon_{\bf k} + 2 n_0 v({\bf k})\big]}.
\label{Eq:resEk}
\end{equation}
Using Eq. (\ref{Eq:mu_0}) into Eq. (\ref{Eq:uv}), one finds:
\begin{equation}
u_{\bf k}^2 = \frac{1}{2}\Big(\frac{\varepsilon_{\bf k} + n_0v({\bf k})}{{E}_{\bf k}} + 1\Big), 
\quad 
v_{\bf k}^2 = \frac{1}{2}\Big(\frac{\varepsilon_{\bf k} + n_0v({\bf k})}{{E}_{\bf k}} - 1\Big).
\label{Eq:uv_2}
\end{equation}
The Bogoliubov ground state (BGS) of the interacting bose gas, 
which we shall denote by the symbol 
$|\Psi_B\rangle$, is defined as the vacuum state for the $\alpha_{\bf k}$ operators, 
and satisfies the condition:
\begin{equation}
\alpha_{\bf k} |\Psi_B\rangle = 0, \quad \mbox{for all } {\bf k} \neq 0.
\label{Eq:defGS}
\end{equation}
With this definition, it is easy to verify that $\tilde{E}_2$ is the expectation value of 
the Hamiltonian $\tilde{H}_2$ in the BGS, i.e. $\tilde{E}_2 = \langle\Psi_B|\tilde{H}_2|\Psi_B\rangle$.

\addvspace{1.5mm}

We now pause for a moment to emphasize the distinction which is usually done in Bogoliubov's theory 
between the operators $a_{\bf k}^\dagger$ and $a_{\bf k}$ on one hand, and the operators 
$\alpha_{\bf k}^\dagger$ and $\alpha_{\bf k}$ on the other. For while the former create and 
annihilate actual bosons, the latter are thought to create and annihilate ``quasiparticles", 
which are identified as collective ``phonon" modes, mainly because the corresponding excitation 
spectrum ${E}_{\bf k}$ has a linear behavior at long wavelengths: 
\begin{equation}
{E}_{\bf k}\sim s\hbar k,
\label{Eq:SpeedSound1}
\end{equation} 
where 
\begin{equation}
s=\sqrt{n_Bv({\bf 0})/m} 
\label{Eq:SpeedSound2}
\end{equation}
is identified as the speed of sound (here and in the rest of this paper,
$n_B=N/V$ is the density of bosons). We will come back to this identification in Sec. \ref{Sec:GoldstoneModes} below. 
For the moment, we want to have a more detailed look at a number of problematic features of Bogoliubov's theory. 
These being more or less known features of this method, our main objective here 
is to provide a motivation for the variational formulations of Bogoliubov's theory that will be studied 
in Secs. \ref{Sec:VariationalMethod} through \ref{Sec:DiagFullH}. This will be the subject of the following Section.

\section{Issues with Bogoliubov's theory}
\label{Sec:Issues}

As mentioned above, in this Section we want to discuss 
a few questionable aspects of the number non-conserving formulation
of Bogoliubov's method as a way to motivate the variational treatment that will
be presented in Secs. \ref{Sec:VariationalMethod} and \ref{Sec:DiagFullH}.
We will start by looking at the ground state expectation values
of quadratic combinations of creation and annihilation operators.

\subsection{Divergence of the depletion $N_{\bf k}=\langle a_{\bf k}^\dagger a_{\bf k}\rangle$
and of the anomalous average $\langle a_{\bf k} a_{-\bf k}\rangle$ in the $k\to 0$ limit}
\label{Sub:depletionBog}

The first quantity we will discuss is the expectation value:
\begin{equation}
N_{\bf k} = \langle a_{\bf k}^\dagger a_{\bf k}\rangle 
=\langle \Psi_B|a_{\bf k}^\dagger a_{\bf k}|\Psi_B\rangle,
\end{equation}
which gives the average number of bosons in the single-particle state of momentum ${\bf k}$.
In the standard formulation of Bogoliubov's theory, this quantity is given by:
\begin{equation}
N_{\bf k} = v_{\bf k}^2 = \frac{1}{2}\Big(\frac{\varepsilon_{\bf k} + n_0v({\bf k})}{E_{\bf k}} - 1\Big).
\label{Eq:resNkBog}
\end{equation}
Using the expression of $E_{\bf k}$ given in Eq. (\ref{Eq:resEk}), it is not difficult to see that $N_{\bf k}$
diverges like $1/k$ as $k\to 0$, which is of course unphysical, 
given that we started from a system of fixed number $N$ of bosons. 

\addvspace{1.5mm}

The divergence of $N_{\bf k}$ as $k\to 0$ will have important ramifications, 
as it will affect the expectation value of one-body operators. Indeed,  
if we consider a given physical observable which is described by a one-body operator
$\hat{O}$ of the form $\hat{O}=\sum_{\bf k}O_{\bf k}a_{\bf k}^\dagger a_{\bf k}$, then the expectation
value of $\hat{O}$ in the Bogoliubov ground state is given by:
\begin{equation}
\langle\hat{O}\rangle = \sum_{\bf k} O_{\bf k} N_{\bf k}.
\label{Eq:avgOpO}
\end{equation}
Since the result (\ref{Eq:resNkBog}) for $N_{\bf k}$
is unphysical near $k=0$, we see that the result (\ref{Eq:avgOpO}) for the expectation value
$\langle\hat{O}\rangle$ includes terms whose contribution is unphysical near $k=0$ as well. 
As an example, let us consider the Fock part of the Hamiltonian
$\hat{H}_F=\sum_{\bf k\neq 0}n_0v({\bf k})a_{\bf k}^\dagger a_{\bf k}$. The contribution of any particular ${\bf k}$ mode in
this last expression to the ground state energy is given by $n_0v({\bf k})N_{\bf k}$, which diverges like
$1/k$ as $k\to 0$. This divergence cannot of course be taken at face value, because it originates in the unphysical
divergence of the quantity $N_{\bf k}$, which should be
finite and bounded by the total number $N$ 
of bosons in the system for all values of the wavevector $k$.

\addvspace{1.5mm}

A closely related problematic feature of Bogoliubov's theory
has to do with the divergence of the anomalous averages 
$\langle a_{\bf k}a_{-\bf k} \rangle=\langle \Psi_B|a_{\bf k}a_{-\bf k} |\Psi_B\rangle$ 
and 
$\langle a_{\bf k}^\dagger a_{-\bf k}^\dagger \rangle
=\langle\Psi_B| a_{\bf k}^\dagger a_{-\bf k}^\dagger|\Psi_B \rangle$ 
in the $k\to 0$ limit.
Following an argument made by de Gennes for the BCS ground state of superconductors,\cite{deGennes}
one can argue that an average such as $\langle a_{\bf k}a_{-\bf k}\rangle$ can be given a sense, 
from a number-conserving perspective, if it is understood as: 
\begin{equation}
\langle a_{\bf k}a_{-\bf k} \rangle = \langle \Psi(N-2) | a_{\bf k}a_{-\bf k}|\Psi(N)\rangle,
\label{Eq:aa_interpretation}
\end{equation}
where we denote by $|\Psi(N)\rangle$ the {\em normalized} ground state wavefunction of a system of 
$N$ bosons. Written in this form, $\langle a_{\bf k}a_{-\bf k} \rangle$ can be interpreted \cite{deGennes}
as the probability amplitude for finding the system in the ground state $|\Psi(N-2)\rangle$
when a pair of bosons $({\bf k},-{\bf k})$ is removed from the ground state $|\Psi(N)\rangle$.
Following this line of thought, if we think of $\langle a_{\bf k}a_{-\bf k} \rangle$ 
as a probability amplitude, then it should be a bounded quantity, and
should not diverge for any value of the wavevector ${\bf k}$.
Unfortunately, in the standard Bogoliubov theory, the expression
of the anomalous average is given by: 
\begin{equation}
\langle a_{\bf k}a_{-\bf k} \rangle = \langle a_{\bf k}^{\dagger}a_{-\bf k}^\dagger\rangle 
= -\frac{n_0v(k)}{2E_{\bf k}}  \longrightarrow -\infty \mbox{ as } k\to 0,
\label{Eq:an_avg_Bog}
\end{equation}
and diverges (negatively) as $k\to 0$, which is of course incompatible
with an interpretation of $\langle a_{\bf k}^{\dagger}a_{-\bf k}^\dagger\rangle$ 
in terms of probabilities. The divergence of the anomalous average $\langle a_{\bf k}a_{-\bf k}\rangle$
will also have major consequences on the evaluation of expectation values of quantities involving
quadratic products of operators of the form $a_{\bf k}a_{-\bf k}$ or $a_{\bf k}^\dagger a_{-\bf k}^\dagger$.
Again, as an example, if we consider the so-called ``pairing" part
of the Bogoliubov Hamiltonian, 
$\hat{H}_P=\frac{1}{2}\sum_{\bf k\neq 0}n_0v({\bf k})(a_{\bf k}^\dagger a_{-\bf k}^\dagger
+ a_{\bf k}a_{-\bf k})$, then the contribution of any particular term
in $\hat{H}_P$ corresponding to a given wavevector $\bf k$ 
to the ground state energy is given by $-\big[n_0v({\bf k})\big]^2/2E_{\bf k}$, 
and again diverges like $-1/k$ as $k\to 0$.
This divergence is again not to be taken too seriously, since it originates solely from the unphysical divergence
of the anomalous average $\langle a_{\bf k}a_{-\bf k} \rangle$ in Eq. (\ref{Eq:an_avg_Bog}).
We will look at these divergences in more detail below (see Sec. \ref{Sec:VariationalMethod}).
For the moment, we want to have a closer look at the physical meaning of the operator $\hat{H}$
in Bogoliubov's method.

\subsection{Physical meaning of the operator $\hat{H}$ in Bogoliubov's method}
\label{Sub:meaningH}

We now want to discuss the physical meaning of the Hamiltonian $\hat{H}$ in the standard
formulation of Bogoliubov's method in view of the fact that the ground state energy
of the system is the expectation value of the ``grand-canonical" Hamiltonian $\tilde{H}=\hat{H}-\mu \hat{N}$,
and not of the Hamiltonian $\hat{H}$ itself. As an aside, we shall observe that the use of the
term ``grand-canonical Hamiltonian" to describe the operator $\tilde{H}$ is quite misleading,
given that the term ``grand-canonical" has a very specific meaning in ensemble theory of statistical mechanics, and
refers to a very specific statistical ensemble which only has a meaning at nonzero temperatures.
We also would like to remind the reader that, in ensemble theory, various operators have the same
physical meaning in the grand-canonical ensemble as in any other
statistical ensemble. In particular, the operator describing the total energy
in the grand-canonical ensemble is the Hamiltonian $\hat{H}$, {\em not} the combination
$\hat{H} - \mu\hat{N}$, which in fact describes the grand-potential at $T=0$. 
It is therefore extremely surprizing that in Bogoliubov's theory the ground state energy is
found by taking the expectation value of $\tilde{H}=\hat{H}-\mu\hat{N}$ instead of the expectation
value of the Hamiltonian $\hat{H}$ itself.\cite{CommentMeaningH} 

\addvspace{1.5mm}

In what follows, we want to go further and show that the use of the Hamiltonian 
$\hat{H}$ instead of $\tilde{H}$ can even lead,
in certain situations, to nonsensical results,
which will allow us to argue that the Hamiltonian $\hat{H}$ has no physical meaning
in the standard formulation of Bogoliubov's theory.
To this end, let us consider states that are generated by successive action of the ``quasiparticle" operator
$\alpha_{\bf k}^\dagger$ on the BGS. It is easy to verify that these states 
are eigenstates of the ``grand-canonical" Hamiltonian $\tilde{H}_2$. For example, one can show that 
the normalized state defined by:
\begin{equation}
|\Phi_{\bf k}^{(n)}\rangle = \frac{\big(\alpha_{\bf k}^\dagger\big)^n}{\sqrt{n!}}|\Psi_B\rangle,
\label{Eq:quasiparticles_n}
\end{equation}
and corresponding to the excitation of $n$ ``quasiparticles" of wavevector ${\bf k}$
is an eigenstate of $\tilde{H}_2$ with eigenvalue $\big[\tilde{E}_2 + n{E}_{\bf k}\big]$:
\begin{equation}
\tilde{H}_2|\Phi_{\bf k}^{(n)}\rangle = \big[\tilde{E}_2 + n{E}_{\bf k}\big]|\Phi_{\bf k}^{(n)}\rangle.
\end{equation}
Similarly, a state of the form:
\begin{equation}
|\Phi_{{\bf k}_1,\cdots, {\bf k}_n}\rangle = 
\alpha_{{\bf k}_1}^\dagger\cdots\alpha_{{\bf k}_n}^\dagger|\Psi_B\rangle,
\end{equation}
and corresponding to the excitation of $n$ ``quasiparticles" with distinct wavevectors 
${\bf k}_1, {\bf k}_2,\cdots, {\bf k}_n$, is an eigenstate of the grand canonical Hamiltonian 
$\tilde{H}_2$ with eigenvalue $\big[\tilde{E}_2 +{E}_{\bf k_1} + \cdots + {E}_{\bf k_n}\big]$:
\begin{equation}
\tilde{H}_2|\Phi_{{\bf k}_1,\cdots,{\bf k}_n}\rangle = \big[\tilde{E}_2 + {E}_{{\bf k}_1} 
+ \cdots + {E}_{{\bf k}_n}\big]|\Phi_{{\bf k}_1,\cdots,{\bf k}_n}\rangle.
\end{equation}
Let us focus for simplicity on the state with one ``quasiparticle" of momentum ${\bf k}$. This is the 
state given by:
\begin{equation}
|\Phi_{\bf k}\rangle = \alpha_{\bf k}^\dagger|\Psi_B\rangle.
\label{Eq:quasiparticles_1}
\end{equation}
Such a state is believed to have an excitation energy which is gapless in the $k\to 0$ limit. 
This belief originates in Eq. (\ref{Eq:diagtildeH2}), where ${E}_{\bf k}$ is identified 
as the excitation energy of the $\alpha_{\bf k}$ ``quasiparticles". Indeed, if we calculate the 
excess free energy with respect to the Bogoliubov ground state of the state $|\Phi_{\bf k}\rangle$
using the ``grand canonical" Hamiltonian $\tilde{H}_2$, we find:
\begin{equation}
\langle \Phi_{\bf k}|[H_0 + \tilde{H}_2]|\Phi_{\bf k}\rangle 
- \langle \Psi_B|[H_0 + \tilde{H}_2]|\Psi_B\rangle = {E}_{\bf k}.
\label{Eq:excitationtildeH2}
\end{equation}
At this stage, it is useful to remember that the true quadratic part of the original Hamiltonian $\hat{H}$
is $\hat{H}_2$, not $\tilde{H}_2$. Consequently, in evaluating the excitation {\em energy} 
(not the excitation grand potential!) $\Delta{E}_{\bf k}^{ex,\alpha}$ of a ``quasiparticle" $\alpha_{\bf k}^\dagger$,
one should use the Hamiltonian $\hat{H}_2$ instead of $\tilde{H}_2$, and define $\Delta{E}_{\bf k}^{ex,\alpha}$ 
as follows (we here use the superscript $\alpha$ to indicate that 
we are calculating the excitation spectrum of the $\alpha_{\bf k}^\dagger$ ``quasiparticles"):
\begin{equation}
\Delta{E}_{\bf k}^{ex,\alpha} =\langle \Phi_{\bf k}|[H_0 + \hat{H}_2]|\Phi_{\bf k}\rangle 
- \langle \Psi_B|[H_0 + \hat{H}_2]|\Psi_B\rangle.
\label{Eq:excitationH2}
\end{equation}
Performing the calculation, we find:
\begin{subequations}
\begin{align}
\Delta{E}_{\bf k}^{ex,\alpha} & = {E}_{\bf k} + n_0v({\bf 0})[u_{\bf k}^2 + v_{\bf k}^2],
\label{Eq:res_excitationH2a}
\\
& = {E}_{\bf k} + n_0v({\bf 0})\frac{\varepsilon_{\bf k} + n_0v({\bf k})}
{\sqrt{\varepsilon_{\bf k}[\varepsilon_{\bf k} + 2n_0v({\bf k})]}}.
\label{Eq:res_excitationH2b}
\end{align}
\label{Eq:res_excitationH2}
\end{subequations}
As it can be seen, the energy cost to excite a ``quasiparticle" $\alpha_{\bf k}^\dagger$, when calculated using $\hat{H}_2$, is no longer
given by the gapless expression ${E}_{\bf k}$. There is now an additional term, which furthermore diverges like $1/k$ as $k\to 0$.
We are therefore faced with the bizarre situation where the operator $\hat{H}_2$ 
has {\em no} physical meaning of its own whatsoever, and where
the energy observable rather mysteriously is no longer represented by this last operator, but
by the combination $\tilde{H}_2 = \hat{H}_2 - \mu\hat{N}_1$, which, in principle,
represents the grand-potential of the system at zero temperature. 
Because of the non-conservation of particle number,
the use of the true energy operator $\hat{H}_2$ in the formulation of Bogoliubov's theory
given in Sec. \ref{Sec:StandardBogoliubov} can lead to non-sensical results, as we have seen above 
with the excitation spectrum of the $\alpha_{\bf k}$ ``quasiparticles".

\subsection{{\em Ad hoc} nature of the canonical transformation from the $a_{\bf k}$'s to the $\alpha_{\bf k}$'s}

We now turn our attention to another questionable aspect of Bogoliubov's method, having to do
with the physical content of the canonical transformation from the $a_{\bf k}$'s to the $\alpha_{\bf k}$'s,
and the rather {\em ad hoc} fashion in which the commutation relations between the 
excitation operators $\alpha_{\bf k}$ and $\alpha_{\bf k}^\dagger$ are introduced.
Taking note of the fact that these operators, in the standard formulation of Bogoliubov's theory,
do not conserve the number of bosons, below we will show that, in a number-conserving
approach, the canonical commutation relations between the $\alpha_{\bf k}$'s 
are not at all exact, but only approximate, and are only valid
if the depletion of the ground state is small. For situations where this condition is violated,
such as for liquid Helium at low temperatures, where the number of bosons in the condensate is only about
$10\%$ of the total number of bosons in the system, the canonical commutation relations
are no longer valid, and should be replaced by more involved expressions. 
For more details, the reader is referred to Sec. \ref{Sec:VariationalMethod} where the commutation
relations between the $\alpha_{\bf k}$'s will be obtained from first principles, and not imposed {\em a priori}, 
within a number-conserving variational approximation.

\subsection{Diagonalizing the various Hamiltonians $\hat{H}_{\bf k}$ corresponding to different momentum modes
independently from one another}

Perhaps one of the less obvious criticisms one can make of Bogoliubov's method, and possibly one
with the most far-ranging ramifications, has to do with the fact that, 
in this theory, the total ``grand-canonical" Hamiltonian $\tilde{H} = \hat{H} - \mu\hat{N}$
is not diagonalized as a whole, but rather, each momentum 
contribution $\tilde{H}_{\bf k}$ to $\tilde{H}$ is diagonalized separately and independently from 
the other momentum contributions. One way to see this is by writing 
the total Hamiltonian $\hat{H}$ as a sum of contributions from different wavevectors ${\bf k}$
of the form $\hat{H} = \sum_{\bf k\neq 0} \hat{H}_{\bf k}$. Then it is easy to convince oneself
that the canonical transformation method of Sec. \ref{Sec:StandardBogoliubov} is in fact
a diagonalization of each one of the Hamiltonians $\hat{H}_{\bf k}$ {\em independently} from 
the other Hamiltonians $\hat{H}_{\bf k' (\neq k)}$.
This process of diagonalizing each of the single-mode Hamiltonians $\tilde{H}_{\bf k}$ separately
would be absolutely sound if the Hilbert spaces in which these Hamiltonians are diagonalized
were totally disjoint from one another. This is unfortunately not the case here, with all these
Hilbert spaces having the ${\bf k}=0$ state in common. Of course, from the perspective of Bogoliubov's method, 
where the dynamic properties of the condensate are ignored and the ${\bf k}=0$ state is summarily removed
from the Hilbert space, this is thought to be a perfectly legitimate way to proceed. However, in this paper, we shall
challenge this view, and explicitly show that, when the conservation of particle number is properly taken into
account, and the ${\bf k}=0$ is restored as an {\em essential} part of the Hilbert space used to describe the system,
then the direct diagonalization of the full Hamiltonian $\hat{H}$ gives very different results from
diagonalizing each of the momentum components $\hat{H}_{\bf k}$ separately, as is done
in the standard Bogoliubov formulation.

\addvspace{1.5mm}

Since the non-conservation of particle number in the standard formulation of Bogoliubov's theory 
is at the core of most of the conceptual difficulties discussed throughout this Section,
it seems worthwhile to try to eliminate these difficulties by reformulating 
the theory within a number-conserving framework. 
This will be done in Sections \ref{Sec:VariationalMethod} through \ref{Sec:DiagFullH}. But, before we do
so, we want to present a rather detailed derivation of Bogoliubov's Hamiltonian in a number
conserving approach. This will be done next.

\section{Derivation of Bogoliubov's Hamiltonian within a number-conserving approach}
\label{Sec:DerivationH_Bog}

We now want to derive, within a number-conserving framework, 
a simplified version of the Hamiltonian (\ref{Eq:HFourierSpace})
which is simple enough to allow for a straightforward evaluation of expectation values of physical observables, and
at the same time captures the essential physics of  a translationally invariant 
system of interacting bosons at zero temperature.
Throughout this paper, it will be assumed, without loss of generality, that the bosons are confined in a cubic
box of size $L$, and we shall use periodic boundary conditions, implying that momenta will be
quantized with the following wavevectors:
\begin{equation}
{\bf k} = \frac{2\pi}{L}(n_x,n_y,n_z),
\end{equation}
where the $n_i$'s are integers such that $-\infty\leq n_x,\, n_y,\, n_z \leq \infty$.

We shall start by specifying the Hilbert space we will use to describe our system. 
In the occupation number representation, a system of $N$ bosons
can in general be in any one of the states
\begin{align}
|\psi_{n_1\cdots n_\infty}^{m_1\cdots m_\infty}\rangle = |N-\sum_{i=1}^\infty (n_i+m_i);n_1,m_1;
\cdots;n_\infty, m_\infty\rangle,
\label{Eq:generalket}
\end{align}
having $n_i$ bosons in the single-particle state of wavevector
${\bf k}_i$ and $m_i$ bosons in the single-particle state of wavevector $-{\bf k}_i$ (for all
$i$ such that $1\leq i\leq \infty$), and
$\big[N-\sum_{i=1}^\infty(n_i+m_i)\big]$ bosons in the single-particle state with wavevector ${\bf k}=0$. 
More formally, if we denote by $|0\rangle$ the vacuum state for bosons, then 
$|\psi_{n_1\cdots n_\infty}^{m_1\cdots m_\infty}\rangle$ is the ket defined by:
\begin{align}
|\psi_{n_1\cdots n_\infty}^{m_1\cdots m_\infty}\rangle = 
\frac{\big(a_0^\dagger\big)^{N-\sum_{i=1}^\infty(n_i+m_i)}}{\sqrt{[N-\sum_{i=1}^\infty(n_i+m_i)]!} }
\prod_{i=1}^\infty\frac{\big(a_{{\bf k}_i}^\dagger\big)^{n_i}}{\sqrt{n_i!}}
\frac{\big(a_{-{\bf k}_i}^\dagger\big)^{m_i}}{\sqrt{m_i!}}
|0\rangle.
\end{align}
It then follows that the most general expression of the wavefunction $|\Psi(N)\rangle$ 
of a system of $N$ bosons in the occupation number representation is given by:
\begin{align}
|\Psi(N)\rangle & = \sum_{n_1,m_1}\cdots\sum_{n_\infty,m_\infty} 
C_{n_1,\ldots,n_\infty}^{m_1,\ldots,m_\infty}|\psi_{n_1\cdots n_\infty}^{m_1\cdots m_\infty}\rangle.
\label{Eq:PsiGeneralForm}
\end{align}
In the above equation, the $C_{n_1,\ldots,n_\infty}^{m_1,\ldots,m_\infty}$'s
are complex numbers to be determined by diagonalization of the Hamiltonian $\hat{H}$
of the system. The summations extend over the range $0 \leq n_i, m_i \leq N$,
subject to the constraint that the number of bosons in the state ${\bf k}={\bf 0}$ in any ket
of the basis must be positive:
\begin{equation}
N - (n_1+m_1) \cdots - (n_\infty + m_\infty) \geq 0.
\end{equation}
Now, it can be easily verified that the total momentum operator 
$\hat{\bf P}=\sum_{\bf k} \hbar{\bf k}a_{\bf k}^\dagger a_{\bf k}$
commutes with the Hamiltonian $\hat{H}$. This implies that $\hat{H}$
and $\hat{\bf P}$ can be diagonalized simultaneously, and that eigenstates
of $\hat{H}$ can be labeled by a definite value of the total momentum
operator $\hat{\bf P}$. In particular, for a system of bosons at rest,
the ground state is the translationally
invariant state corresponding to the eigenvalue $P=0$ of the total momentum:\cite{FW}
\begin{equation}
\hat{\bf P}|\Psi(N)\rangle = 0.
\label{Eq:ConstraintMomentum}
\end{equation}
This condition imposes a rather cumbersome constraint on the coefficients 
$C_{n_1,\ldots,n_\infty}^{m_1,\ldots,m_\infty}$, namely:
\begin{align}
\sum_{n_1,m_1}\cdots\sum_{n_\infty,m_\infty} 
C_{n_1,\ldots,n_\infty}^{m_1,\ldots,m_\infty}\big[\hbar{\bf k}_1(n_1-m_1) +\cdots
+ \hbar{\bf k}_\infty(n_\infty-m_\infty)
\big]|\psi_{n_1,\cdots, n_\infty}^{m_1,\cdots,m_\infty}\rangle = 0.
\end{align}
The easiest way to satisfy the above constraint is to restrict the Hilbert space
to one where there are as many bosons with momentum ${\bf k}_i$ as there are bosons 
with momentum $-{\bf k}_i$, {\em i.e.} we choose to work in the restricted Hilbert space such that $n_i=m_i$,
for all values of the index $i$. Eq. (\ref{Eq:PsiGeneralForm}) then becomes: 
\begin{align}
|\Psi(N)\rangle = \sum_{n_1}\cdots\sum_{n_\infty} C_{n_1,\ldots,n_\infty}
|\psi_{n_1,\cdots,n_\infty}\rangle
\label{Eq:PsiGeneralForm2}
\end{align}
where now $|\psi_{n_1,\cdots,n_\infty}\rangle$ is given by:
\begin{subequations}
\begin{align}
|\psi_{n_1,\cdots,n_\infty}\rangle & = |N-2n_1 \cdots - 2n_\infty;
n_1, n_1;\ldots;n_\infty,n_\infty\rangle,
\\
& = \frac{\big(a_0^\dagger\big)^{N-2\sum_{i=1}^\infty n_i}}{\sqrt{[N-2\sum_{i=1}^\infty n_i]!} }
\prod_{i=1}^\infty\frac{\big(a_{{\bf k}_i}^\dagger\big)^{n_i}}{\sqrt{n_i!}}
\frac{\big(a_{-{\bf k}_i}^\dagger\big)^{n_i}}{\sqrt{n_i!}}
|0\rangle.
\end{align}
\label{Eq:PsiGeneralForm3}
\end{subequations}

We now want to isolate those terms in the Hamiltonian which will give significant contributions to the ground state
energy in states of the form (\ref{Eq:PsiGeneralForm2}). For definiteness, let us rewrite the 
interaction part $\hat{V}$ of the many-body Hamiltonian, which is given by:
\begin{equation}
\hat{V} = \frac{1}{2V}\sum_{\bf k,k'}\sum_{\bf q} v({\bf q}) a_{\bf k + q}^\dagger a_{\bf k' - q}^\dagger
a_{\bf k'}a_{\bf k}.
\end{equation}
In the above expression, we shall first separate the Hartree terms, corresponding to 
${\bf k}+{\bf q}={\bf k}$ and ${\bf k}'-{\bf q}={\bf k}'$, {\em i.e.} to ${\bf q}=0$;
and the Fock (or exchange) terms, which correspond to 
${\bf k}+{\bf q}={\bf k}'$ and ${\bf k}'-{\bf q}={\bf k}$, {\em i.e.} to ${\bf q}={\bf k}'-{\bf k}$.
Hence, we shall write the interaction term $\hat{V}$ in the form:
\begin{equation}
\hat{V} = \hat{V}_H + \hat{V}_F + 
\frac{1}{2V}\sum_{\bf k,k'}\sum_{\bf q\neq 0,{\bf k}'-{\bf k}} v({\bf q}) 
a_{\bf k + q}^\dagger a_{\bf k' - q}^\dagger a_{\bf k'}a_{\bf k},
\label{Eq:DecompositionV}
\end{equation}
where the Hartree and Fock contributions, $\hat{V}_H$ and $\hat{V}_F$ respectively, are given by:
\begin{subequations}
\begin{align}
\hat{V}_H & = \frac{v({\bf 0})}{2V}\sum_{\bf k,\bf k'} a_{\bf k}^\dagger a_{\bf k'}^\dagger
a_{\bf k'}a_{\bf k},
\\
& = \frac{v({\bf 0})}{2V}\hat{N}(\hat{N}-1),
\\
\hat{V}_F & = \frac{1}{2V}\sum_{\bf k}\sum_{\bf k'(\neq k)} v({\bf k}'-{\bf k}) 
a_{\bf k'}^\dagger a_{\bf k}^\dagger a_{\bf k'} a_{\bf k}.
\end{align}
\end{subequations}
We next consider the so-called ``pairing" terms. These are obtained by letting ${\bf k}'=-{\bf k}$
both in $\hat{V}_F$ and in the last term of Eq. (\ref{Eq:DecompositionV}). This allows us to write:
\begin{align}
\hat{V} = \hat{V}_H + \hat{V}_F' + \hat{V}_P
+\frac{1}{2V}\sum_{\bf k}\sum_{\bf k'(\neq -k)}\sum_{\bf q\neq 0,{\bf k}'-{\bf k}} v({\bf q}) 
a_{\bf k + q}^\dagger a_{\bf k' - q}^\dagger a_{\bf k'}a_{\bf k},
\label{Eq:DecompositionV2}
\end{align}
where we defined:
\begin{subequations}
\begin{align}
&\hat{V}_F' = \frac{1}{2V}\sum_{\bf k}\sum_{\bf k'(\neq k,-k)} v({\bf k}'-{\bf k}) 
a_{\bf k'}^\dagger a_{\bf k}^\dagger a_{\bf k'} a_{\bf k},
\label{Eq:V_F'}
\\
&\hat{V}_P = \frac{1}{2V}\sum_{\bf k}\sum_{\bf q\neq 0} v({\bf q}) 
a_{\bf k + q}^\dagger a_{-\bf (k+q)}^\dagger a_{\bf k} a_{-\bf k}.
\end{align}
\end{subequations}
It can be shown that the last term on the {\em rhs} of Eq. (\ref{Eq:DecompositionV2}) 
has zero expectation value in the state $|\Psi(N)\rangle$ of Eq. (\ref{Eq:PsiGeneralForm2}), and 
hence it shall be discarded. Moreover, in keeping with the spirit of Bogoliubov's method,
both in $\hat{V}_F'$ and $\hat{V}_P$ we shall only retain terms
in which either ${\bf k}$ or ${\bf k}'$ is zero, an approximation which is believed to be valid
if the depletion of the condensate is small.\cite{Leggett2001} Hence, we shall write:
\begin{equation}
\hat{V} \equiv \hat{V}_H + \hat{V}_F' + \hat{V}_P,
\end{equation}
with:
\begin{subequations}
\begin{align}
& \hat{V}_F'\simeq \frac{1}{V}\sum_{\bf k\neq 0} v({\bf k}) a_0^\dagger a_0 a_{\bf k}^\dagger a_{\bf k}, 
\label{Eq:V_F'0}
\\
& \hat{V}_P \simeq \frac{1}{2V} \sum_{\bf k\neq 0} v({\bf k})\big[
a_0 a_0 a_{\bf k}^\dagger a_{-\bf k}^\dagger + a_0^\dagger a_0^\dagger a_{\bf k} a_{-\bf k}
\big].
\end{align}
\end{subequations}
Now, if we take the origin of energies at the Gross-Pitaevskii value $v(0)N(N-1)/2V$
(which, it should be noted, is an exact eigenvalue of the operator $\hat{V}_H$ in
the state $|\Psi(N)\rangle$ of Eq. (\ref{Eq:PsiGeneralForm2})), then
the Hamiltonian can be written as a sum of independent contributions from different values
of ${\bf k}$:
\begin{align}
\hat{H} \simeq \sum_{\bf k \neq 0} \hat{H}_{\bf k},
\label{Eq:HsumHk}
\end{align}
where:
\begin{align}
\hat{H}_{\bf k} & = \frac{1}{2}\varepsilon_{\bf k}
\big( a_{\bf k}^\dagger a_{\bf k} + a_{-\bf k}^\dagger a_{-\bf k}\big) 
\nonumber\\
&+ \frac{v({\bf k})}{2V}\,\big( a_0^\dagger a_0 a_{\bf k}^\dagger a_{\bf k} 
+a_0^\dagger a_0 a_{-\bf k}^\dagger a_{-\bf k}
+ a_{\bf k}^\dagger a_{-\bf k}^\dagger a_0 a_0
+ a_0^\dagger a_0^\dagger a_{\bf k} a_{-\bf k}\big).
\label{Eq:defHk}
\end{align}
In the next Section, we shall restrict our attention to the single-mode 
Hamiltonian $\hat{H}_{\bf k}$. We shall see that the ground state of $\hat{H}_{\bf k}$
can be found quite easily using a number-conserving variational ansatz,\cite{Leggett2001}
leading to results for the ground state energy and excitation spectrum
that are very similar to those of the standard (number non-conserving) Bogoliubov
theory. The implications of the diagonalization of $\hat{H}_{\bf k}$
on the diagonalization of the full many-body Boson Hamiltonian $\hat{H}$
will be discussed in more detail in Sec. \ref{Sec:DiagFullH}.

\section{Number-conserving approach: variational formulation of Bogoliubov's theory}
\label{Sec:VariationalMethod}

In view of the conceptual difficulties of the number non-conserving formulation of Bogoliubov's 
theory reviewed in Section \ref{Sec:Issues}, we now want to study an alternative formulation of this theory in which 
the number of bosons is conserved, and which therefore should be free of many of the problems encountered in the
number non-conserving version. This formulation, which has already been presented in slightly different form
in the literature,\cite{LeeHuangYang1957,Leggett2001}
consists in using a variational ansatz to find the ground state wavefunction
of each single-mode Hamiltonian $\hat{H}_{\bf k}$ independently 
from the Hamiltonians $\hat{H}_{\bf k' (\neq k)}$
of the other momentum modes and writing the ground state wavefunction of the total
Hamiltonian $\hat{H}$ as a tensor product of all these separate single-mode ground states. 
However, since the previous expositions of this method are not widely in use, 
we here shall present a comprehensive review, which will allow us to assess 
the validity of several aspects of the standard, number non-conserving 
Bogoliubov formulation.

\subsection{Ground state energy of the single-mode Hamiltonian $\hat{H}_{\bf k}$}
\label{Sub:GSHkBog}

We now proceed to diagonalize the Hamiltonian $\hat{H}_{\bf k}$ {\em independently} of the remaining
contributions $\hat{H}_{\bf k'(\neq k)}$ to the total Hamiltonian $\hat{H}$ of the system; 
in other words, we consider a fictitious system
where bosons are only allowed to be in one of the three single particle 
states with wavevector ${\bf 0}$, $+{\bf k}$ or $-\bf k$.
We shall therefore restrict our attention to trial states of the form (without loss
of generality, throughout this paper it will be assumed that the total
number of bosons $N$ is even):
\begin{equation}
|\psi_{\bf k}\rangle = \sum_{n=0}^{N/2} C_n|N-2n,n,n\rangle,
\label{Eq:defpsik}
\end{equation}
where each basis state $|N-2n,n,n\rangle$ has $(N-2n)$ bosons in the condensate, 
and $n$ bosons in each one of the two momentum states $\pm\bf k$.
To clarify what we mean exactly by the notation $|N-2n,n,n\rangle$ in the present context, 
if $|0\rangle$ designates the vacuum state for bosons,
the normalized general state $|N-n-m,n,m\rangle$ can be defined as:
\begin{equation}
|N-n-m,n,m\rangle = \frac{(a_0^\dagger)^{N-n-m}}{\sqrt{(N-n-m)!}}\frac{(a_{\bf k}^\dagger)^n}{\sqrt{n!}}
\frac{(a_{-\bf k}^\dagger)^m}{\sqrt{m!}}|0\rangle.
\end{equation}

The expectation value of $\hat{H}_{\bf k}$ in the un-normalized state $|\psi_{\bf k}\rangle$ of
Eq. (\ref{Eq:defpsik}) is given by:
\begin{equation} 
\langle \hat{H}_{\bf k}\rangle_{\bf k}=\frac{\langle\psi_{\bf k}|\hat{H}_{\bf k}|\psi_{\bf k}\rangle}
{\langle\psi_{\bf k}|\psi_{\bf k}\rangle}.
\label{Eq:Avg_h}
\end{equation}
Let us calculate the above expectation value. To simplify the notation, in the following we shall denote
by $|n\rangle$ the ket $|N-2n,n,n\rangle$. We then can write:
\begin{subequations}
\begin{align}
a_{\bf k}^\dagger a_{\bf k}|n\rangle & = a_{-\bf k}^\dagger a_{-\bf k}|n\rangle = n |n\rangle,
\\
a_{0}^\dagger a_{0} a_{\bf k}^\dagger a_{\bf k} |n\rangle & = 
a_{0}^\dagger a_{0} a_{-\bf k}^\dagger a_{-\bf k} |n\rangle = n(N-2n)|n\rangle,
\\
a_{\bf k}^\dagger a_{-\bf k}^\dagger a_0 a_0 |n\rangle & = (n+1)
\sqrt{(N-2n)(N-2n-1)}|n+1\rangle,
\\
a_{\bf k} a_{-\bf k} a_0^\dagger a_0^\dagger |n\rangle & = n
\sqrt{(N-2n+1)(N-2n+2)}|n-1\rangle,
\end{align}
\label{Eq:opspsik}
\end{subequations}
Using the above equations, one can easily show, after a few manipulations, that:
\begin{subequations}
\begin{align}
\hat{H}_{\bf k}|\psi_{\bf k}\rangle & = \sum_{n=0}^{N/2} 
C_n \Big[n\varepsilon_{\bf k} + \frac{v(k)}{V}n(N-2n)\Big]|n\rangle
\nonumber\\
&+ \frac{v(k)}{2V}\sum_{m=1}^{N/2}mC_{m-1}\sqrt{(N-2m+2)(N-2m+1)}|m\rangle
\nonumber\\
& + \frac{v(k)}{2V}\sum_{l=0}^{(N/2)-1}(l+1)C_{l+1}
\sqrt{(N-2l-1)(N-2l)}|l\rangle.
\end{align}
\end{subequations}
Now, it is not difficult to see that 
the second sum on the {\em rhs} of the above equation can be extended to $m=0$, 
since the factor $mC_{m-1}$ in the summand will make the extra term vanish. Also, the third
sum can be extended to $l=N/2$, because the factor $(N-2l)$ inside the square root will
make this term vanish. 
This allows us to write:
\begin{align}
\hat{H}_{\bf k}|\psi_{\bf k}\rangle & = \sum_{n=0}^{N/2}\Big\{
C_n\Big[n\varepsilon_{\bf k} + \frac{v(k)}{V}n(N-2n)\Big]
\nonumber\\
&+\frac{v(k)}{2V}nC_{n-1}\sqrt{(N-2n+2)(N-2n+1)}
\nonumber\\
&+\frac{v(k)}{2V}(n+1)C_{n+1}\sqrt{(N-2n-1)(N-2n)}
\Big\}|n\rangle.
\end{align}
Using this last result, we obtain that the expectation value $\langle\psi_{\bf k}|\hat{H}_{\bf k}|\psi_{\bf k}\rangle$ 
is given by:
\begin{align}
\langle\psi_{\bf k}|\hat{H}_{\bf k}|\psi_{\bf k}\rangle & = \sum_{n=0}^{N/2}\Big\{
|C_n|^2\Big[n\varepsilon_{\bf k} + \frac{v(k)}{V}n(N-2n)\Big]
\nonumber\\
&+\frac{v(k)}{2V}nC_n^*C_{n-1}\sqrt{(N-2n+2)(N-2n+1)}
\nonumber\\
&+\frac{v(k)}{2V}(n+1)C_n^*C_{n+1}\sqrt{(N-2n-1)(N-2n)}
\Big\}.
\label{Eq:PsihPsi1}
\end{align}
Let us assume, for simplicity, that the coefficients $C_n$ are real. 
Then, for $v(k)>0$, we see that the expectation value 
$\langle\psi_{\bf k}|\hat{H}_{\bf k}|\psi_{\bf k}\rangle$
will be lowered if the coefficients $C_n$ have alternating positive and 
negative signs. In this case, the terms on the second and third line will be negative, 
making the expectation value lower than what one would obtain if consecutive
coefficients have the same sign, in which case products of the form
$C_nC_{n-1}$ and $C_nC_{n+1}$ will be positive. Below, we will show that
Bogoliubov's theory corresponds to the following geometric
ansatz\cite{LeeHuangYang1957,Leggett2001} for the coefficients $C_n$:
\begin{equation}
C_n = C_0 (-c_{\bf k})^n,
\label{Eq:ansatz_Cn}
\end{equation}
where the constant $c_{\bf k}$ is to be determined variationally ($C_0$ will turn out to be an
overall constant which cancels out in the normalization of $|\psi_{\bf k}\rangle$
and whose value is hence unimportant for the evaluation of expectation values
of physical observables). Note that, for the $C_n$'s to have alternating positive and
negative signs, $c_{\bf k}$ has to be positive. On the other hand, we expect on physical grounds
that the coefficients $C_n$ will decrease with increasing values of $n$, 
or, in other words, that the probability amplitude of states
$|n\rangle$ with a large number $n\gg 1$ of bosons outside the condensate will be small. This implies
that the constant $c_{\bf k}$ must be less than unity. It then follows that $c_{\bf k}$
is subject to the following constraint:
\begin{equation}
0 < c_{\bf k} < 1.
\label{Eq:constraintck}
\end{equation}

Inserting the variational ansatz (\ref{Eq:ansatz_Cn}) into Eq. (\ref{Eq:PsihPsi1}),
and making use of the approximation:
\begin{equation}
\sqrt{N(N+1)} = N\sqrt{1+\frac{1}{N}}\simeq N + \frac{1}{2},
\label{Eq:ApproxSqrt}
\end{equation}
which is valid for $N\gg 1$, we can rewrite Eq. (\ref{Eq:PsihPsi1}) in the form:
\begin{align}
\langle\psi_{\bf k}|\hat{H}_{\bf k}|\psi_{\bf k}\rangle &\simeq C_0^2\sum_{n=0}^{N/2}\Big\{
(c_{\bf k})^{2n}\big[n\varepsilon_{\bf k} + \frac{v(k)}{V}n(N-2n)\big]
+\frac{v(k)}{2V}n(c_{\bf k})^{2n-1}\Big(N-2n+\frac{3}{2}\Big)
\nonumber\\
&+\frac{v(k)}{2V}(n+1)(c_{\bf k})^{2n+1}\Big(N-2n-\frac{1}{2}\Big)
\Big\}.
\label{Eq:PsihPsi2}
\end{align}
The summations in the above equation can be calculated analytically 
by taking successive derivatives with respect to the variable $x$ of the 
finite geometric sum:
\begin{align}
\sum_{n=0}^{N/2} x^n & = \frac{1-x^{\frac{N}{2}+1}}{1-x},
\label{Eq:xn}
\end{align}
hence obtaining the following results:
\begin{subequations}
\begin{align}
\sum_{n=1}^{N/2} nx^n & = x\left[
\frac{1-x^{\frac{N}{2}+1}}{(1-x)^2} - \Big(\frac{N}{2}+1\Big)\frac{x^{N/2}}{1-x}
\right],
\label{Eq:nxn}
\\
\sum_{n=1}^{N/2} n^2x^n & = x\Bigg[
\frac{1-x^{\frac{N}{2}+1}}{(1-x)^2} - \Big(\frac{N}{2}+1\Big)\frac{x^{N/2}}{1-x}
\Bigg]
\nonumber\\
&+x^2\Bigg[
\frac{2(1-x^{1+N/2})}{(1-x)^3}
-2\Big(\frac{N}{2}+1\Big)\frac{x^{N/2}}{(1-x)^2}
\nonumber\\
&-\frac{N}{2}\Big(\frac{N}{2}+1\Big)\frac{x^{\frac{N}{2}-1}}{1-x}
\Bigg].
\label{Eq:n2xn}
\end{align}
\label{Eq:geo_sums}
\end{subequations}
Now, it would be highly impractical to use the full expressions on the {\em rhs} of  
Eqs. (\ref{Eq:geo_sums}) in the evaluation of the sums on the {\em rhs} of Eq. (\ref{Eq:PsihPsi2}). Luckily,
these expressions simplify considerably for $N\to \infty$ and $0<x<1$, 
in which case we can write:
\begin{subequations}
\begin{align}
\sum_{n=1}^{N/2} nx^n &\simeq \frac{x}{(1-x)^2}, 
\label{Eq:nxn_asymptotic}
\\
\sum_{n=1}^{N/2} n^2x^n &\simeq \frac{x}{(1-x)^2} + \frac{2x^2}{(1-x)^3}. 
\label{Eq:n2xn_asymptotic}
\end{align}
\label{Eq:asymptoticseries}
\end{subequations}
Using these last two equations in Eq. (\ref{Eq:PsihPsi2}), we obtain
(we remind the reader that $n_B=N/V$ is the density of bosons in the system):
\begin{subequations}
\begin{align}
\langle\psi_{\bf k}|\hat{H}_{\bf k}|\psi_{\bf k}\rangle &= |C_0|^2\Big[
\varepsilon_{\bf k} + v(k)n_B\Big(1-\frac{1}{c_{\bf k}}\Big)
\Big]\sum_{n=0}^{N/2}nc_{\bf k}^{2n},
\\
&=|C_0|^2\Big\{
\frac{c_{\bf k}^2}{(1-c_{\bf k}^2)^2}\big[\varepsilon_{\bf k}+v(k)n_B\big] 
- v(k)n_B\frac{c_{\bf k}}{(1-c_{\bf k}^2)^2}
\Big\}.
\label{Eq:PsihPsi3}
\end{align}
\end{subequations}
On the other hand, from Eq. (\ref{Eq:defpsik}), we easily see that the norm $\langle\psi_{\bf k}|\psi_{\bf k}\rangle$ of the wavefunction 
$|\psi_{\bf k}\rangle$ is given by:
\begin{subequations}
\begin{align}
\langle\psi_{\bf k}|\psi_{\bf k}\rangle & = \sum_{n=0}^{N/2}|C_n|^2
= |C_0|^2\sum_{n=0}^{N/2} c_{\bf k}^{2n},
\\
& \simeq |C_0|^2 \frac{1}{1-c_{\bf k}^2},
\label{Eq:NormPsik}
\end{align}
\end{subequations}
where, in going from the first to the second line, use has been made of the
asymptotic form of Eq. (\ref{Eq:xn}), namely:
\begin{align}
\sum_{n=0}^{N/2} x^n \simeq \frac{1}{1-x}, \quad N\to \infty,\; 0<x<1.
\label{Eq:xn_asymptotic}
\end{align}

\begin{figure}[tb]
\centerline{\includegraphics[width=8.89cm, height=5.5cm]{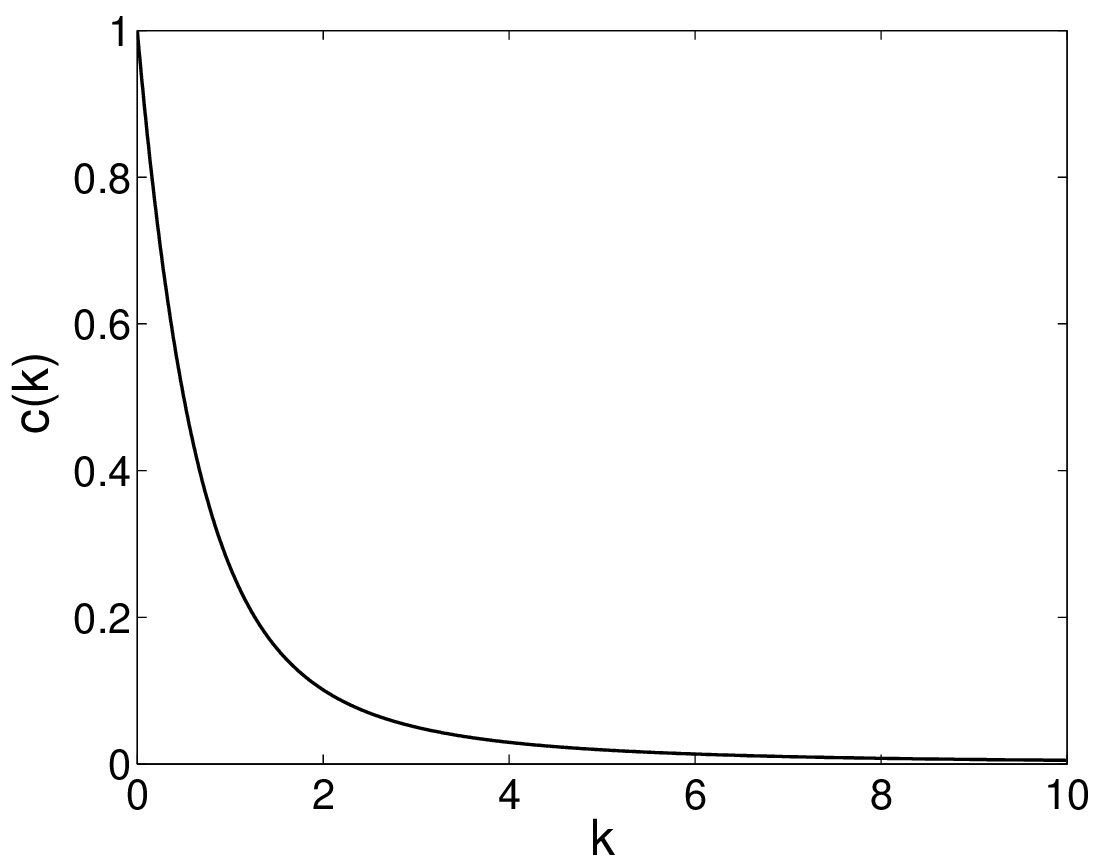}}
\caption[]{Plot of the constant $c_{\bf k}$ of Eq. (\ref{Eq:resultck}) vs. wavevector $k$.
Here, $k$ is the dimensionless combination $\tilde{k}=\hbar k/\sqrt{2mv(k)n_B}$.
}\label{Fig:plotcofk}
\end{figure}

Now, if we divide Eq. (\ref{Eq:PsihPsi3}) by Eq. (\ref{Eq:NormPsik}), we can write for the
expectation value $\langle \hat{H}_{\bf k}\rangle_{\bf k} $ of Eq. (\ref{Eq:Avg_h})
the following expression (notice that $|C_0|^2$ cancels out in the division):
\begin{align}
\langle\hat{H}_{\bf k}\rangle_{\bf k} \simeq 
\frac{c_{\bf k}^2}{1-c_{\bf k}^2}\big[\varepsilon_{\bf k}+v(k)n_B\big] - v(k)n_B\frac{c_{\bf k}}{1-c_{\bf k}^2}.
\label{Eq:expectation_h_min}
\end{align}
Minimization of the above expectation value with respect to $c_{\bf k}$
leads to the following quadratic equation:\cite{LeeHuangYang1957,Leggett2001}
\begin{equation}
c_{\bf k}^2 - 2\left(\frac{\cal{E}_{\bf k}}{v(k)n_B}\right) c_{\bf k} + 1 = 0,
\label{Eq:quadratic_ck}
\end{equation}
where we defined 
\begin{equation}
{\cal{E}_{\bf k}}=\varepsilon_{\bf k} + v(k)n_B.
\end{equation} 
The above quadratic equation has the following
two roots:
\begin{equation}
c_{\bf k}^\pm = \left(\frac{\cal{E}_{\bf k}}{v(k)n_B}\right) 
\pm \sqrt{\left(\frac{\cal{E}_{\bf k}}{v(k)n_B}\right)^2 - 1}.
\end{equation}
Of these two roots, only the one with the minus sign obeys the constraint $0<c_{\bf k}<1$ 
of Eq. (\ref{Eq:constraintck}) for arbitrary
values of ${\cal E}_{\bf k}$. We shall therefore write:\cite{LeeHuangYang1957,Leggett2001}
\begin{equation}
c_{\bf k} = \frac{1}{v(k)n_B}\left[
{\cal E}_{\bf k} - \sqrt{{\cal E}_{\bf k}^2-v(k)^2n_B^2}\right].
\label{Eq:resultck}
\end{equation}
This result for the constant $c_{\bf k}$, which is plotted as a function of the wavevector $k$ 
in Fig. \ref{Fig:plotcofk},
fully determines the coeffcients $C_n=C_0(-c_{\bf k})^n$
of the variational ground state wavefunction $|\psi_{\bf k}\rangle$ of the Hamiltonian $\hat{H}_{\bf k}$
(apart, of course, from the overall constant $C_0$).
This, in turn, will allow us to determine the expectation values of physical observables 
in the ground state $|\psi_{\bf k}\rangle$. In particular, the expectation value 
of $\hat{H}_{\bf k}$ in the ground state $|\psi_{\bf k}\rangle$ can readily be found
if we use the result (\ref{Eq:resultck}) for $c_{\bf k}$ in Eq. (\ref{Eq:expectation_h_min}), 
upon which we obtain:
\begin{align}
\langle\hat{H}_{\bf k}\rangle_{\bf k} = -\frac{1}{2}\Big(
\varepsilon_{\bf k} +n_Bv(k) - E_{\bf k}\Big) < 0,
\label{Eq:avgHk}
\end{align}
with the definition:
\begin{equation}
E_{\bf k} = \sqrt{\varepsilon_{\bf k}\big[\varepsilon_{\bf k} + 2n_Bv(k)\big]}.
\label{Eq:newdefEk}
\end{equation}
The result (\ref{Eq:avgHk}) is exactly the result one obtains in the standard, number non-conserving
Bogoliubov approach for the expectation value of a given contribution $\hat{H}_{\bf k}$ to
the Bogoliubov ground state energy. 
We shall discuss the meaning of this observation in more detail below.
For the moment, we want to use our result for the variational wavefunction $|\psi_{\bf k}\rangle$
to explore the properties of the ground state.

\subsection{Variational result for the depletion of the condensate}
\label{Sub:vardepletion}

\begin{figure}[tb]
\centerline{\includegraphics[width=8.89cm, height=5.5cm]{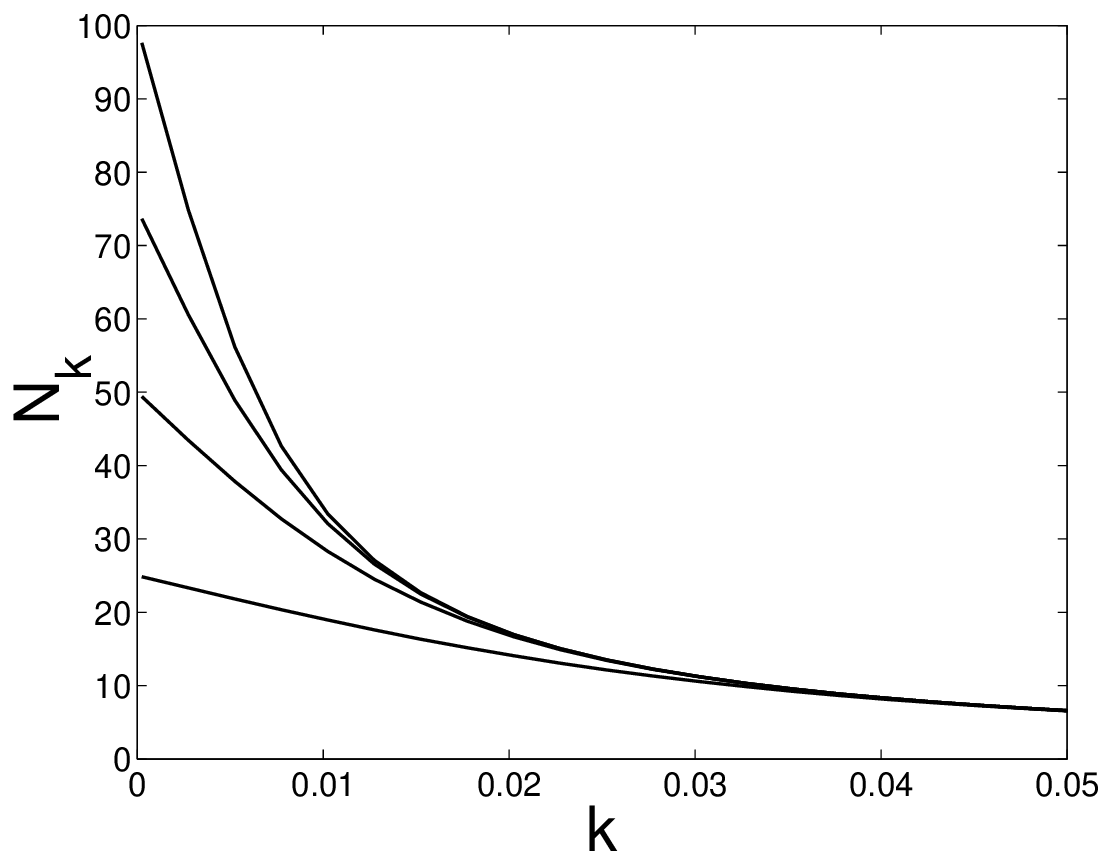}}
\caption[]{Depletion $N_k=\langle a_{\bf k}^\dagger a_{\bf k}\rangle$
as obtained from the variational approach, Eq. (\ref{Eq:fullNkvariational}), for various values of $N$.
The various curves are for $N=100,\,200,\,300$ and $400$ bosons, from bottom to top. As it can be seen,
the depletion $N_{\bf k}\to N/4$ as $k\to 0$. Here again, $k$ is the dimensionless combination 
$\tilde{k}=\hbar k/\sqrt{2mv(k)n_B}$.
}\label{Fig:VariationalDepletionVariousN}
\end{figure}

We now turn our attention to the depletion of the condensate. 
The average number of bosons $N_{\bf k}$ in the single-particle state of momentum ${\bf k}$ is given by:
\begin{equation}
N_{\bf k} \equiv \langle a_{\bf k}^\dagger a_{\bf k}\rangle_{\bf k}
= \frac{\langle\psi_{\bf k}|a_{\bf k}^\dagger a_{\bf k}|\psi_{\bf k}\rangle}
{\langle\psi_{\bf k}|\psi_{\bf k}\rangle}.
\end{equation}
Now, it is easy to see that:
\begin{subequations}
\begin{align}
\langle \psi_{\bf k}|a_{\bf k}^\dagger a_{\bf k}|\psi_{\bf k}\rangle 
& = |C_0|^2 \sum_{n=0}^{N/2} n|C_n|^2 = |C_0|^2\sum_{n=0}^{N/2} n c_{\bf k}^{2n},
\\
& = |C_0|^2 \frac{c_{\bf k}^2}{(1-c_{\bf k}^2)^2},
\end{align}
\end{subequations}
where, in going from the first to the second line, use has been made of 
Eq. (\ref{Eq:nxn_asymptotic}). Then, using expression (\ref{Eq:NormPsik}), we finally obtain:
\begin{equation}
\langle a_{\bf k}^\dagger a_{\bf k}\rangle_{\bf k} = \frac{c_{\bf k}^2}{1-c_{\bf k}^2}.
\label{Eq:single-mode-depletion}
\end{equation}
If we now use the expression of $c_{\bf k}$ in Eq. (\ref{Eq:resultck}), we find after a few manipulations:
\begin{equation}
N_{\bf k} = \frac{1}{2}\Big[
\frac{\varepsilon_{\bf k} + n_Bv(k)}{E_{\bf k}} - 1
\Big].
\label{Eq:resultNk}
\end{equation}
This is again the same result one obtains in Bogoliubov's standard, number non-conserving approach. Unfortunately, 
and as we have already discussed in Sec. \ref{Sub:depletionBog},
the above expression of $N_{\bf k}$ diverges for $k\to 0$, which does not make much sense for a system of $N$
bosons.  It is indeed quite easy to see that, $N_{\bf k}$ being the ratio of two {\em finite} sums:
\begin{align}
N_{\bf k} = \frac{\sum_{n=1}^{N/2}n(c_{\bf k}^2)^n}{\sum_{n=0}^{N/2}(c_{\bf k}^2)^n},
\end{align}
with $0<c_k<1$, the ratio {\em must} remain finite for all values of $c_k$ in the range $(0,1)$.
Luckily, the origin of the divergence in Eq. (\ref{Eq:resultNk}) can easily be elucidated, 
and has to do with our use of the approximate limiting expressions of the sums 
(\ref{Eq:nxn_asymptotic}) and (\ref{Eq:xn_asymptotic}) (which both diverge as $c_{\bf k}\to 1$)
when evaluating the average $\langle a_{\bf k}^\dagger a_{\bf k}\rangle_{\bf k}$. If we use the full expressions
(\ref{Eq:nxn}) and (\ref{Eq:xn}) instead, we find the result:
\begin{equation}
N_{\bf k} = \frac{c_{\bf k}^2}{1-c_{\bf k}^2} 
- \Big(\frac{N}{2} + 1\Big)\frac{c_{\bf k}^{2+N}}{1-c_{\bf k}^{2+N}}.
\label{Eq:fullNkvariational}
\end{equation} 
It can be verified that the above result, which
to the best of the author's knowledge has not been derived previously, 
is finite for all values of $c_{\bf k}$.
In particular, as $k\to 0$, $c_{\bf k}\to 1$, and a Taylor expansion in $x=c_{\bf k}^2$
near $x=1$ shows that $N_{\bf k}\to N/4$.
This behaviour is shown in Fig. \ref{Fig:VariationalDepletionVariousN} where we plot the 
depletion $N_{\bf k}$ from Eq. (\ref{Eq:fullNkvariational}) 
as a function of the wavevector $k=|{\bf k}|$ for a number of values of the total 
number of bosons $N$.

\addvspace{1.5mm}

The result of Eq. (\ref{Eq:fullNkvariational}) allows us to obtain more meaningful results
for expectation values of one-body operators of the form $\hat{O}=\sum_{\bf k}O_{\bf k}a_{\bf k}^\dagger a_{\bf k}$
than were obtained within the standard, number non-conserving Bogoliubov approximation. To revisit the example given in
Sec. \ref{Sub:depletionBog}, let us consider the Fock part of the Hamiltonian 
$\hat{H}_F \approx \frac{1}{2}\sum_{\bf k\neq 0}n_0v({\bf k})(a_{\bf k}^\dagger a_{\bf k} 
+ a_{-\bf k}^\dagger a_{-\bf k})$, where for simplicity we made use of Bogoliubov's approximation
$a_0=a_0^\dagger\approx\sqrt{N}_0$. While in the standard Bogoliubov approach the expectation
value of a given mode ${\bf k}$ of $\hat{H}_F$ in the Bogoliubov ground state diverges like $1/k$ as $k\to 0$
(which, as we discussed earlier, is problematic), in the
variational Bogoliubov method this expectation value is finite for all values of the wavevector ${\bf k}$, and is
given by the usual expression involving the product of the number $N_0$ of bosons 
in the mode ${\bf k}=0$ by the number $N_{\bf k}$ of
depleted bosons in each of the modes ${\bf k}$ and $-{\bf k}$ (since these are the only modes kept in  $\hat{H}_F$), 
times the interaction energy $v({\bf k})/V$. The fact that all quantities involved are well-defined and finite within the
variational approach is reassuring, and constitutes a significant improvement with respect to the standard, number
non-conserving Bogoliubov approximation.

\subsection{Variational result for the anomalous averages $\langle a_{\bf k}a_{-\bf k}\rangle$ 
and $\langle a_{\bf k}^{\dagger}a_{-\bf k}^\dagger\rangle$}
\label{Sub:AnomalousAverages}

We now want to investigate the anomalous average $\langle a_{\bf k}a_{-\bf k}\rangle$ 
in the variational formulation of Bogoliubov's method. In order to do so, 
we note that for $N\gg 1$ the coefficients $C_n$ of the variational wavefunction
$|\psi_{\bf k}(N)\rangle = \sum_{n=0}^{N/2}C_n|n\rangle$ do not explicitly depend 
on the number of bosons $N$, and are given to a very good approximation 
by the ansatz expression $C_n=C_0(-c_{\bf k})^n$, regardless of the specific
value of $N$ (in other words, the constant $c_{\bf k}$ is independent of $N$ for $N\to\infty$). 
Under these circumstances, we shall write for the wavefunction of a system of $(N-2)$ bosons
the following variational ansatz:
\begin{align}
|\psi_{\bf k}(N-2)\rangle = \sum_{n=0}^{(N-2)/2}C_n|(N-2)-2n,n,n\rangle.
\end{align}
It is now easy to verify that:
\begin{subequations}
\begin{align}
a_{\bf k}^\dagger a_{-\bf k}^\dagger|\psi_{\bf k}(N-2)\rangle & =
\sum_{n=0}^{N/2-1} (n+1)C_n
|N-2(n+1),n+1,n+1\rangle,
\\
& = \sum_{n=1}^{N/2} n C_{n-1}|n\rangle,
\end{align}
\end{subequations}
and hence that:
\begin{subequations}
\begin{align}
\langle\psi_{\bf k}(N)|a_{\bf k}^\dagger a_{-\bf k}^\dagger|\psi_{\bf k}(N-2)\rangle & =
\sum_{n=1}^{N/2} nC_{n-1}C_n^*,
\\
& = -\frac{C_0^2}{c_{\bf k}}\sum_{n=1}^{N/2} n (c_{\bf k}^2)^n.
\end{align}
\end{subequations}
Using the limiting form (\ref{Eq:nxn_asymptotic}) for the geometric sum as $N\to \infty$, we obtain:
\begin{align}
\langle\psi_{\bf k}(N)|a_{\bf k}^\dagger a_{-\bf k}^\dagger|\psi_{\bf k}(N-2)\rangle \simeq
-C_0^2\frac{c_{\bf k}}{\big(1-c_{\bf k}^2\big)^2}.
\label{Eq:psiNpsiN-2}
\end{align}
On the other hand, the limiting form (\ref{Eq:xn}) as $N\to \infty$ of the norm of the wavefunctions 
$|\psi_{\bf k}(N)\rangle$ and $|\psi_{\bf k}(N-2)\rangle$ is given by:
\begin{equation}
\langle\psi_{\bf k}(N)|\psi_{\bf k}(N)\rangle = \langle\psi_{\bf k}(N - 2)|\psi_{\bf k}(N-2)\rangle
= C_0^2\frac{1}{1-c_{\bf k}^2}.
\label{Eq:normPsiNN-2}
\end{equation}
If we use the {\em normalized} ground state
wavefunctions $|\widetilde\psi_{\bf k}(N)\rangle$ such that:
\begin{equation}
|\widetilde\psi_{\bf k}(N)\rangle = \frac{|\psi_{\bf k}(N)\rangle}
{\sqrt{\langle\psi_{\bf k}(N)|\psi_{\bf k}(N)\rangle}},
\end{equation}
then the anomalous average $\langle a_{\bf k}^\dagger a_{-\bf k}^\dagger\rangle_{\bf k}$ 
can be defined in the following way:
\begin{align}
\langle a_{\bf k}^\dagger a_{-\bf k}^\dagger\rangle_{\bf k} =
\langle\widetilde\psi_{\bf k}(N)|a_{\bf k}^\dagger a_{-\bf k}^\dagger|\widetilde\psi_{\bf k}(N-2)\rangle.
\end{align}
Using Eqs. (\ref{Eq:psiNpsiN-2}) and (\ref{Eq:normPsiNN-2}), we find:
\begin{align}
\langle a_{\bf k}^\dagger a_{-\bf k}^\dagger\rangle_{\bf k} \simeq - \frac{c_{\bf k}}{1-c_{\bf k}^2}.
\end{align}
Given the expression (\ref{Eq:resultck}) of $c_{\bf k}$, we obtain, after a few manipulations:
\begin{equation}
\langle a_{\bf k}^\dagger a_{-\bf k}^\dagger\rangle_{\bf k} = 
-\frac{n_Bv({\bf k})}{2E_{\bf k}},
\label{Eq:AnAvgBog}
\end{equation}
which is nothing but the standard Bogoliubov result of Eq. (\ref{Eq:an_avg_Bog}) above, which
diverges negatively like $-1/k$ as $k\to 0$.

\begin{figure}[tb]
\centerline{\includegraphics[width=8.89cm, height=5.5cm]{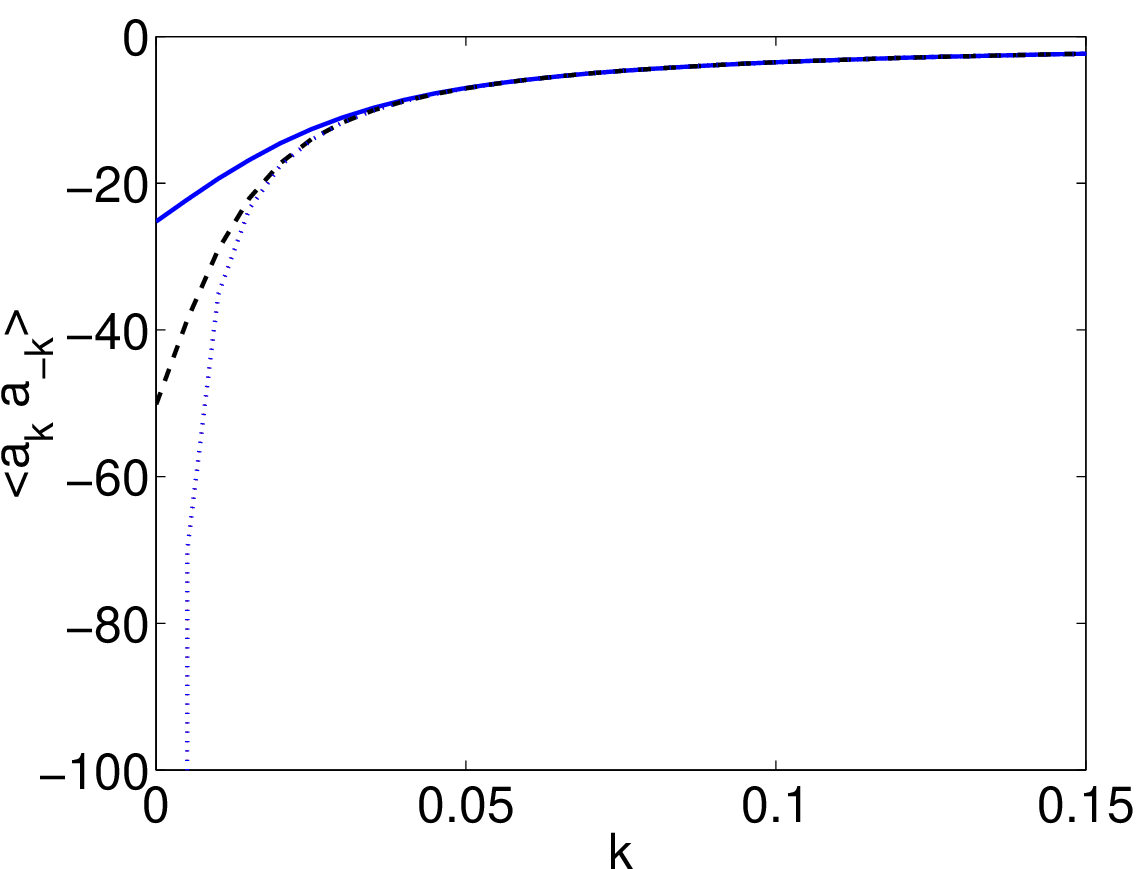}}
\caption[]{Anomalous average $\langle a_{\bf k}a_{-\bf k}\rangle$
as obtained from the variational approach, Eq. (\ref{Eq:anomavg2}).
The Bogoliubov result is shown as the dotted line, and diverges negatively in the $k\to 0$ limit.
The solid line is the result of the variational approach for $N=100$ bosons, and
the dashed line is the variational result for $N=200$ bosons.
As it can be seen, in the variational method the quantity 
$\langle a_{\bf k}a_{-\bf k}\rangle$ goes to a finite limit as $k\to 0$, as it should.
(As in Figs. \ref{Fig:plotcofk} and \ref{Fig:VariationalDepletionVariousN}, 
$k$ in this figure is the dimensionless combination $\tilde{k}=\hbar k/\sqrt{2mv(k)n_B}$.)
}\label{Fig:AnomalousAvg}
\end{figure}

\addvspace{1.5mm}

We now show that the divergence of the {\em rhs} of Eq. (\ref{Eq:AnAvgBog}) 
as $k\to 0$ is also a direct consequence of using asymptotic forms as $N\to\infty$
of geometric sums, and that, if the finite $N$ results are used throughout, then a 
completely diffrent result will be obtained which remains finite for all values of the wavevector $k$. 
Indeed, using the notation $x=c_{\bf k}^2$,
we can write:
\begin{align}
\langle\psi_{\bf k}(N)|a_{\bf k}^\dagger a_{-\bf k}^\dagger|\psi_{\bf k}(N-2)\rangle
&= - \frac{C_0^2}{c_{\bf k}}\sum_{n=1}^{N/2}nx^n,
\nonumber\\
& = - \frac{C_0^2}{c_{\bf k}}x\Big[
\frac{1-x^{1+N/2}}{(1-x)^2} - \Big(\frac{N}{2}+1\Big)\frac{x^{N/2}}{1-x}
\Big].
\label{Eq:anomavg1}
\end{align}
In a similar fashion, we obtain for the norm of the wavefunctions 
$|\psi_{\bf k}(N)\rangle$ and $|\psi_{\bf k}(N-2)\rangle$ the following expressions:
\begin{subequations}
\begin{align}
\langle\psi_{\bf k}(N)|\psi_{\bf k}(N)\rangle &= C_0^2\frac{1-x^{1+N/2}}{1-x},
\\
\langle\psi_{\bf k}(N - 2)|\psi_{\bf k}(N-2)\rangle
& = C_0^2\frac{1-x^{N/2}}{1-x}.
\end{align}
\label{Eq:norms}
\end{subequations}
Combining Eqs. (\ref{Eq:anomavg1}) and (\ref{Eq:norms}) allows us to write the final expression:
\begin{align}
&\langle\widetilde\psi_{\bf k}(N)|a_{\bf k}^\dagger a_{-\bf k}^\dagger|\widetilde\psi_{\bf k}(N-2)\rangle
=-\frac{c_{\bf k}}{\sqrt{(1-c_{\bf k}^N)(1-c_{\bf k}^{N+2})}}
\Big[
\frac{1-c_{\bf k}^{N+2}}{1-c_{\bf k}^2} -\Big(\frac{N}{2}+1\Big)c_{\bf k}^N
\Big],
\label{Eq:anomavg2}
\end{align}
an expression which again has not been derived before, and
which can be shown to be bounded for all values of $c_{\bf k}$ in the range $(0,1)$ 
(see Fig. \ref{Fig:AnomalousAvg}).

\addvspace{1.5mm}

The result of Eq. (\ref{Eq:anomavg2}) allows us again to obtain more meaningful results
for expectation values of one-body operators of the form $\hat{O}=\sum_{\bf k}O_{\bf k}a_{\bf k} a_{-\bf k}$
or $\hat{O}=\sum_{\bf k}O_{\bf k}a_{\bf k}^\dagger a_{-\bf k}^\dagger$
than were obtained within the standard, number non-conserving Bogoliubov approximation. 
To revisit the example given in
Sec. \ref{Sub:depletionBog}, let us consider the pairing part of the Hamiltonian 
$\hat{H}_P \approx \frac{1}{2}\sum_{\bf k\neq 0}n_0v({\bf k})(a_{\bf k}^\dagger a_{-\bf k}^\dagger 
+ a_{-\bf k} a_{\bf k})$, where for simplicity we again made use of Bogoliubov's approximation
$a_0=a_0^\dagger\approx\sqrt{N}_0$. While in the standard Bogoliubov approximation the expectation
value of a given mode ${\bf k}$ of $\hat{H}_P$ in the Bogoliubov ground state 
is given by $-\big[n_Bv({\bf k})\big]^2/(2E_{\bf k})$ and
diverges like $-1/k$ as $k\to 0$,
in the variational Bogoliubov 
method this expectation value is finite for all values of the wavevector ${\bf k}$,
and this even in lower spatial dimensions.

\subsection{Elementary excitations: variational approach for the single-mode Hamiltonian $\hat{H}_{\bf k}$}
\label{Sub:ExcOneMode}

We now introduce the following operator:\cite{Leggett2001}
\begin{equation}
\alpha_{\bf k} = \tilde{u}_{\bf k} a_{\bf k} a_0^\dagger + \tilde{v}_{\bf k} a_0 a_{-\bf k}^\dagger.
\label{Eq:defalphak}
\end{equation}
We want to choose the constants $\tilde{u}_{\bf k}$ and 
$\tilde{v}_{\bf k}$ in such a way that acting on $|\psi_{\bf k}\rangle$
with $\alpha_{\bf k}^\dagger$ creates an excited state of the Hamiltonian $\hat{H}_{\bf k}$. A necessary condition
for that to happen is that $\alpha_{\bf k}^\dagger|\psi_{\bf k}\rangle$ 
has to be orthogonal to $|\psi_{\bf k}\rangle$, {\em i.e.}, that:
\begin{equation}
\langle\psi_{\bf k}|\Big(\alpha_{\bf k}^\dagger|\psi_{\bf k}\rangle\Big) = 0.
\end{equation}
Taking the adjoint of the above equation, we obtain the condition
\begin{equation}
\langle\psi_{\bf k}|\alpha_{\bf k}|\psi_{\bf k}\rangle = 0,
\end{equation}
which will {\em always} be satisfied if we require that
\begin{equation}
\alpha_{\bf k}|\psi_{\bf k}\rangle = 0.
\label{Eq:condalphak}
\end{equation}
Let us calculate the effect of the action of $\alpha_{\bf k}$ on $|\psi_{\bf k}\rangle$. We have:
\begin{align}
\alpha_{\bf k}|\psi_{\bf k}\rangle & = \sum_{n=0}^{N/2} C_n\Big[
\tilde{u}_{\bf k}\sqrt{n(N-2n+1)}|N-2n+1,n-1,n\rangle
\nonumber\\
& + \tilde{v}_{\bf k} \sqrt{(n+1)(N-2n)} |N-2n-1,n,n+1\rangle
\Big].
\end{align}
Noting that the $n=0$ term in the first summation and the $n=N/2$ term in the second summation vanish,
we relabel $n\to n-1$ in the second sum, hence obtaining:
\begin{align}
\alpha_{\bf k}|\psi_{\bf k}\rangle &= \sum_{n=1}^{N/2} \Big[
C_n\tilde{u}_{\bf k}\sqrt{n(N-2n+1)} 
\nonumber\\
&+ C_{n-1}\tilde{v}_{\bf k} \sqrt{n(N-2n+2)} \Big]|N-2n+1,n-1,n\rangle.
\end{align}
Requiring that $\alpha_{\bf k}|\psi_{\bf k}\rangle=0$ leads to the following condition:
\begin{equation}
\frac{C_n}{C_{n-1}} = - \frac{\tilde{v}_{\bf k}}{\tilde{u}_{\bf k}}\sqrt{\frac{N-2n+2}{N-2n+1}}.
\label{Eq:ratioCnCn-1}
\end{equation}
For $N\gg 1$, the square root is very close to unity for most values of $n$ such that $1\leq n\leq N/2$, 
and so we obtain that, to a very good approximation,
the condition $\alpha_{\bf k}|\psi_{\bf k}\rangle=0$ is equivalent to:
\begin{equation}
\frac{\tilde{v}_{\bf k}}{\tilde{u}_{\bf k}} = - \frac{C_n}{C_{n-1}}.
\end{equation}
Using the fact that $C_n=C_0(-c_{\bf k})^n$, we finally obtain:
\begin{equation}
\frac{\tilde{v}_{\bf k}}{\tilde{u}_{\bf k}} = c_{\bf k}.
\label{Eq:ratioukvk} 
\end{equation}
Hence, we obtain that the operator $\alpha_{\bf k}$ annihilates the variational ground
state $|\psi_{\bf k}\rangle$ of the Hamiltonian $\hat{H}_{\bf k}$
if the constants $\tilde{v}_{\bf k}$ and $\tilde{u}_{\bf k}$ 
have the ratio specified in Eq. (\ref{Eq:ratioukvk}). 
This result is quite important, given that the equation 
$\alpha_{\bf k}|\psi_{\bf k}\rangle=0$ in effect ensures that the state 
$\alpha^\dagger_{\bf k}|\psi_{\bf k}\rangle$ is orthogonal to 
$|\psi_{\bf k}\rangle$, and is hence an excited state as we discussed earlier in this Subsection.

\addvspace{1.5mm}

We now want to establish that the operator $\alpha_{\bf k}^\dagger$
creates an excited state of momentum ${\bf k}$.
To this end, let us find the average momentum in the state $\alpha_{\bf k}^\dagger|\psi_{\bf k}\rangle$,
which is defined as:
\begin{equation}
\langle\hat{\bf P}\rangle_{\bf k}^{exc} = \frac{\langle\psi_{\bf k}|
\alpha_{\bf k}\hat{\bf P}\alpha_{\bf k}^\dagger|\psi_{\bf k}\rangle}
{\langle\psi_{\bf k}|\alpha_{\bf k}\alpha_{\bf k}^\dagger|\psi_{\bf k}\rangle},
\end{equation}
with $\hat{\bf P}=\sum_{\bf k}\hbar {\bf k}a_{\bf k}^\dagger a_{\bf k}$ the total momentum operator.
Noting that the commutator $[\hat{\bf P},\alpha_{\bf k}^\dagger]$ is given by:
\begin{equation}
[\hat{\bf P},\alpha_{\bf k}^\dagger] = \hbar{\bf k}\alpha_{\bf k}^\dagger,
\end{equation}
and that $\hat{\bf P}|\psi_{\bf k}\rangle=0$, we can easily write:
\begin{equation}
\langle\psi_{\bf k}|\alpha_{\bf k}\hat{\bf P}\alpha_{\bf k}^\dagger|\psi_{\bf k}\rangle=
\hbar{\bf k}\langle\psi_{\bf k}|\alpha_{\bf k}\alpha_{\bf k}^\dagger|\psi_{\bf k}\rangle,
\end{equation}
from which we finally obtain:
\begin{equation}
\langle\hat{\bf P}\rangle_{\bf k}^{exc} = \hbar{\bf k},
\end{equation}
thus proving that the state $\alpha_{\bf k}^\dagger|\psi_{\bf k}\rangle$
has an average momentum $\hbar{\bf k}$. 

\addvspace{1.5mm}

Let us recapitulate what we have done so far. We have defined a new operator $\alpha_{\bf k}$ such that the state
$\alpha_{\bf k}^\dagger|\psi_{\bf k}\rangle$ is {\it (i)} orthogonal to $|\psi_{\bf k}\rangle$ and 
{\it (ii)} has an average momentum of $\hbar{\bf k}$. 
These two properties allow us to identify the state $\alpha_{\bf k}^\dagger|\psi_{\bf k}\rangle$ 
as an an excited state of the Hamiltonian $\hat{H}_{\bf k}$ with momentum $\hbar{\bf k}$. 
At this point, we make the observation that, while the requirement 
$\alpha_{\bf k}|\psi_{\bf k}\rangle=0$ allowed us to determine the ratio
of $\tilde{v}_{\bf k}$ to $\tilde{u}_{\bf k}$, Eq. (\ref{Eq:ratioukvk}), these constants are otherwise
still arbitrary. In order to determine their actual values, we shall require that the excited state 
$\alpha^\dagger_{\bf k}|\widetilde\psi_{\bf k}\rangle$
be normalized to unity, {\em i.e.} that:
\begin{equation}
\langle\widetilde\psi_{\bf k}|\alpha_{\bf k}\alpha_{\bf k}^\dagger|\widetilde\psi_{\bf k}\rangle = 1,
\label{Eq:normcond}
\end{equation}
where we remind the reader that $|\widetilde{\psi}_{\bf k}\rangle$ denotes the {\em normalized}
ground state of the single-mode Hamiltonian $\hat{H}_{\bf k}$.
Using the fact that $\alpha_{\bf k}\alpha_{\bf k}^\dagger=[\alpha_{\bf k},\alpha_{\bf k}^\dagger]
+\alpha_{\bf k}^\dagger\alpha_{\bf k}$, the condition (\ref{Eq:normcond}) becomes:
\begin{equation}
\langle\widetilde\psi_{\bf k}|[\alpha_{\bf k},\alpha_{\bf k}^\dagger]|\widetilde\psi_{\bf k}\rangle = 1,
\label{Eq:condcommutator}
\end{equation}
where we used the fact that $\alpha_{\bf k}|\widetilde\psi_{\bf k}\rangle=0$.
An immediate way to satisfy this last condition is to require that the 
commutator $[\alpha_{\bf k},\alpha_{\bf k}^\dagger]$ be equal to unity:
\begin{equation}
[\alpha_{\bf k},\alpha_{\bf k}^\dagger] = 1.
\label{Eq:comm_var}
\end{equation}
We thus see that the canonical commutation relation $[\alpha_{\bf k},\alpha_{\bf k}^\dagger] = 1$,
which is imposed {\em a priori} in the standard textbook formulation of Bogoliubov's theory,
emerges as a natural requirement that we need to satisfy in order for the excited state 
$\alpha_{\bf k}^\dagger|\widetilde\psi_{\bf k}\rangle$ to be normalized to unity.

Now, if we calculate the commutator $[\alpha_{\bf k},\alpha_{\bf k}^\dagger]$, 
given the definition (\ref{Eq:defalphak}), we find, after a few manipulations:
\begin{equation}
[\alpha_{\bf k},\alpha_{\bf k}^\dagger] = (\tilde{u}_{\bf k}^2-\tilde{v}_{\bf k}^2)a_0^\dagger a_0
- \tilde{u}_{\bf k}^2 a_{\bf k}^\dagger a_{\bf k} + \tilde{v}_{\bf k}^2 a_{-\bf k}^\dagger a_{-\bf k}.
\label{Eq:commutatoralphas1}
\end{equation}
While in Bogoliubov's theory the commutator $[\alpha_{\bf k},\alpha_{\bf k}^\dagger]$ is a c-number,
the fact that the {\em rhs} of the above equation is an operator makes it impossible to satisfy
the constraint $[\alpha_{\bf k},\alpha_{\bf k}^\dagger]=1$ exactly in the number-conserving formalism. 
However, we can try to satisfy this constraint in an averaged sense in
the ground state $|\widetilde\psi_{\bf k}\rangle$
by imposing the condition given in Eq. (\ref{Eq:condcommutator}), which
can be rewritten in the form:
\begin{equation}
\langle[\alpha_{\bf k},\alpha_{\bf k}^\dagger]\rangle_{\bf k} = 
(\tilde{u}_{\bf k}^2-\tilde{v}_{\bf k}^2)\Big[
\langle a_0^\dagger a_0\rangle_{\bf k}
-\langle a_{\bf k}^\dagger a_{\bf k}\rangle_{\bf k}\Big],
\end{equation}
where we denote by $\langle\cdots\rangle_{\bf k}$ the quantum average 
$\langle\widetilde\psi_{\bf k}|\cdots|\widetilde\psi_{\bf k}\rangle$, and where we used the fact that 
$\langle a_{\bf k}^\dagger a_{\bf k}\rangle_{\bf k} = \langle a_{-\bf k}^\dagger a_{-\bf k}\rangle_{\bf k}$.
Now, the normalization condition (\ref{Eq:condcommutator})
leads to the following constraint on the constants
$\tilde{u}_{\bf k}$ and $\tilde{v}_{\bf k}$:
\begin{equation}
\tilde{u}_{\bf k}^2 - \tilde{v}_{\bf k}^2 = \gamma_{\bf k}^2,
\label{Eq:diffukvk}
\end{equation}
where, for compactness of notation, we defined the quantity $\gamma_{\bf k}$ such that:
\begin{subequations}
\begin{eqnarray}
\gamma_{\bf k}^2 & = & \frac{1}{\langle a_0^\dagger a_0\rangle_{\bf k} 
- \langle a_{\bf k}^\dagger a_{\bf k}\rangle_{\bf k}}
\label{Eq:defgammaka}
\\
& \simeq & \frac{1}{\langle a_0^\dagger a_0\rangle_{\bf k}},
\label{Eq:defgammakb}
\end{eqnarray}
\label{Eq:defgammak}
\end{subequations}
In going from the first to the second line of Eq. (\ref{Eq:defgammak}) we assumed that 
$\langle a_{\bf k}^\dagger a_{\bf k}\rangle_{\bf k}\ll \langle a_0^\dagger a_0\rangle_{\bf k}$, 
{\em i.e.} that the depletion of the condensate into the single-particle state with momentum ${\bf k}$
is small. Note that this is not always the case, as in the standard formulation of Bogoliubov's theory
the quantity $N_{\bf k}=\langle a_{\bf k}^\dagger a_{\bf k}\rangle_{\bf k}$
diverges in the $k\to 0$ limit. This, incidentally, gives us another reason
why it is important to be able to formulate a version of the theory
in which the expectation value $\langle a_{\bf k}^\dagger a_{\bf k}\rangle$ 
always remains finite. As mentioned above, in the number non-conserving formulation of Bogoliubov's theory,
$\langle a_{\bf k}^\dagger a_{\bf k}\rangle$ diverges as $k\to 0$, and taking this result at face value
would lead to a completely different result for $\gamma_{\bf k}^2$ in Eq. (\ref{Eq:defgammakb}),
which would in turn affect the evaluation of the excitation energies given below, Eq. (\ref{Eq:Hcanonical}).

\addvspace{1.5mm}

Using Eqs. (\ref{Eq:ratioukvk}) and (\ref{Eq:diffukvk}), we now can determine the
values of the constants $\tilde{u}_{\bf k}$ and $\tilde{v}_{\bf k}$. These can be written in the form:
\begin{subequations}
\begin{align}
&\tilde{u}_{\bf k} = \gamma_{\bf k} u_{\bf k}, \quad \tilde{v}_{\bf k} = \gamma_{\bf k} v_{\bf k},
\\
& u_{\bf k} = \frac{1}{\sqrt{1-c_{\bf k}^2}}, \quad v_{\bf k} = \frac{c_{\bf k}}{\sqrt{1-c_{\bf k}^2}}.
\end{align}
\label{Eq:resulttildeukvk}
\end{subequations}
Using the result (\ref{Eq:resultck}) for $c_{\bf k}$, we finally obtain:
\begin{subequations}
\begin{align}
u_{\bf k}^2 & = \frac{1}{2}\Big(\frac{\varepsilon_{\bf k}+n_Bv({\bf k})}{E_{\bf k}} + 1\Big),
\\
v_{\bf k}^2 & = \frac{1}{2}\Big(\frac{\varepsilon_{\bf k}+n_Bv({\bf k})}{E_{\bf k}} - 1\Big).
\end{align}
\label{Eq:resultukvk}
\end{subequations}
Note that these are the same expressions for the quantities $u_{\bf k}$ and $v_{\bf k}$ defined 
in the conventional formulation of Bogoliubov's theory.
The above results, Eqs. (\ref{Eq:resulttildeukvk}) and (\ref{Eq:resultukvk}), 
allow us to rewrite the expression of $\alpha_{\bf k}$, 
Eq. (\ref{Eq:defalphak}), in the form:
\begin{equation}
\alpha_{\bf k} = \gamma_{\bf k}\Big(u_{\bf k} a_{\bf k} a_0^\dagger + v_{\bf k} a_0 a_{-\bf k}^\dagger\Big).
\label{Eq:defalphak2}
\end{equation}
If in the above expression we replace $a_0$ and $a_0^\dagger$ by $\sqrt{N_0}$, and then use
the fact that $\gamma_{\bf k}\simeq 1/\sqrt{N_0}$ as indicated by Eq. (\ref{Eq:defgammakb}), 
we recover the usual expression, Eq. (\ref{Eq:defalphas0}), of $\alpha_{\bf k}$ in terms of 
$a_{\bf k}$ and $a^\dagger_{-\bf k}$ used in the standard, number non-conserving formulation of Bogoliubov's method.

\addvspace{1.5mm}

We now are finally in a position to find the energy of the excited state $\alpha_{\bf k}^\dagger$, which
is given by the expression:
\begin{subequations}
\begin{align}
\Delta E_{exc}^{(1)}({\bf k}) & = \frac{\langle\widetilde\psi_{\bf k}|\alpha_{\bf k}
\hat{H}_{\bf k}\alpha_{\bf k}^\dagger|\widetilde\psi_{\bf k}\rangle}
{\langle\widetilde\psi_{\bf k}|\alpha_{\bf k}\alpha_{\bf k}^\dagger|\widetilde\psi_{\bf k}\rangle}
- \langle\widetilde\psi_{\bf k}|
\hat{H}_{\bf k}|\widetilde\psi_{\bf k}\rangle,
\label{Eq:defEkexc_a}
\\
& = \langle\widetilde\psi_{\bf k}|\alpha_{\bf k}
\hat{H}_{\bf k}\alpha_{\bf k}^\dagger|\widetilde\psi_{\bf k}\rangle
- \langle\widetilde\psi_{\bf k}|
\hat{H}_{\bf k}|\widetilde\psi_{\bf k}\rangle.
\label{Eq:defEkexc_b}
\end{align}
\label{Eq:defEkexc}
\end{subequations}
Here, in going from the first to the second line, we used the fact that the excited state
$\alpha_{\bf k}^\dagger|\widetilde\psi_{\bf k}\rangle$ is normalized to unity,
{\em i.e.} that ${\langle\widetilde\psi_{\bf k}|\alpha_{\bf k}\alpha_{\bf k}^\dagger|\widetilde\psi_{\bf k}\rangle}=1$.
In Appendix \ref{AppendixA} we show that, for weak perturbation, the Hamiltonian $\hat{H}_{\bf k}$ 
can be written in the form:\cite{Leggett2001}
\begin{align}
\hat{H}_{\bf k} \simeq -\frac{1}{2}\big[\varepsilon_{\bf k} + n_Bv({\bf k}) - E_{\bf k}\big] +
\frac{1}{2}E_{\bf k}\big(\alpha_{\bf k}^\dagger\alpha_{\bf k} + \alpha^\dagger_{-\bf k}\alpha_{-\bf k}\big),
\label{Eq:Hcanonical}
\end{align}
where $E_{\bf k}$ is the Bogoliubov spectrum given in Eq. (\ref{Eq:newdefEk}).
This last result, when used in conjunction with Eq. (\ref{Eq:defEkexc})
and the commutation relations for the $\alpha_{\bf k}$'s, Eqs. (\ref{Eq:App:commutator1b})-(\ref{Eq:App:commutator3})
of Appendix \ref{AppendixA}, leads to the conclusion that the excitation energy of the Hamiltonian $\hat{H}_{\bf k}$
from the ground state $|\widetilde\psi_{\bf k}\rangle$
to the excited state $\alpha_{\bf k}^\dagger|\widetilde\psi_{\bf k}\rangle$ 
of momentum $+{\bf k}$ is given by the following result:
\begin{equation}
\Delta E_{exc}^{(1)}({\bf k}) = \frac{1}{2}E_{\bf k} 
=\frac{1}{2}\sqrt{\varepsilon_{\bf k}\big[\varepsilon_{\bf k}+2n_Bv(k)\big]}.
\label{Eq:Eexc1}
\end{equation}
We here would like to emphasize that this is the excitation energy of the Hamiltonian $\hat{H}_{\bf k}$.
The full Hamiltonian $\hat{H} = \sum_{\bf k\neq 0}\hat{H}_{\bf k}$ contains identical contributions from
$\hat{H}_{\bf k}$ and $\hat{H}_{-\bf k}$, and hence one can easily see that the total energy cost 
$\Delta E_{exc}^{tot}({\bf k})$ of
the excitation $\alpha_{\bf k}^\dagger|\widetilde\psi_{\bf k}\rangle$, 
accounting for contributions from $\hat{H}_{\bf k}$ as well
as $\hat{H}_{-\bf k}$, is twice the amount given by Eq. (\ref{Eq:Eexc1}), and is given by the standard 
Bogoliubov result, namely: 
\begin{equation}
\Delta E_{exc}^{tot}({\bf k}) = \Delta E_{exc}^{(1)}({\bf k})+ \Delta E_{exc}^{(1)}(-{\bf k}) = E_{\bf k}.
\end{equation} 

\addvspace{1.5mm}

Before we end this Section, we briefly comment on why we define the excitation operator
$\alpha_{\bf k}$ by the expression (\ref{Eq:defalphak}), and why
we find it useful to impose the condition (\ref{Eq:condalphak}) 
as a way to fix the value of the constants $\tilde{u}_{\bf k}$ and $\tilde{v}_{\bf k}$.
As a matter of fact, it is not difficult to verify that the operator $a_{\bf k}^\dagger a_0$
also creates a state which is orthogonal to the ground state $|\psi_{\bf k}\rangle$
of the operator $\hat{H}_{\bf k}$, and could therefore be a possible choice for an excited state
of the Hamiltonian $\hat{H}_{\bf k}$.
Indeed, the action of the operator $a_{\bf k}^\dagger a_0$ on $|\psi_{\bf k}\rangle$ gives:
\begin{equation}
a_{\bf k}^\dagger a_0|\psi_{\bf k}\rangle = \sum_{n=0}^{N/2} C_n\sqrt{(N-2n)(n+1)}|N-2n-1,n+1,n\rangle.
\end{equation}
Now, the {\em rhs} of the above equation belongs to the Hilbert space spanned by kets of the form
$|N-2n-1,n+1,n\rangle$, which is totally disjoint from the Hilbert space spanned by states of the form
$|N-2n,n,n\rangle$ and to which the ground state belongs, hence the
orthogonality of the state $a_{\bf k}^\dagger a_0|\psi_{\bf k}\rangle$ 
and the ground state $|\psi_{\bf k}\rangle$ of the Hamiltonian $\hat{H}_{\bf k}$. 
By the same token, $\langle\widetilde\psi_{\bf k}|\alpha_{\bf k}^\dagger|\widetilde\psi_{\bf k}\rangle$
is automatically zero, and so $\alpha_{\bf k}^\dagger|\widetilde\psi_{\bf k}\rangle$
is automatically orthogonal to $|\widetilde\psi_{\bf k}\rangle$.
So, why do we choose the special combination 
$\alpha_{\bf k}=\tilde{u}_{\bf k}a_{\bf k}a_0^\dagger + v_{\bf k}a_{-\bf k}^\dagger a_0$
instead of simply $\alpha_{\bf k}=a_{\bf k}a_0^\dagger$? 
And why do we go through the trouble of requesting that $\alpha_{\bf k}|\widetilde\psi_{\bf k}\rangle$ itself
has to vanish?

\addvspace{1.5mm}

To answer the first question, we note that the operator 
$\alpha_{\bf k}^\dagger = \tilde{u}_{\bf k}a_{\bf k}^\dagger a_0 + \tilde{v}_{\bf k}a_{-\bf k}a_0^\dagger$ 
increases the momentum of the system by an amount $+\hbar{\bf k}$. The expression of 
$\alpha_{\bf k}^\dagger$ simply emphasizes the fact that this can be done 
in two different ways. The first way consists in removing a boson from the condensate and adding 
a boson in the single-particle state of wavevector $\bf k$, hence the $\tilde{u}_{\bf k}a_{\bf k}^\dagger a_0$ 
contribution to $\alpha_{\bf k}^\dagger$. The second way consists in removing a boson of wavevector $-{\bf k}$
from the system and adding a boson to the condensate, 
hence the term $\tilde{v}_{\bf k}a_{-\bf k}a_0^\dagger$ in the expression of $\alpha_{\bf k}^\dagger$.
The use of the special combination $\tilde{u}_{\bf k}a_{\bf k}^\dagger a_0 
+ \tilde{v}_{\bf k}a_{-\bf k}a_0^\dagger$ allows us to combine
the two types of processes into one excitation operator, and also allows us to choose 
values for the constants $\tilde{u}_{\bf k}$ and $\tilde{v}_{\bf k}$
such that the ket $\alpha_{\bf k}^\dagger|\widetilde\psi_{\bf k}\rangle$ is automatically normalized to unity.
As for the second question, the reason  we impose the condition (\ref{Eq:condalphak}) is in fact to ensure
that {\em all} states obtained by successive application of the operator $\alpha_{\bf k}^\dagger$ are orthogonal to the
ground state. Indeed, without the condition (\ref{Eq:condalphak}), there is no guarantee that a state of the form 
$(\alpha_{\bf k}^\dagger)^2|\psi_{\bf k}\rangle$ satisfies the orthogonality condition 
$\langle\psi_{\bf k}|(\alpha_{\bf k}^\dagger)^2|\psi_{\bf k}\rangle=0$, even though 
the state $\alpha_{\bf k}^\dagger|\psi_{\bf k}\rangle$ satisfies 
$\langle\psi_{\bf k}|\alpha_{\bf k}^\dagger|\psi_{\bf k}\rangle=
\langle\psi_{\bf k}|\alpha_{\bf k}|\psi_{\bf k}\rangle=0$. 
Equation (\ref{Eq:condalphak}) is necessary because it ensures that all states of the form $(\alpha_{\bf k}^\dagger)^n|\psi_{\bf k}\rangle$
are orthogonal to the ground state $|\psi_{\bf k}\rangle$ of the Hamiltonian $\hat{H}_{\bf k}$, 
a fact that can be easily verified by taking the adjoint of the expression 
$\langle\psi_{\bf k}|(\alpha_{\bf k}^\dagger)^n|\psi_{\bf k}\rangle$.\cite{FieldTheoreticFormulation}
(It can in fact be shown that the condition (\ref{Eq:condalphak}) ensures
that states of the form $(\alpha_{\bf k}^\dagger)^n|\psi_{\bf k}\rangle$
and $(\alpha_{\bf k}^\dagger)^m|\psi_{\bf k}\rangle$ are orthogonal to each other
if $n\neq m$. The proof, which is quite straightforward, will not be presented here.)

\section{Exact numerical diagonalization of the single-mode Hamiltonian $\hat{H}_{\bf k}$}
\label{Sec:ExactDiagOneMode}

Having discussed a number-conserving variational approach to the single-mode Hamiltonian
$\hat{H}_{\bf k}$ of Eq. (\ref{Eq:defHk}),
we now proceed to perform an exact numerical diagonalization of this Hamiltonian. 
Such a numerical treatment, which to the best of the author's knowledge has not been attempted before, 
will allow us to assess the validity of the variational treatment of the previous Section, and possibly
point to areas where the variational approach may need improvement. 
We shall again use the complete basis $|n\rangle \equiv |N-2n,n,n\rangle$  
formed by states with $(N-2n)$ bosons in the ${\bf k}={\bf 0}$ state, 
and $n$ bosons in each of the states ${\bf k}$ and $-{\bf k}$. 
Since $n$ can only take values between $0$ and $N/2$, the basis $\{|n\rangle\}$ 
has $(1+N/2)$ kets, and hence diagonalizing $\hat{H}_{\bf k}$
requires the diagonalization of an $(1+N/2)\times(1+N/2)$ matrix.
For a system of 1000 bosons, this would be a $501\times 501$ matrix, which is not a prohibitively large matrix size to diagonalize numerically given the capabilities of modern day computers.

\addvspace{1.5mm}

Using Eqs. (\ref{Eq:opspsik}), we can easily write the following matrix elements:
\begin{align}
\langle m|\hat{H}_{\bf k}|n\rangle &= \Big[n\varepsilon_{\bf k}
+\frac{v({\bf k})}{V} n(N-2n)
\Big]\delta_{nm}
\nonumber\\
&+\frac{v({\bf k})}{2V}\Big[
(n+1)\sqrt{(N-2n)(N-2n-1)}\delta_{m,n+1}
\nonumber\\
&+ n\sqrt{(N-2n+1)(N-2n+2)}\delta_{m,n-1}
\Big].
\label{Eq:mHn1}
\end{align}

We now introduce dimensionless units, where we measure energies in units of 
$v({\bf 0})n_B = v({\bf 0})N/V$. Hence, we shall write:
\begin{align}
\frac{\langle m|\hat{H}_{\bf k}|n\rangle}{v({\bf 0})n_B} &= 
\Big[n\tilde\varepsilon_{\bf k} + {\tilde v}({\bf k})\frac{n}{N}(N-2n)\Big]\delta_{nm}
\nonumber\\
& + \frac{1}{2}{\tilde v}({\bf k})\Big[
\frac{(n+1)}{N}\sqrt{(N-2n)(N-2n-1)}\delta_{m,n+1}
\nonumber\\
&+ \frac{n}{N}\sqrt{(N-2n+1)(N-2n+2)}\delta_{m,n-1}
\Big],
\label{Eq:mHn2}
\end{align}
where we denote by $\tilde\varepsilon_{\bf k}$ and ${\tilde v}({\bf k})$ the dimensionless quantities:
\begin{subequations}
\begin{align}
& \tilde\varepsilon_{\bf k} = \frac{\varepsilon_{\bf k}}{v({\bf 0})n_B},
\\
& \tilde{v}({\bf k}) = \frac{v({\bf k})}{v({\bf 0})}.
\end{align}
\end{subequations}
Having in mind an interaction potential of the form $v({\bf r})= g\delta({\bf r})$,
whereby $v({\bf k})= g$, throughout this Section we shall present numerical results
using for $\tilde{v}({\bf k})$ the value
\begin{equation}
\tilde{v}({\bf k}) = 1.
\end{equation}
Now, using the dimensionless matrix elements in Eq. (\ref{Eq:mHn2}),
it is a relatively easy task to diagonalize the Hamiltonian $\hat{H}_{\bf k}$
for a given value of the total number of bosons $N$ and 
for various values of the dimensionless variable $\tilde\varepsilon_{\bf k}$.
In the following, we shall present plots of physical quantities of interest as a function 
of the dimensionless wavevector $\tilde{k}$ such that:
\begin{equation}
\tilde{k} = \frac{k}{k_0},
\label{Eq:deftildek}
\end{equation}
where we defined:
\begin{equation}
k_0=\frac{\sqrt{2mn_Bv({\bf 0})}}{\hbar}.
\label{Eq:defk0}
\end{equation}

\subsection{Ground state energy}

\begin{figure}[tb]
\centerline{\includegraphics[width=8.89cm, height=5.5cm]{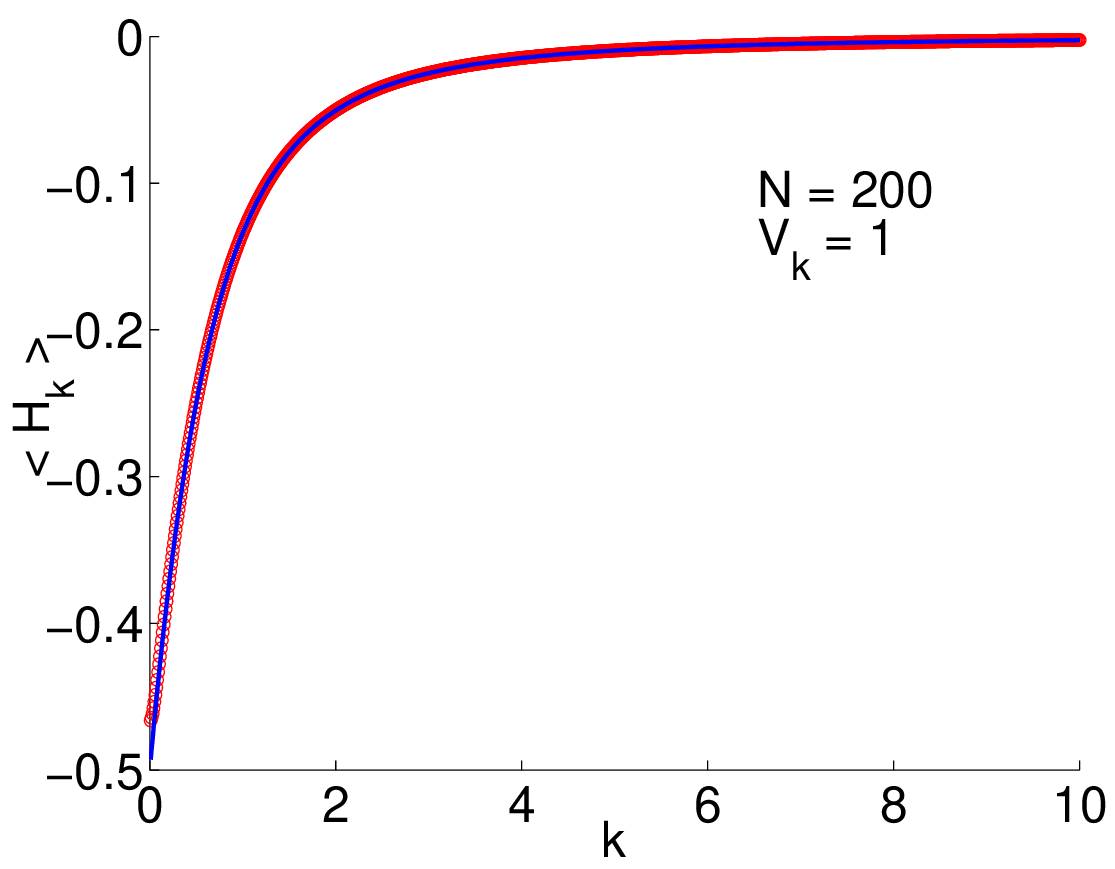}}

\centerline{\includegraphics[width=8.89cm, height=5.5cm]{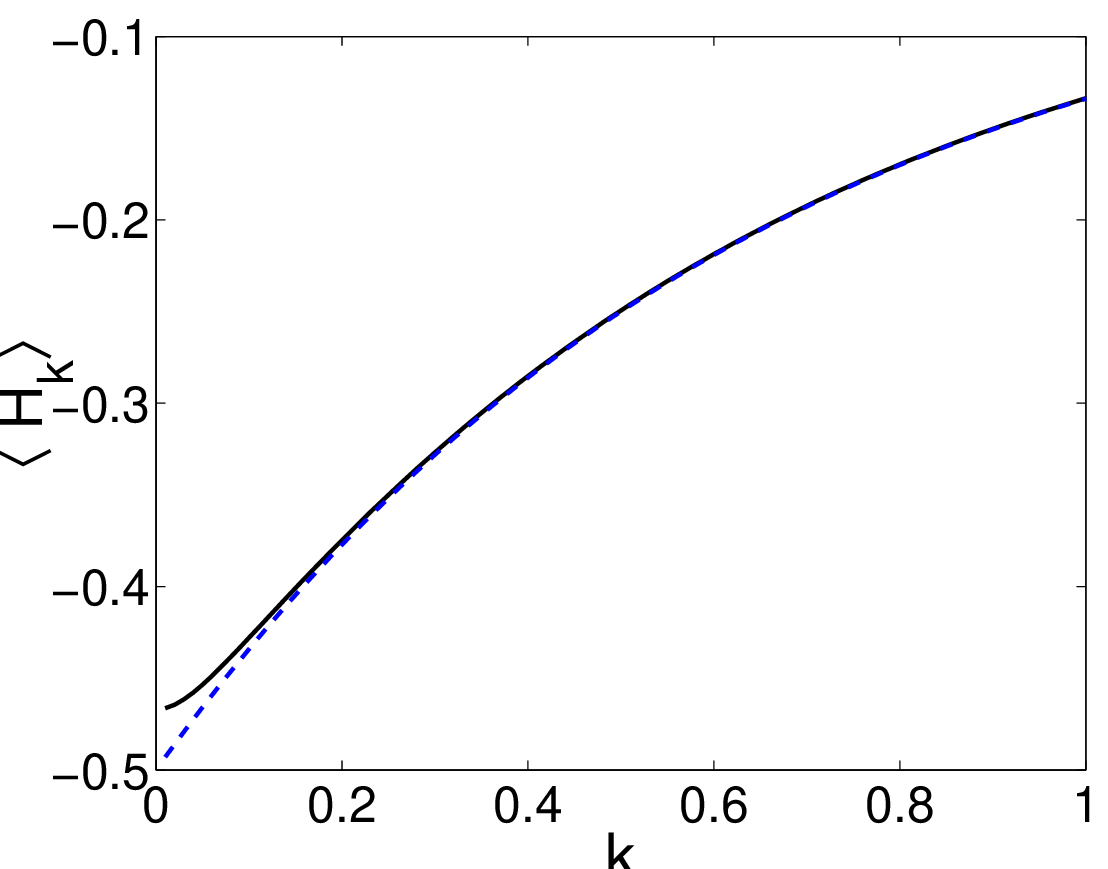}}
\caption[]{Upper panel: Ground state energy of the Hamiltonian $\hat{H}_{\bf k}$ for $N=200$ bosons
as a function of the wavevector $k$. The solid line represents the exact result obtained 
by numerical diagonalization of the Hamiltonian $\hat{H}_{\bf k}$,
while the dashed line is the result of the standard and of the variational Bogoliubov formulations. 
Lower panel: Detail from the above figure at small wavevectors showing that
the exact numerical result (solid line) very slightly deviates from the results of Bogoliubov's theory
(dashed line) near $k=0$.
}\label{Fig:GroundStateEnergySingleMode}
\end{figure}

The first quantity we shall be interested in is the ground state energy of the Hamiltonian $\hat{H}_{\bf k}$.
Numerically, the ground state energy $E_0^{\bf k}(N)$ of the single-mode Hamiltonian $\hat{H}_{\bf k}$
is defined as the smallest eigenvalue of $\hat{H}_{\bf k}$. In Bogoliubov's method, the ground state energy
of $\hat{H}_{\bf k}$ is the expectation value of this last quantity in the
normalized ground state $|\widetilde\psi_{\bf k}\rangle$, 
and is given by Eq. (\ref{Eq:avgHk}). Fig. \ref{Fig:GroundStateEnergySingleMode}
shows plots of $E_0^{\bf k}(N)$ 
obtained by numerical diagonalization of $\hat{H}_{\bf k}$
for $N=200$ bosons on one hand, and by using the Bogoliubov expression 
for $\langle\hat{H}_{\bf k}\rangle_{\bf k}$, Eq. (\ref{Eq:avgHk}), on the other. 
The agreement between Bogoliubov's method and the exact numerical treatment is quite impressive. 
The Bogoliubov results are practically indistinguishable 
from the exact numerical ones for all values of ${k}$, except in a narrow interval near $k=0$, where there is a small
deviation between the two results. Here, we shall not spend too much time 
trying to understand this (rather small) deviation. Suffice it to
say that it is near $k=0$ where $c_{\bf k}\to 1$ that we previously found 
that the standard Bogoliubov theory breaks down, leading to
unphysically diverging results for the averages $\langle a_{\bf k}^\dagger a_{\bf k}\rangle_{\bf k}$ and
$\langle a_{\bf k} a_{-\bf k}\rangle_{\bf k}$. 
Here, we speculate that the deviation of Bogoliubov's result from
the exact one for $E_0^{\bf k}(N)$ may be due to similar reasons; 
in other words, this small deviation may be due to the ground state energy being
evaluated using the truncated version of the sums (\ref{Eq:nxn_asymptotic})-(\ref{Eq:n2xn_asymptotic}) 
and (\ref{Eq:xn_asymptotic}) instead of the complete ones in Eqs. (\ref{Eq:nxn})-(\ref{Eq:n2xn}) and (\ref{Eq:xn}).

\subsection{Ground state wavefunction and depletion of the condensate}

\begin{figure}[tb]
\centerline{\includegraphics[width=8.89cm, height=5.5cm]{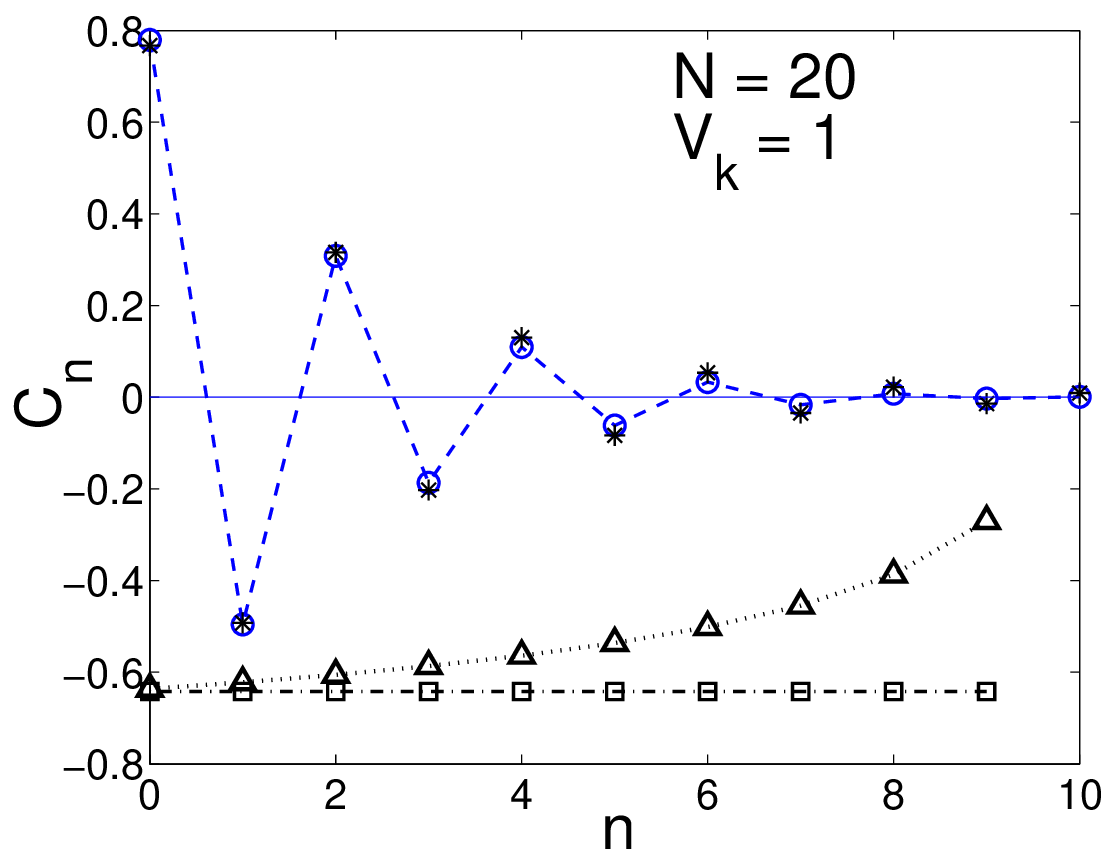}}

\centerline{\includegraphics[width=8.89cm, height=5.5cm]{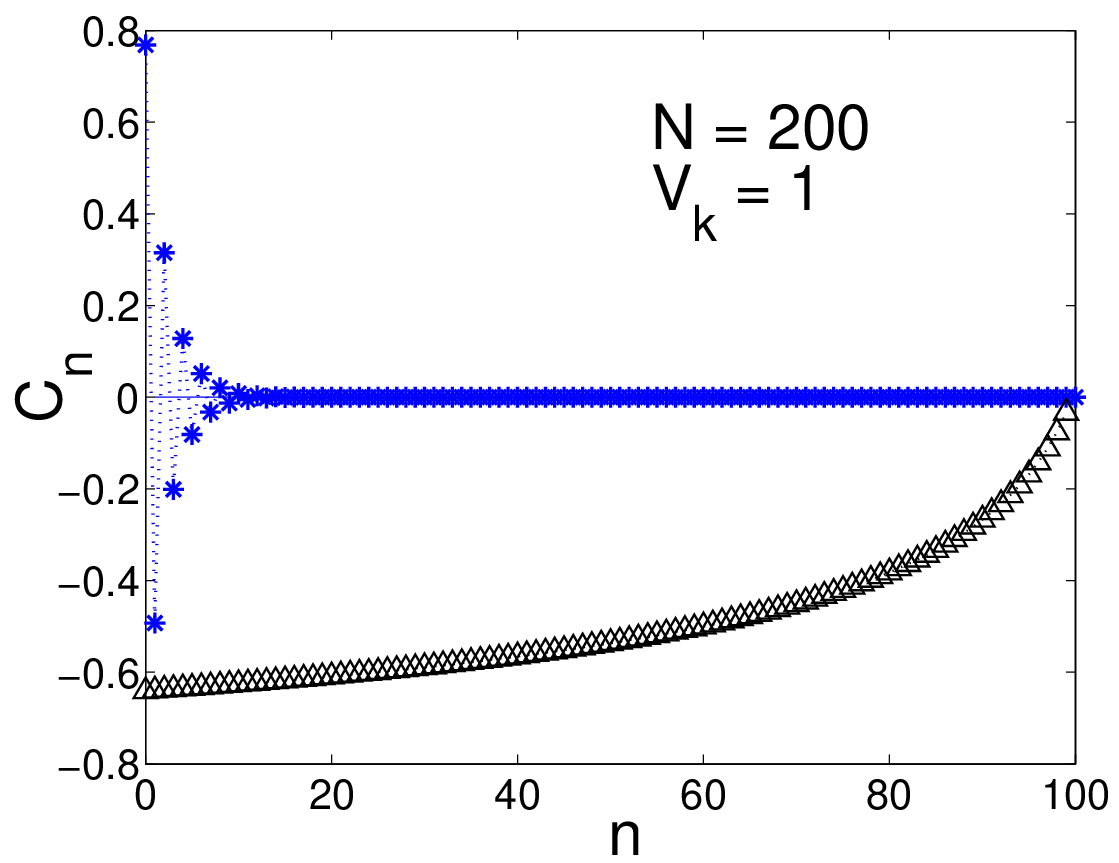}}
\caption[]{Upper panel: Coefficients $\widetilde{C}_n$ of the {\em normalized}
ground state wavefunction $|\widetilde\psi_{\bf k}\rangle$ of the Hamiltonian
$\hat{H}_{\bf k}$ {\em vs.} the index $n$ for $N=20$ bosons and $k=0.1$. The circles
are the results of the exact numerical diagonalization, and the stars are the result
of the variational approach. The triangles show the ratio 
$\widetilde{C}_{n+1}/\widetilde{C}_n$ {\em vs.} $n$. Contrarily to
the variational formulation of Bogoliubov's theory where this ratio is found to be a constant 
given by $c_{\bf k}$ (see the constant line of squares in the figure), 
in the exact treatment the ratio $\widetilde{C}_{n+1}/\widetilde{C}_n$ is an increasing function 
of $n$ and goes to zero as $n\to N/2$. Lower panel: Same as the upper panel for $N=200$ bosons.
}\label{Fig:coefficients1mode}
\end{figure}

While Bogoliubov's theory gives results for the ground state energy $E_0^{\bf k}(N)$
that are in excellent agreement with the exact diagonalization of the single-mode Hamiltonian
$\hat{H}_{\bf k}$, there is only limited agreement between the exact results for the properties
of the ground state wavefunction itself and the results of Bogoliubov's method. 
This is a well-known feature of variational
methods in general, which may, with a judicious choice of trial wavefunctions, give excellent approximations to the 
ground state energy of the system, but quite often only produce tentative agreement as far as the spatial
properties of the wavefunction itself are concerned. In the case at hand, and as it can be seen
in Fig. \ref{Fig:coefficients1mode}, we find that
the coefficients $\widetilde{C}_n$ of the {\em normalized} ground state function $|\widetilde\psi_{\bf k}\rangle$, 
as obtained by exact diagonalization of $\hat{H}_{\bf k}$, do alternate in sign
between positive and negative values as predicted by the variational formulation of Bogoliubov's method.
However, the ratio $\widetilde{C}_{n+1}/\widetilde{C}_{n}$ does not assume a constant value as the latter theory predicts, 
which indicates that Bogoliubov's method produces good qualitative but only approximate quantitative
agreement with the exact treatment as far as the ground state wavefunction 
$|\widetilde\psi_{\bf k}\rangle$ itself is concerned. 
More specifically, we find that the ratio $\widetilde{C}_{n+1}/\widetilde{C}_{n}$
goes to zero at large values of the index $n$ in the numerical 
treatment, which indicates that in the exact solution the highly depleted states $|N-2n,n,n\rangle$ with $n\sim N/2$ are much
more strongly suppressed than in the variational Bogoliubov method. This in turn will have a pronounced effect on the
depletion of the condensate. 

\begin{figure}[tb]
\centerline{\includegraphics[width=8.89cm, height=5.5cm]{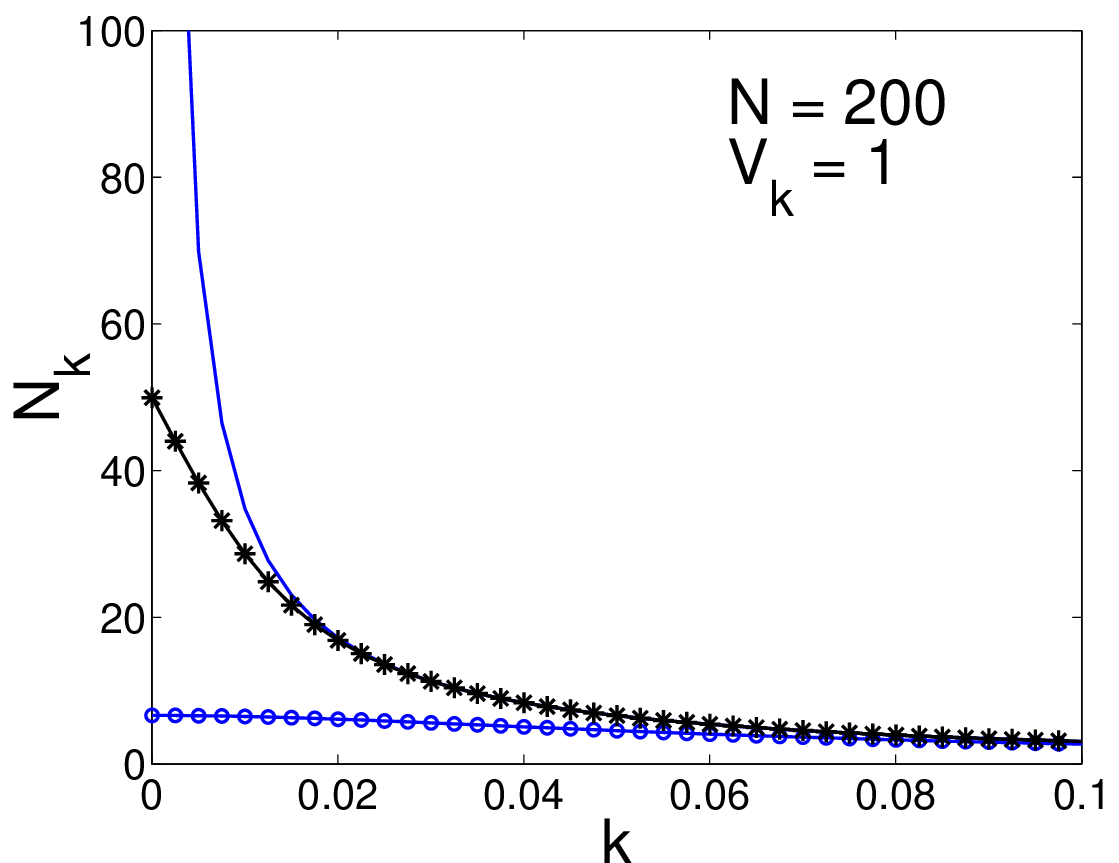}}
\caption[]{Comparison between the depletion of the condensate 
$N_{\bf k}=\langle a_{\bf k}^\dagger a_{\bf k}\rangle_{\bf k}$ vs. wavevector $k$ for $N=200$ bosons, as obtained from
the exact numerical diagonalization of $\hat{H}_{\bf k}$ (circles), the variational number-conserving formulation of
Bogoliubov's method (stars), and the standard number non-conserving Bogoliubov theory (solid line).
}\label{Fig:Depletion200BosonsSingleModeVskThreeMethods}
\end{figure}

\addvspace{1.5mm}

In Fig. \ref{Fig:Depletion200BosonsSingleModeVskThreeMethods}, we plot the depletion
$N_{\bf k}$ vs. wavevector $k$ that is obtained for $N=200$ bosons using three different methods. 
The circles show the exact result:
\begin{equation}
N_{\bf k} = \langle a_{\bf k}^\dagger a_{\bf k}\rangle_{\bf k} = \sum_{n=1}^{N/2} n|\widetilde{C}_n|^2.
\label{Eq:deplk}
\end{equation}
where the $\widetilde{C}_n$'s are obtained by exact numerical diagonalization of $\hat{H}_{\bf k}$. 
The stars on the other hand, show the result obtained from the number-conserving variational formulation 
of Bogoliubov's method, Eq. (\ref{Eq:fullNkvariational}), 
while the solid line shows the result of the standard (number non-conserving) 
Bogoliubov treatment, Eq. (\ref{Eq:resultNk}). It is seen that, 
while the latter result diverges near $k\to 0$ and is therefore unphysical, the variational result is
finite for all values of $k$, and goes to the limiting value $N/4=50$ near $k=0$. 
It is also seen that the variational method overestimates the depletion of the condensate near $k=0$ 
by a factor of about $7$
for the particular value of $N$ chosen. As we mentioned earlier in this paragraph, this is due to the overestimation in
Bogoliubov's variational treatment of the coefficients $\widetilde{C}_n$ for large values of the index $n$, which also contribute to 
the calculation of the depletion of the condensate according to Eq. (\ref{Eq:deplk}).

\addvspace{1.5mm}

\begin{figure}[tb]
\centerline{\includegraphics[width=8.89cm, height=5.5cm]{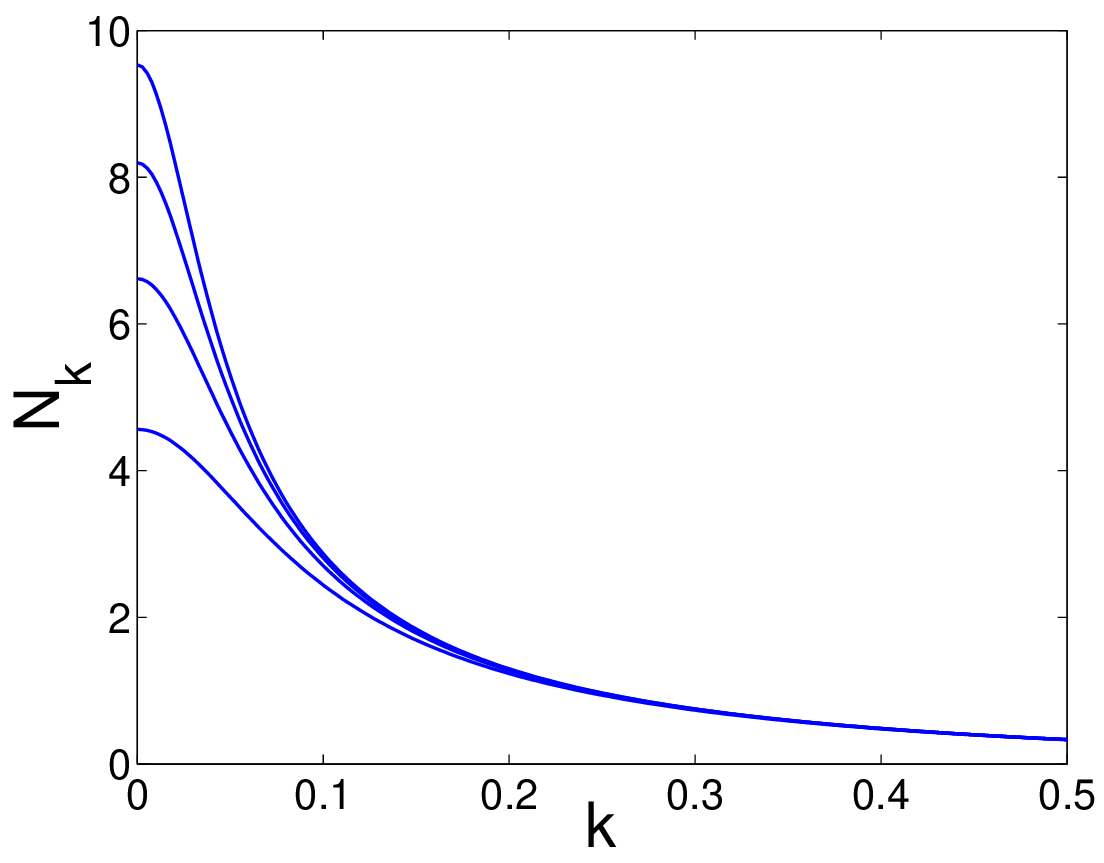}}
\caption[]{Depletion of the condensate $N_{\bf k}=\langle a_{\bf k}^\dagger a_{\bf k}\rangle_{\bf k}$
as obtained from the exact numerical diagonalization of the Hamiltonian $\hat{H}_{\bf k}$ for $N=400,\,300,\,200$ and
$100$ bosons, from top to bottom.
}\label{Fig:ExactDepletionSingleModeVsk}
\end{figure}

In Fig. \ref{Fig:ExactDepletionSingleModeVsk}, we plot the depletion
that is obtained by exact numerical diagonalization of the Hamiltonian $\hat{H}_{\bf k}$
for a number of values of the total number of bosons $N$.
Again, comparison with the results shown in Fig. \ref{Fig:VariationalDepletionVariousN}
shows that the variational method consistently overestimates the depletion near $k=0$ for all
the values of $N$ considered by about one order of magnitude. 
The depletion obtained by exact numerical diagonalization
of the Hamiltonian $\hat{H}_{\bf k}$ is seen to be extremely small, and is in fact much smaller than
what the standard Bogoliubov theory predicts in the $|{\bf k}|\to 0$ limit.

\subsection{Elementary excitations}

The exact numerical diagonalization of $\hat{H}_{\bf k}$ described in the two previous Subsections
gives access not only to the ground state
energy, but also to all excited eigenvalues and eigenfunctions of the system in the Hilbert space
spanned by the kets $|n\rangle=|N-2n,n,n\rangle$, where $0\leq n\leq N/2$. Such excited states can typically
be generated by successive application of the operator $\alpha_{\bf k}^\dagger\alpha_{-\bf k}^\dagger$ 
(to be specific, here the $\alpha_{\bf k}$'s are the operators defined 
in the number-conserving version of Bogoliubov's theory, 
Eq. (\ref{Eq:defalphak})). Indeed, if we calculate the quantity 
$\alpha_{\bf k}^\dagger\alpha_{-\bf k}^\dagger|\psi_{\bf k}\rangle$, where $|\psi_{\bf k}\rangle$ is of the form
$|\psi_{\bf k}\rangle=\sum_{n=0}^{N/2}C_n|N-2n,n,n\rangle$, we find successively:
\begin{subequations}
\begin{align}
& \alpha_{\bf k}^\dagger|\psi_{\bf k}\rangle = \sum_{n=1}^{N/2}A_n|N-2n+1,n,n-1\rangle,
\\
& A_n = C_{n-1}\tilde{u}_{\bf k}\sqrt{n(N-2n+2)} + C_n\tilde{v}_{\bf k}\sqrt{n(N-2n+1)},
\end{align}
\end{subequations}
then:
\begin{subequations}
\begin{align}
&\alpha_{-\bf k}^\dagger\alpha_{\bf k}^\dagger|\psi_{\bf k}\rangle  = \sum_{n=0}^{N/2}B_n|N-2n,n,n\rangle,
\label{Eq:alphasqrd}\\
& B_n = A_{n}\tilde{u}_{\bf k}\sqrt{n(N-2n+1)} 
+ A_{n+1}\tilde{v}_{\bf k}\sqrt{(n+1)(N-2n)},
\end{align}
\end{subequations}
where, in the last equation, it is understood that $A_{1+N/2}=0$. Eq. (\ref{Eq:alphasqrd})
shows that the doubly excited state $\alpha_{\bf k}^\dagger\alpha_{-\bf k}^\dagger|\psi_{\bf k}\rangle$
belongs to the Hilbert space spanned by the kets $|n\rangle\equiv|N-2n,n,n\rangle$,
and hence is accessible through the numerical diagonalization of the Hamiltonian $\hat{H}_{\bf k}$.

\addvspace{1.5mm}

Let us step back for a moment and try to calculate the analytical expression of the 
energy required to excite the system from the ground state $|\psi_{\bf k}\rangle$
to the doubly excited state 
$\alpha_{\bf k}^\dagger \alpha_{-\bf k}^\dagger|\psi_{\bf k}\rangle$. This is the quantity:
\begin{equation}
\Delta E_{exc}^{(2)}({\bf k}) = \frac{\langle\widetilde\psi_{\bf k}|\alpha_{\bf k}\alpha_{-\bf k}\hat{H}_{\bf k}
\alpha_{\bf k}^\dagger \alpha_{-\bf k}^\dagger|\widetilde\psi_{\bf k}\rangle}
{\langle\widetilde\psi_{\bf k}|\alpha_{\bf k}\alpha_{-\bf k}
\alpha_{\bf k}^\dagger \alpha_{-\bf k}^\dagger|\widetilde\psi_{\bf k}\rangle}
- \langle\widetilde\psi_{\bf k}|\hat{H}_{\bf k}|\widetilde\psi_{\bf k}\rangle.
\label{Eq:Eexc2}
\end{equation}
Using the diagonalized expression of $\hat{H}_{\bf k}$ in terms of the $\alpha_{\bf k}$'s, 
Eq. (\ref{Eq:Hcanonical}), we find, after a straightforward calculation (in which the approximation
$[\alpha_{\bf k},\alpha_{\bf k}^\dagger]=\gamma_{\bf k}^2a_0^\dagger a_0\approx 1$,
see Eq. (\ref{Eq:App:commutator3}) from Appendix \ref{AppendixA}, is used throughout)
that $\Delta E_{exc}^{(2)}({\bf k})$ is given by:
\begin{equation}
\Delta E_{exc}^{(2)}({\bf k}) = E_{\bf k}.
\label{Eq:doubleexcE}
\end{equation}
This is not at all surprising. Indeed, since the excitation energy for a singly excited state 
with respect to the Hamiltonian $\hat{H}_{\bf k}$ was found to be 
$E_{\bf k}/2$, here for a doubly excited state we expect an excitation energy $2\times(E_{\bf k}/2)=E_{\bf k}$.

\addvspace{1.5mm}

\begin{figure}[tb]
\centerline{\includegraphics[width=8.89cm, height=5.5cm]{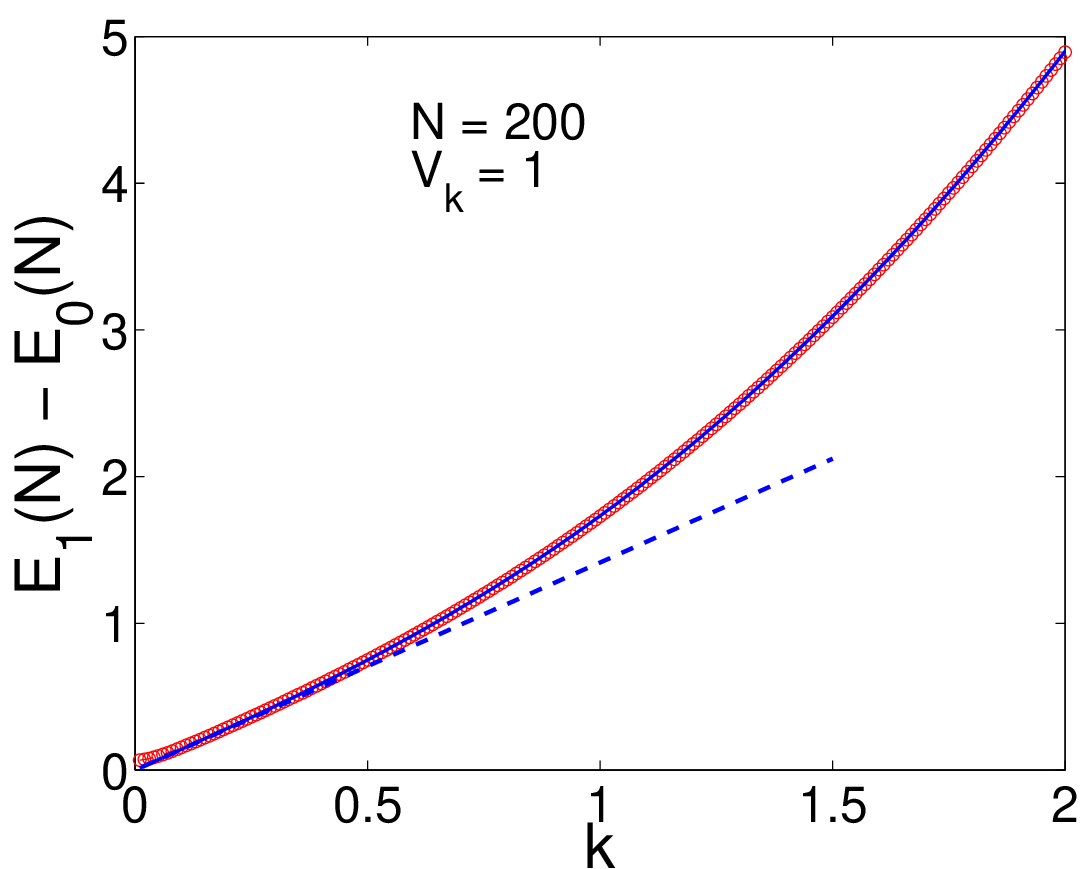}}

\centerline{\includegraphics[width=8.89cm, height=5.5cm]{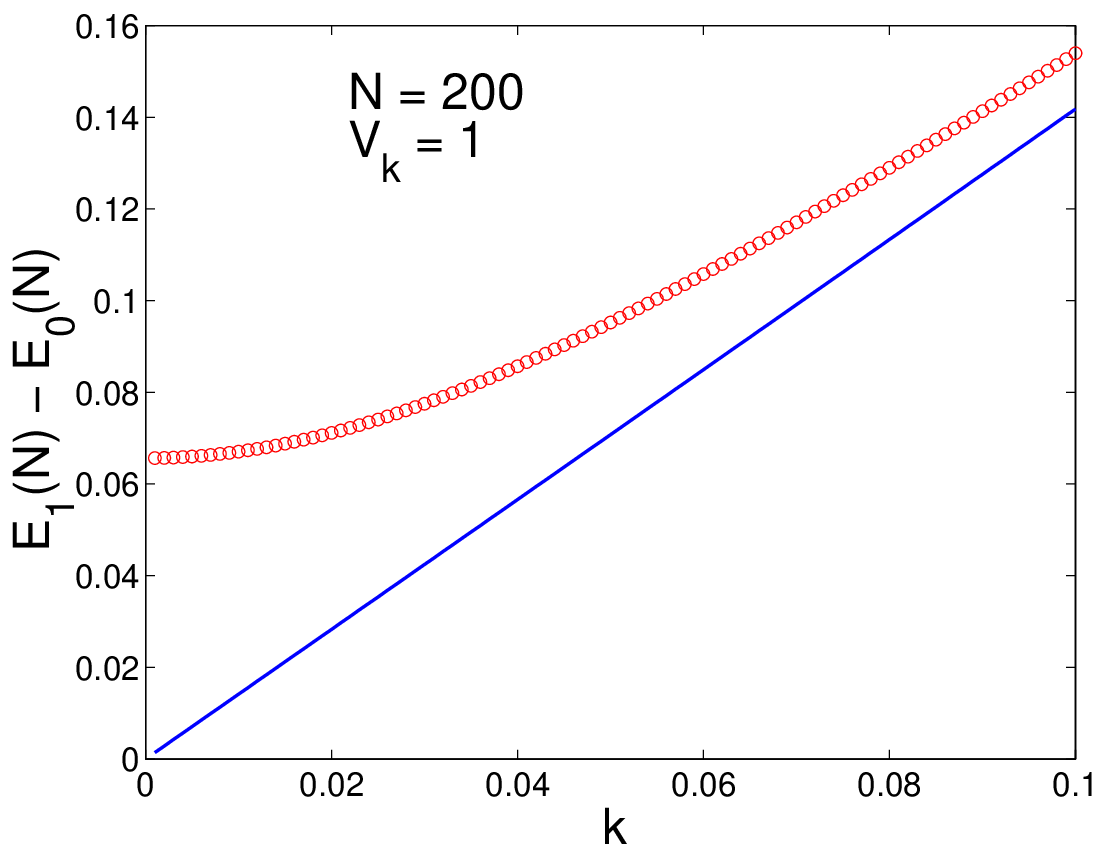}}
\caption[]{Upper panel: Plot of the excitation energy $\Delta E_{exc}^{(2)}({\bf k})$ vs. 
wavevector $k$ for $N=200$ bosons. The solid line is the exact diagonalization result, 
and the dashed line is the result of the variational Bogoliubov method for the Hamiltonian $\hat{H}_{\bf k}$. 
For the most part, these two lines are indistinguishable.
The dotted line indicates the asymptotic form of Bogoliubov's result, $\sqrt{2}\tilde{k}$ 
in dimensionless units, as $k\to 0$.
Lower panel: Detail of the region near $k=0$, where $\Delta E_{exc}^{(2)}({\bf k})$ 
goes to a finite limiting value as $k\to 0$, signaling the existence
of a small energy gap at low momenta, which is most likely due to finite size effects.
}\label{Fig:ExcitationEnergy200BosonsSingleModeVsk}
\end{figure}

\begin{figure}[tb]
\centerline{\includegraphics[width=8.89cm, height=5.5cm]{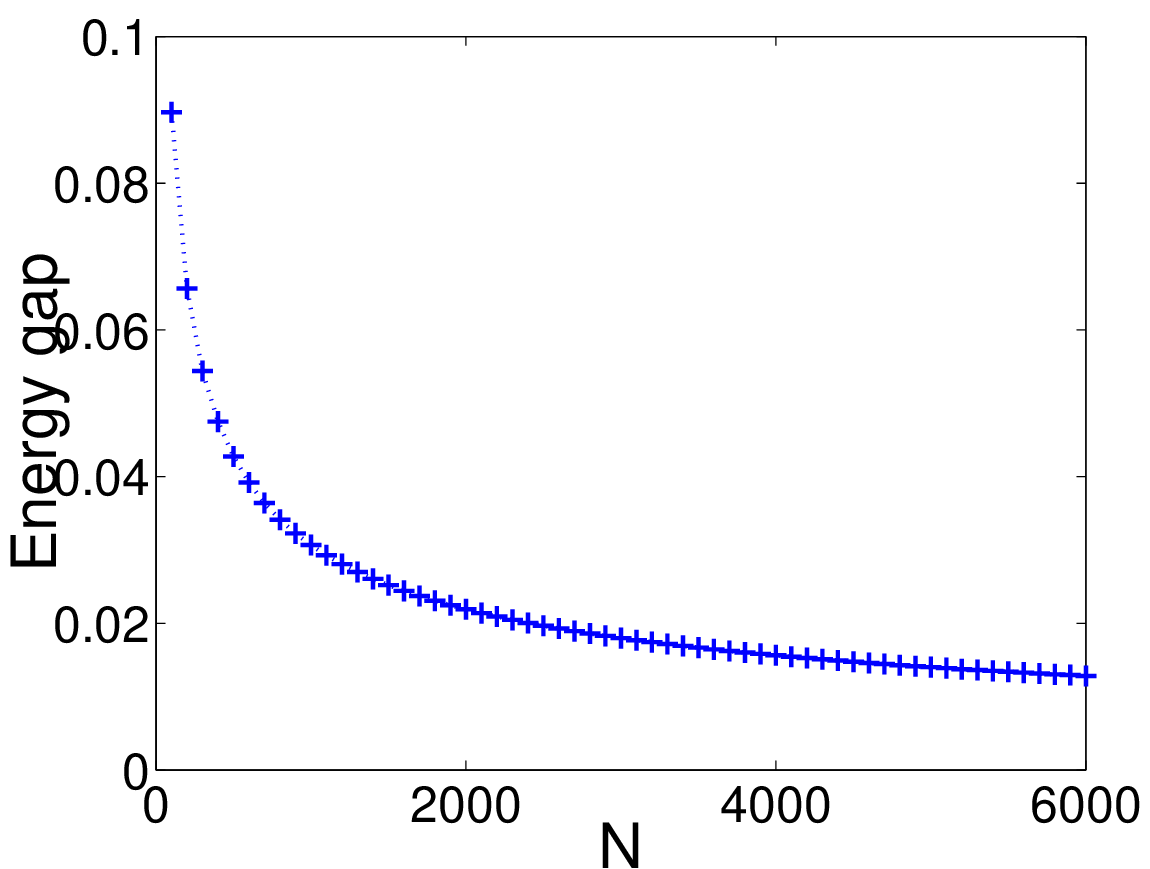}}
\caption[]{Plot of the excitation energy $\Delta E_{exc}^{(2)}({\bf k})$ 
at $\tilde{k}=0.001$ obtained from
the exact numerical diagonalization of the single-mode Hamiltonian $\hat{H}_{\bf k}$
as a function of the number of bosons $N$. 
The fact that the excitation energy decreases with increasing values of $N$ suggests that
the small gap seen in the lower panel of Fig. \ref{Fig:ExcitationEnergy200BosonsSingleModeVsk}
is due to finite-size effects and vanishes in the thermodynamic limit $N\to\infty$.
}\label{Fig:GapSingleModeVsN}
\end{figure}

We now go back to the numerics, and show that the energy of the first excited
state in our numerical treatment coincides with the energy of the double excitation given in
Eq. (\ref{Eq:doubleexcE}). As we mentioned in the opening paragraph of this Subsection, 
dealing with doubly excited states 
of the form $\alpha_{\bf k}^\dagger \alpha_{-\bf k}^\dagger|\psi_{\bf k}\rangle$ has the advantage
of keeping us inside the same Hilbert space we used to diagonalize the Hamiltonian $\hat{H}_{\bf k}$.
Hence, in our numerical diagonalization procedure, if we extract the ground state energy $E_{\bf k}^{(0)}(N)$
and the energy of the first excited state $E_{\bf k}^{(1)}(N)$, then the difference 
$\Delta E_{exc}^{(2)}({\bf k}) = E_{\bf k}^{(1)}(N) - E_{\bf k}^{(0)}(N)$ 
should correspond to the energy of a doubly excited state
of the form $\alpha_{\bf k}^\dagger \alpha_{-\bf k}^\dagger|\psi_{\bf k}\rangle$
and should be given by the {\em rhs} of Eq. (\ref{Eq:doubleexcE}).
The upper panel of Fig. \ref{Fig:ExcitationEnergy200BosonsSingleModeVsk}
shows a comparison between the Bogoliubov result (\ref{Eq:doubleexcE}) and the excitation energy 
$\Delta E_{exc}^{(2)}({\bf k})$ as a function of ${k}$ for $N=200$ bosons. The agreement is
again excellent for most values of $\tilde{k}$, except for small $\tilde{k}$ where the lower panel
of that same figure reveals a deviation from linear behavior and the existence of a small energy
gap near $k=0$, where the excitation energy goes to a finite value as $k\to 0$. 
To probe whether this gap, which it should be noted is much smaller
than the gap predicted by Hartree-Fock theory, is due to a finite-size effect, in Fig. \ref{Fig:GapSingleModeVsN}
we plot the excitation energy obtained numerically at $\tilde{k}=0.001$ for increasing values of the total number
of bosons $N$ in the system. The results obtained show that the value of the gap monotonically
decreases as the number $N$ of bosons increases, which in turn suggests that the gap will go to zero as $N$ grows
infinitely large in the thermodynamic limit.

\section{Generalization: full Bogoliubov Hamiltonian}
\label{Sec:DiagFullH}

Having studied in detail the number-conserving variational formulation
of Bogoliubov's method for the single-mode Hamiltonian $\hat{H}_{\bf k}$,
we now are in a position to tackle the more involved case
of the full Hamiltonian $\hat{H}=\sum_{\bf k\neq 0} \hat{H}_{\bf k}$
of the interacting Bose system. However, before we do so, we
want to examine the relashionship between what we did so far
and previously published work on this system. This will be done next.

\subsection{Connection to previous work}

Up til now, we have been discussing the variational approach to the Hamiltonian
$\hat{H}_{\bf k}$ describing the interaction of the condensate with states of momentum ${\bf k}$ and $-{\bf k}$. 
At this point, we will observe that it is quite remarkable 
that most of the results we derived by diagonalizing $\hat{H}_{\bf k}$ 
perfectly coincide with the results of the standard Bogoliubov theory, implying that the latter 
is effectively a theory in which each one of the Hamiltonians $\hat{H}_{\bf k}$ which contribute
to the total Hamiltonian $\hat{H}=\sum_{\bf k\neq 0}\hat{H}_{\bf k}$ is diagonalized independently
from the other Hamiltonians $\hat{H}_{\bf k'(\neq k)}$ in essentially disjoint Hilbert spaces.
For example, the ground state energy of the system in Bogoliubov's theory
is given by (note that we still take the origin of energies to be the Gross-Pitaevskii value $N(N-1)v(0)/2V$):
\begin{equation}
E_{Bog} = - \frac{1}{2}\sum_{\bf k\neq 0}\big(\varepsilon_{\bf k} + n_Bv(k) - E_{\bf k}\big),
\label{Eq:EBog1}
\end{equation}
and it can be verified that this quantity is nothing but:
\begin{equation}
E_{Bog} = \sum_{\bf k\neq 0}\langle\widetilde\psi_{\bf k}(N)|\hat{H}_{\bf k}|\widetilde\psi_{\bf k}(N)\rangle,
\label{Eq:EBog2}
\end{equation}
where we remind the reader that $|\widetilde\psi_{\bf k}(N)\rangle$ is the normalized ground state
of the {\em single-mode} Hamiltonian $\hat{H}_{\bf k}$. This is rather surprizing, since
we expect the ground state energy of the system to be given by an expression of the form:
\begin{subequations}
\begin{align}
E_{Bog} & = \langle\Psi(N)|\hat{H}|\Psi(N)\rangle,
\label{Eq:EBog3a}
\\
& = \sum_{\bf k\neq 0}\langle\Psi(N)|\hat{H}_{\bf k}|\Psi(N)\rangle,
\label{Eq:EBog3b}
\end{align}
\label{Eq:EBog3}
\end{subequations}
where $|\Psi(N)\rangle$ is the normalized ground state wavefunction of the {\em total} Hamiltonian $\hat{H}$.
In what follows, we want to examine how, within the standard formulation of Bogoliubov's theory,
one can go from Eqs. (\ref{Eq:EBog3}) to the result in Eq. (\ref{Eq:EBog2}), 
which will entail uncovering how the ground state wavefunction 
$|\Psi(N)\rangle$ for the total Hamiltonian $\hat{H}$ has been constructed in the literature
from the ground state functions $|\widetilde\psi_{\bf k}(N)\rangle$ 
of the single-mode Hamiltonians $\hat{H}_{\bf k}$.

\addvspace{1.5mm}

The main argument allowing one to construct the ground state $|\Psi(N)\rangle$ of the total Hamiltonian $\hat{H}$
from the single-mode ground state wavefunctions $|\widetilde\psi_{\bf k}(N)\rangle$ 
is based on the following reasoning. 
As we have already seen in Sec. \ref{Sec:StandardBogoliubov}, in the standard formulation of Bogoliubov's method the
creation and annihilation operators of bosons in the condensate, $a_0^\dagger$ and $a_0$, are replaced by the c-number
$\sqrt{N_0}\approx\sqrt{N}$. After this replacement is made, 
the various Hilbert spaces where the single-mode Hamiltonians $\hat{H}_{\bf k}$ act are
totally disjoint from each other.
This in turn implies that the single-mode Hamiltonians $\{\hat{H}_{\bf k}\}$ can be diagonalized
independently from one another, and that the ground state wavefunction of the system can be written 
as the {\em product} of the ground state wavefunctions of each of the single-mode Hamiltonians
$\hat{H}_{\bf k}$: 
\begin{equation}
|\Psi(N)\rangle = \prod_{i=1}^\infty|\widetilde\psi_{{\bf k}_i}(N)\rangle.
\label{Eq:ProductPsik}
\end{equation}
Now, it can be shown that, even though the operators $a_0$ and $a_0^\dagger$ have been replaced by $\sqrt{N}$
in the expression of the Hamiltonian $\hat{H}$, minimization of the expectation 
value of the single-mode Hamiltonian $\hat{H}_{\bf k}$ in the single-mode ground state
$|\widetilde\psi_{\bf k}(N)\rangle \simeq\sqrt{1-c_{\bf k}^2}\sum_{n=0}^{N/2}(-c_{\bf k})^n|n\rangle$
leads to the same variational coefficients $c_{\bf k}$ as in Eq. (\ref{Eq:resultck}).
This leads, in the notation of the classic work of Lee, Huang and Yang \cite{LeeHuangYang1957} 
to the following expression of the normalized ground state $|\Psi(N)\rangle$ of the Hamiltonian $\hat{H}$ 
(see also Ref. \citen{HuangBook}):
\begin{align}
|\Psi(N)\rangle  = Z\sum_{n_1=0}^\infty \ldots\sum_{n_\infty=0}^\infty
(-c_{{\bf k}_1})^{n_1}\cdots (-c_{{\bf k}_\infty})^{n_\infty}
|n_1,n_1;\ldots;n_\infty,n_\infty\rangle,
\label{Eq:fullPsi(N)0}
\end{align}
where:
\begin{equation}
|n_1,n_1;\ldots;n_\infty,n_\infty\rangle = \prod_{i=1}^\infty \frac{\big(a_{{\bf k}_i}^\dagger)^{n_i}}{\sqrt{n_i!}}
\frac{\big(a_{{-\bf k}_i}^\dagger)^{n_i}}{\sqrt{n_i!}}|0\rangle.
\label{Eq:standardBasis}
\end{equation}
In the above equations, $n_i$ is the number of bosons in the states $\pm{\bf k}_i$,
and the coefficient 
$C_{n_1,\ldots,n_\infty}=(-c_{{\bf k}_1})^{n_1} (-c_{{\bf k}_2})^{n_2}\cdots (-c_{{\bf k}_\infty})^{n_\infty}$ 
represents the probability amplitude of finding the system in the many-body 
state $|n_1,n_1;\cdots;n_\infty,n_\infty\rangle$. 
In principle, we would like the coefficients $C_{n_1,\ldots,n_\infty}$ 
of the wavefunction $|\Psi(N)\rangle$ to vanish if $n_1 + n_2 +\cdots + n_\infty \ge N/2$:
\begin{equation}
C_{n_1,\ldots,n_\infty} = 0 \quad \mbox{if}\quad n_1 + n_2 +\cdots + n_\infty \ge N/2.
\label{Eq:constraintCn}
\end{equation}
However, enforcing the above constraint is not expected to give rise
to any significant change in the value of the ground state energy in the thermodynamic $N\to\infty$ limit.
Going back to Eq. (\ref{Eq:fullPsi(N)0}), the quantity $Z$ defined in that equation is the normalization constant: 
\cite{LeeHuangYang1957}
\begin{equation}
Z \simeq \prod_{i=1}^\infty \sqrt{1 - c_{ {\bf k}_i }^2}.
\label{Eq:defZ}
\end{equation}
With the definitions (\ref{Eq:fullPsi(N)0}) and (\ref{Eq:defZ}), one can verify that the ground state
energy $E_0(N) = \langle\Psi(N)|\hat{H}|\Psi(N)\rangle$ is given by the {\em rhs} of Eq. (\ref{Eq:EBog1}),
which is again the same as the {\em rhs} of Eq. (\ref{Eq:EBog2}).
One can also verify that the depletion of the condensate is still given by an expression
of the form (\ref{Eq:single-mode-depletion}), namely:
\begin{equation}
\langle\Psi(N)|a_{\bf k}^\dagger a_{\bf k}|\Psi(N)\rangle = \frac{c_{\bf k}^2}{1 - c_{\bf k}^2}.
\end{equation}
Finally, {\em if} we forego the number conserving expression of the operators $\alpha_{\bf k}$ 
and $\alpha_{\bf k}^\dagger$ derived in Sec. \ref{Sub:ExcOneMode} above, 
and use the number non-conserving approximations of Eq. (\ref{Eq:defalphas0}),
\begin{equation}
\alpha_{\bf k} = u_{\bf k}a_{\bf k} + v_{\bf k}a^\dagger_{-\bf k},
\quad
\alpha_{\bf k}^\dagger = u_{\bf k}a_{\bf k}^\dagger + v_{\bf k}a_{-\bf k},
\label{Eq:defalphas01}
\end{equation}
with $u_{\bf k}$ and $v_{\bf k}$ the quantities defined in Eq. (\ref{Eq:uv_2}),
one can easily verify that the action of the operator $\alpha_{{\bf k}}$ on $|\Psi(N)\rangle$ yields zero,
and that the excitation spectrum generated by the $\alpha_{\bf k}^\dagger$ operator is indeed given
by the Bogoliubov spectrum $E_{\bf k} = \sqrt{\varepsilon_{\bf k}(\varepsilon_{\bf k} + 2n_Bv({\bf k}))}.$

\addvspace{1.5mm}

At this juncture, we would like to attract the reader's attention to the fact that 
in the expression (\ref{Eq:fullPsi(N)0}) of the wavefunction $|\Psi(N)\rangle$,
the number of bosons in the ${\bf k}=0$ is not explicitly specified
(this is indeed how Lee {\em et al.} write the ground state wavefunction
in their work, Ref. \citen{LeeHuangYang1957}). One may correctly observe that 
the number of bosons in the ${\bf k}=0$ state needs not be written explicitly, because once we specify how many bosons
$n_i$ are in each state with wavevector $\pm{\bf k}_i$, then the number of condensed bosons is
automatically known and is given by $(N-2\sum_i n_i)$. But in fact the reason Lee {\em et al.}
do not specify the number of bosons
in the ${\bf k}=0$ state is a more trivial one, and has to do with the fact that these authors use Bogoliubov's
approximation of replacing the operators $a_0$ and $a_0^\dagger$ by the c-number $\sqrt{N}$ (one place where this
is made obvious is the expressions of the excitation operators they use, $\alpha_{\bf k}$ and $\alpha_{\bf k}^\dagger$,
which does not contain any trace of the operators $a_0$ and $a_0^\dagger$, unlike to the $\alpha_{\bf k}$'s used by
Leggett in Ref. \citen{Leggett2001}, which are similar to those we used in Sec. \ref{Sub:ExcOneMode} above), 
and hence their calculation is one where the ${\bf k}=0$ state 
is removed altogether from the Hilbert space used to describe the system.
A major question that therefore arises is to know if and how the standard results that were derived using
the expression (\ref{Eq:fullPsi(N)0}) of $|\Psi(N)\rangle$ will change if we 
put the ${\bf k}=0$ back into the Hilbert space, and we keep a more accurate tally of how many bosons 
are present in this ${\bf k}=0$ state, hence enforcing the conservation of boson number.
This will be the subject of the next Subsection.

\subsection{Variational treatment of the full Hamiltonian $\hat{H}=\sum_{\bf k\neq 0}\hat{H}_{\bf k}$}
\label{Sub:GeneralizationFullH}

We now want to generalize the variational approach of Sec. \ref{Sec:VariationalMethod} to treat
the {\em full} Hamiltonian $\hat{H}=\sum_{\bf k\neq 0}\hat{H}_{\bf k}$ of the interacting Bose system.
To this end, we shall use for $|\Psi(N)\rangle$ an expression of the
form (we here denote by $M$ the total number of momentum
modes kept in the calculation, which will eventually be sent to infinity):
\begin{align}
|\Psi(N)\rangle & = Z\sum_{n_1=0}^{n_{1max}} \ldots\sum_{n_M=0}^{n_{Mmax}}
C_{n_1}C_{n_2}\ldots C_{n_M}
|N-2\sum_{i=1}^M n_i; n_1,n_1;\ldots;n_M,n_M\rangle,
\label{Eq:fullPsi(N)}
\end{align}
where the normalized basis wavefunctions are given by (compare with Eq. (\ref{Eq:standardBasis})):
\begin{align}
|N-2\sum_{i=1}^M n_i;n_1,n_1;\ldots;n_M,n_M\rangle  = 
\frac{\big(a_0^\dagger\big)^{N-2\sum_{i=1}^M n_i}}{\sqrt{[N-2\sum_{i=1}^M n_i]!}}
\prod_{i=1}^M \frac{\big(a_{{\bf k}_i}^\dagger)^{n_i}}{\sqrt{n_i!}}
\frac{\big(a_{{-\bf k}_i}^\dagger)^{n_i}}{\sqrt{n_i!}}|0\rangle.
\end{align}
Note that the ground state wavefunction in Eq. (\ref{Eq:fullPsi(N)}) is {\em not} a simple product of 
ground state wavefunctions of the single-mode Hamiltonians $\hat{H}_{\bf k}$, 
as in Eq. (\ref{Eq:ProductPsik}), and that, even
though the $\hat{H}_{\bf k}$'s may seem to be decoupled,
the presence of all the $n_i$'s in the number of condensed bosons $\big(N-2\sum_{i=1}^Mn_i\big)$
acts like an implicit and rather nontrivial coupling between all these single-mode Hamiltonians.
Note also that the summations over the $n_i$'s in Eq. (\ref{Eq:fullPsi(N)}) extend from $0$ to a value $n_{i,max}$,
emphasizing the constraint that the number of bosons in the ${\bf k}=0$ state in each
of the basis wavefunctions has to be greater than or equal to zero, {\em i.e.}
$\sum_{i=1}^Mn_i\leq N/2$ (see Eq. (\ref{Eq:constraintCn})). A possible (but, by no means, unique) 
choice for the values $n_{i,max}$ is given by:
\begin{subequations}
\begin{align}
&n_{1,max} = N/2,
\\
&n_{2,max} = N/2 - n_1,
\\
&\ldots\nonumber
\\
&n_{M,max} = N/2 - n_1 -\cdots - n_{M-1}.
\end{align}
\label{Eq:constraints_nmax}
\end{subequations}
It can be verified that the above parametrization of the $n_{i,max}$'s
exhausts all possible states compatible with the constraint (\ref{Eq:constraintCn}). 
In what follows, however, we shall place ourselves in the thermodynamic limit $N\to \infty$ where it can be shown that it is 
not important to keep track of the constraints (\ref{Eq:constraints_nmax}),
and in the evaluation of physical expectation values we shall for simplicity extend the summations
to infinity, as is done in the standard Bogoliubov theory of Eq. (\ref{Eq:fullPsi(N)0}). 
However, we shall find it essential to keep track of the number of bosons in the ${\bf k}=0$ state, as this
significantly modifies some important results of the standard Bogoliubov approach, changing
the nature of the excitation sprectrum from a gapless one to one which has a finite energy gap
as $k\to 0$.

\addvspace{1.5mm}

As we mentioned above, although it may appear at first glance that the wavefunction given in Eq. (\ref{Eq:fullPsi(N)})
should allow us to reproduce the salient features of Bogoliubov's theory, a careful analysis shows
that, in fact, different results are obtained for the ground state energy and for the energy
of elementary excitations, since now the number of particles in the condensate is not given
by $(N-2n_i)$ as was the case in the variational approach for the single-mode Hamiltonian $\hat{H}_{{\bf k}_i}$, 
but is given by the more complicated expression
$\big(N-2\sum_{i=1}^M n_i\big)$ involving {\em all} momentum modes. In particular, 
in Appendix \ref{AppendixB} we show that the expectation value
of the Hamiltonian $\hat{H}_{{\bf k}_j}$ in the state $|\Psi(N)\rangle$
is no longer given by Eq. (\ref{Eq:expectation_h_min}), but by the following expression:
\begin{align}
\langle\Psi(N)|\hat{H}_{{\bf k}_j}|\Psi(N)\rangle \simeq 
\frac{c_{{\bf k}_j}^2}{1-c_{{\bf k}_j}^2}\Big[\varepsilon_{{\bf k}_j}+\bar{v}(k_j)n_B\Big] 
 - \bar{v}(k_j)n_B\frac{c_{{\bf k}_j}}{1-c_{{\bf k}_j}^2},
\label{Eq:expectation_Hk}
\end{align}
where $\bar{v}(k_j)$ is given by:
\begin{align}
\bar{v}(k_j) = v(k_j)\Big( 1 - \frac{2}{N}\sum_{i=1(\neq j)}^M
\frac{c_{{\bf k}_i}^2}{1-c_{{\bf k}_i}^2}\Big).
\label{Eq:defbarv}
\end{align}
If it were not for the term between parenthesis in this last equation, the result in Eq. (\ref{Eq:expectation_Hk})
would be perfectly identical to the expectation value obtained within the single-mode approach, 
Eq. (\ref{Eq:expectation_h_min}). Minimization of the expectation value of the total Hamiltonian
$\langle\Psi(N)|\hat{H}|\Psi(N)\rangle$ with respect to the variational parameters $\{c_{\bf k}\}$
would then give the same values for these parameters
as in the single-mode case, Eq. (\ref{Eq:resultck}), and hence
everything we discussed in Sec. \ref{Sec:VariationalMethod} would remain pretty much unchanged.  
Below we will show that minimization of $\langle\Psi(N)|\hat{H}|\Psi(N)\rangle$ as given
by Eq. (\ref{Eq:expectation_Hk}), {\em i.e.} with $v({\bf k})$ replaced by $\bar{v}({\bf k})$,
leads to a very different solution for the $c_{\bf k}$'s,
leading in turn to some important changes for the ground state properties of 
the interacting Bose system.

\addvspace{1.5mm}

Going back to the result of Eq. (\ref{Eq:expectation_Hk}),
if we now minimize the expectation value of the {\em total} Hamiltonian
$\hat{H}=\sum_{j=1}^M \hat{H}_{{\bf k}_j}$, which is given by:
\begin{align}
\langle\Psi(N)|\hat{H}|\Psi(N)\rangle = \sum_{j=1}^M\Bigg\{
\Big[ 
\varepsilon_{{\bf k}_j} + n_B \bar{v}({\bf k}_j)
\Big]\frac{c_{{\bf k}_j}}{1 - c_{{\bf k}_j}^2}
- n_B \bar{v}({\bf k}_j)
\frac{c_{{\bf k}_j}}{1 - c_{{\bf k}_j}^2}
\Bigg\},
\label{Eq:expvaluefullH}
\end{align}
with respect to the constants $c_{{\bf k}_j}$, 
instead of Eq. (\ref{Eq:quadratic_ck}) we now obtain the following equation
(see Appendix \ref{AppendixC}):
\begin{align}
c_{{\bf k}_j}^2 - 2\Big(\frac{\tilde{\cal E}_{{\bf k}_j}}{n_B\bar{v}({\bf k}_j)}\Big)c_{{\bf k}_j} + 1 =  0.
\label{Eq:Eqckfull}
\end{align}
In the above equation, $\tilde{\cal E}_{\bf k_j}$ now denotes the quantity:
\begin{subequations}
\begin{align}
\tilde{\cal E}_{\bf k_j}  & = \varepsilon_{{\bf k}_j} + n_B \bar{v}({\bf k}_j) + \sigma_{\bf k},
\\
\mbox{with}\quad\sigma_{{\bf k}_j} & = \frac{2}{N}\sum_{i=1(\neq j)}^M n_Bv(k_i)
\frac{c_{{\bf k}_i}}{1+c_{{\bf k}_i}}.
\label{Eq:defsigmak}
\end{align}
\end{subequations}
Solving Eq. (\ref{Eq:Eqckfull}) for $c_{\bf k}$, we obtain:
\begin{align}
c_{\bf k} = \Big(\frac{\tilde{\cal E}_{\bf k}}{n_B\bar{v}({\bf k})}\Big) - 
\sqrt{\Big(\frac{\tilde{\cal E}_{\bf k}}{n_B\bar{v}({\bf k})}\Big)^2-1}.
\label{Eq:newc_k}
\end{align}
where again the sign of the second term has been chosen so that $0<c_{\bf k}<1$.
Expressions (\ref{Eq:expvaluefullH}) for the ground state expectation value of the Hamiltonian 
and (\ref{Eq:newc_k}) for the coefficients of the ground state wavefunction are the main results of this
paper. In the rest of this Section we want to explore how these expressions, which,
to the best of the author's knowledge, have not been studied previously, alter the description
of interacting Bose systems given in the standard Bogoliubov formulation.

\addvspace{1.5mm}

We now rewrite the quantity $\sigma_{\bf k}$ of Eq. (\ref{Eq:defsigmak}) in the form:
\begin{equation}
\sigma_{{\bf k}_j} = \frac{2}{N}\sum_{i=1}^M n_Bv(k_i)
\frac{c_{{\bf k}_i}}{1+c_{{\bf k}_i}} 
- \frac{2}{N} n_Bv(k_j)
\frac{c_{{\bf k}_j}}{1+c_{{\bf k}_j}}. 
\label{Eq:sigmaint}
\end{equation}
Using the fact that $n_B=N/V$, and neglecting the second term on the {\em rhs} of this last equation,
(which is of order $(1/N)$) we obtain:
\begin{equation}
\sigma_{{\bf k}_j} \simeq \frac{2}{V}\sum_{i=1}^M v(k_i)
\frac{c_{{\bf k}_i}}{1+c_{{\bf k}_i}} . 
\label{Eq:sigmaint2}
\end{equation}
Transforming the sum into an integral, we obtain
(notice that the factor of 2 disappears because the summation in Eq. (\ref{Eq:sigmaint2}) is over half
of phase space only -- notice also that the number of momentum modes $M$ has been sent to infinity):
\begin{equation}
\sigma_{\bf k} \simeq \int_{\bf k'} v(k')\frac{c_{\bf k'}}{1+c_{\bf k'}}.
\label{Eq:sigmakintdivg}
\end{equation}
As it can be seen, the quantity on the {\em rhs} of the above equation
does not depend on $k$, and we shall henceforth drop the subscript ${\bf k}$
from $\sigma_{\bf k}$, and rewrite Eq. (\ref{Eq:defsigmak}) in the form:
\begin{subequations}
\begin{align}
\tilde{\cal E}_{\bf k_j}  & = \varepsilon_{{\bf k}_j} + n_B \bar{v}({\bf k}_j) + \sigma,
\\
\mbox{with}\quad\sigma & = \int_{\bf k} v(k)\frac{c_{\bf k}}{1+c_{\bf k}}.
\label{Eq:defsigmak2b}
\end{align}
\label{Eq:defsigmak2}
\end{subequations}

In the particular case where $v({\bf k}) = g$, corresponding to $v({\bf r})=g\delta({\bf r})$ in real space, 
it is not difficult to see that the integral in Eq. (\ref{Eq:defsigmak2b}) 
has an ultraviolet divergence in three dimensions, since
$c_{\bf k}$ as given in Eq. (\ref{Eq:newc_k})
behaves like $1/k^2$ as $k\to \infty$. To circumvent this difficulty, instead of the interaction
potential $v({\bf r})= g\delta({\bf r})$ we shall use:
\begin{equation}
v({\bf r}) = \frac{ge^{-r^2/(2\lambda^2)}}{(2\pi\lambda^2)^{3/2}},
\label{Eq:pot_lambda_RS}
\end{equation}
where $\lambda$ is a positive quantity having the dimensions of length. This expression of the interaction
potential has been chosen to ensure that we recover the form $v({\bf r})=g\delta({\bf r})$
in the limit $\lambda \to 0$. In Fourier space, the interaction potential 
of Eq. (\ref{Eq:pot_lambda_RS}) is given by:
\begin{equation}
v({\bf k}) = ge^{-\frac{1}{2}k^2\lambda^2}.
\label{Eq:pot_lambda_FS}
\end{equation}
With this expression of $v({\bf k})$, the integral on the {\em rhs} of Eq. (\ref{Eq:defsigmak2b})
converges. Going back to this last equation, we see that the integral on the {\em rhs} depends on $\sigma$
through $c_{\bf k}$, and hence we see that Eq. (\ref{Eq:defsigmak2b}) can be seen as a self-consistent
equation for $\sigma$. In the following, we shall solve this self-consitent equation and find the value
of $\sigma$ for a given value of the paramters $g$, $n_B$ and $\lambda$. But, before we do that, we want to
go back and re-examine the quantity $\bar{v}({\bf k})$ we introduced in Eq. (\ref{Eq:defbarv}).

Let us rewrite the {\em rhs} of Eq. (\ref{Eq:defbarv}) in the form:
\begin{align}
\bar{v}(k_j) = v(k_j)\Big( 1 - \frac{2}{N}\sum_{i=1}^M\frac{c_{{\bf k}_i}^2}{1-c_{{\bf k}_i}^2}
+ \frac{2}{N}\frac{c_{{\bf k}_j}^2}{1-c_{{\bf k}_j}^2}\Big).
\end{align}
The last term between brackets in the above equation is of order $(1/N)$, and hence
can be neglected in the thermodynamic $N\to \infty$ limit. On the other hand, the sum
$\sum_{i=1}^M c_{{\bf k}_i}^2/(1-c_{{\bf k}_i}^2)$ is recognized as the expectation value
$\langle\Psi(N)|\sum_{i=1}^M a_{{\bf k}_i}^\dagger a_{{\bf k}_i}|\Psi(N)\rangle$, where the summation
extends over half of the wavevectors ${\bf k}$ in phase space. Hence, we can write:
\begin{align}
\bar{v}({\bf k}) = v({\bf k})\Big(1 - \frac{N_d}{N}\Big),
\label{Eq:vbarv}
\end{align}
where $N_d$ is the total number of depleted bosons:
\begin{subequations}
\begin{align}
N_d & = 2\sum_{i=1}^M \frac{c_{{\bf k}_i}^2}{1 - c_{{\bf k}_i}^2} 
\\
& = \sum_{\bf k\neq 0} \frac{c_{\bf k}^2}{1 - c_{\bf k}^2} 
\end{align}
\end{subequations}
For ease of notation, we shall introduce a new symbol $C_d$, which we shall dub the ``depletion factor", such that:
\begin{equation}
C_d = 1 - \frac{N_d}{N}.
\label{Eq:def_Cd}
\end{equation} 
Then, Eq. (\ref{Eq:vbarv}) can be rewritten in the form:
\begin{align}
\bar{v}({\bf k}) = C_dv({\bf k}),
\label{Eq:vbarv2}
\end{align}

We now are ready to tackle the self-consistent equation for $\sigma$. To this end,
we will again use dimensionless units, where energies are measured in units of $n_Bv({\bf 0})=gn_B$, 
and wavevectors are measured in units of $k_0=\sqrt{2mn_Bg}/\hbar$, and we will find it convenient
to express the interaction strength $g$ in terms of the $s$-wave scattering length $a$, such that:
\begin{equation}
\frac{4\pi a\hbar2}{m} = g.
\label{Eq:def:scatteringlength}
\end{equation}
Then we can write for $c_{\bf k}$
the following expression:
\begin{align}
c_{\bf k} = 1 + \frac{1}{C_d}(\tilde{k}^2 + \tilde\sigma)e^{4\pi n_B a\lambda^2\tilde{k}^2}
- \sqrt{\Big[ 1 + \frac{1}{C_d}(\tilde{k}^2 + \tilde\sigma)e^{4\pi n_B a\lambda^2\tilde{k}^2} \Big]^2 - 1},
\label{Eq:res_ck}
\end{align}
where we denote by $\tilde{k}$ and $\tilde\sigma$ the dimensionless quantities:
\begin{subequations}
\begin{align}
\tilde{k} & = \frac{k}{k_0},
\\
\tilde\sigma & = \frac{\sigma}{gn_B}.
\end{align}
\end{subequations}
If we use this last expression of $c_{\bf k}$ in Eq. (\ref{Eq:defsigmak2b}),
and change the variable of integration from ${\bf k}$ to the dimensionless quantity $\tilde{\bf k}$,
we can write for the dimensionless quantity $\tilde{\sigma}$ the following self-consistent equation:
\begin{align}
\tilde{\sigma} = \frac{8\sqrt{2}}{\sqrt{\pi}}
(n_Ba^3)^{\frac{1}{2}} 
\int_{0}^\infty d\tilde{k} \;\tilde{k}^2e^{-4\pi(n_Ba\lambda^2)\tilde{k}^2}
\frac{1+Q^2 -\sqrt{\big(1+Q^2\big)^2 - 1}}
{2+Q^2 -\sqrt{\big(1+Q^2\big)^2 - 1}},
\label{Eq:self-consistent-sigma}
\end{align}
where we used the shorthand notation:
\begin{equation}
Q^2 = C_d^{-1}(\tilde{k}^2+\tilde\sigma)e^{4\pi(n_Ba\lambda^2)\tilde{k}^2}.
\end{equation}
A numerical solution to the above equation for $\tilde{\sigma}$ can be found by iteration in the following way.
First, one starts with an initial guess for $\tilde\sigma$. Using this initial guess, one computes the ratio
$N_d/N$ using Eq. (\ref{Eq:ratioNdN}) below, which allows us to find the depletion factor $C_d= 1 - N_d/N$. This 
value of $C_d$ is then used to solve Eq. (\ref{Eq:self-consistent-sigma}) for $\tilde\sigma$.
This computed value of $\tilde\sigma$ is then used again as an input to find a better estimate of the ratio
$N_d/N$ using Eq. (\ref{Eq:ratioNdN}), and the process is henceforth repeated until convergence and a stable solution for 
$C_d$ and $\tilde\sigma$ is found.

\addvspace{1.5mm}

In the following, we shall be mostly interested in dilute Bose gases, for which $n_Ba^3\ll 1$.
There indeed seems to be a broad consensus in the physics community\cite{MahanBook} 
that Bogoliubov's theory is not adequate for describing the properties of
denser systems such as, {\em e.g.}, superfluid liquid Helium near zero temperature. For the particular case of liquid Helium,
this belief stems from the fact that Bogoliubov's theory fails to capture a few major features of this system, most
notably the fact that the depletion, even at the lowest temperatures, is quite substantial (the number of bosons in the
condensate not representing more than 10$\%$ of the total number of bosons $N$ in the system), as well as the 
appearance of a roton minimum in the excitation spectrum at higher values of the wavevector ${k}$. 
Given the above, we shall restrict ourselves to the situation of a dilute Bose gas, the ground state properties
of which are believed to be properly captured by the Bogoliubov Hamiltonian.

\addvspace{1.5mm}

\begin{figure}[tb]
\centerline{\includegraphics[width=8.89cm, height=5.5cm]{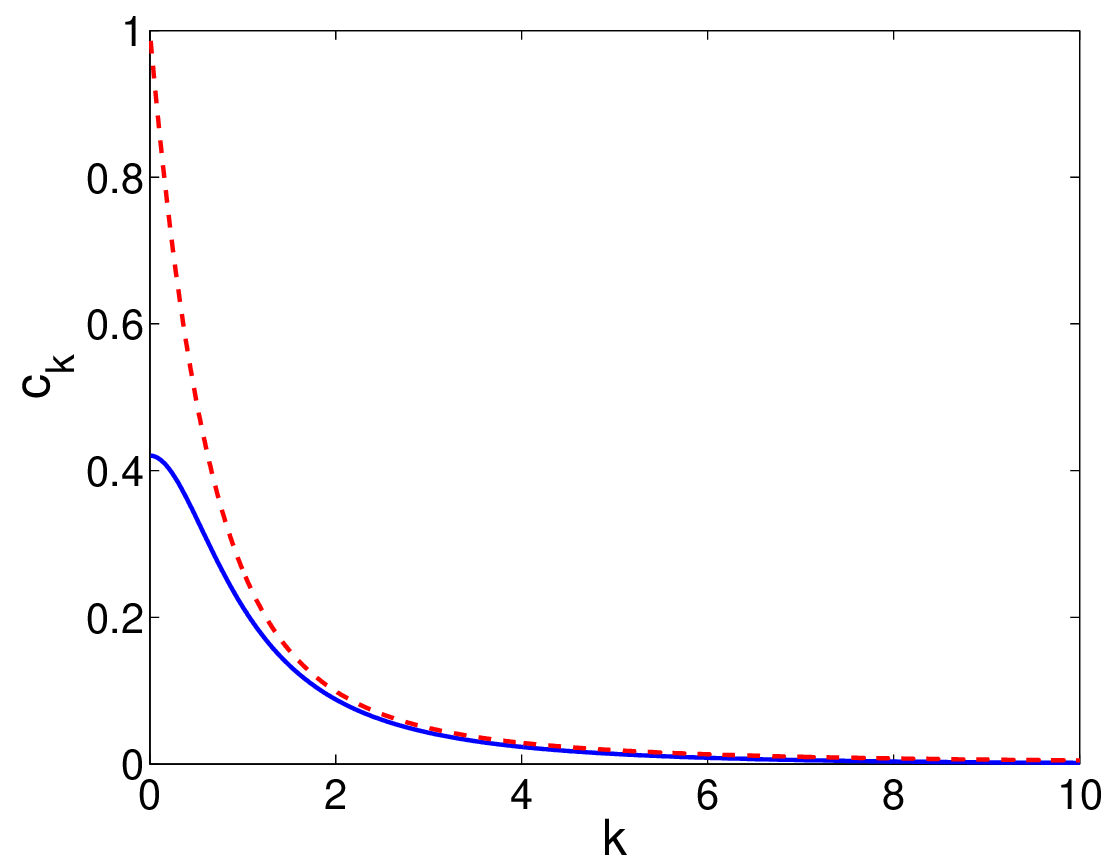}}
\caption[]{Plot of the coefficient $c_{\bf k}$ as a function of the dimensionless wavevector $\tilde{k}$.
The dashed line shows the coefficient $c_{\bf k}$ as given in the single-mode theory, Eq. (\ref{Eq:resultck}).
The solid line, on the other hand, shows the coefficient $c_{\bf k}$
as obtained from the multi-mode variational treatment of the present section,
Eq. (\ref{Eq:newc_k}), with $\tilde\sigma = 0.39$ and $C_d=0.9762$.
}\label{Fig:plotck}
\end{figure}

For simplicity, we shall take the characteristic length scale $\lambda$ which governs the range
of the interaction potential $v({\bf r})$ to be the scattering length $a$. Under these circumstances,
a numerical solution of the self-consistency equation (\ref{Eq:self-consistent-sigma})
for $n_Ba^3=10^{-3}$ yields the following values for the quantities $\tilde\sigma$ and $C_d$
(the reader will note that similar results are obtained using other numerical values of 
the parameter $n_Ba^3$ as well):
\begin{subequations}
\begin{align}
\tilde\sigma & = 0.3899 \simeq 0.39,
\\
C_d & = 0.9762 \simeq 0.98.
\end{align}
\end{subequations}
With the knowledge of $\tilde\sigma$ and $C_d$, we now are in a position to determine the coefficients $c_{\bf k}$,
and hence find the depletion and the ground state energy of the gas.
Going back to Eq. (\ref{Eq:res_ck}), we see that the exponential factor 
$\exp(k^2\lambda^2/2)$ for $\lambda=a$ is given by $\exp(4\pi n_Ba^3 \tilde{k}^2)$,
and for small values of $\tilde{k}$ this quantity can be approximated by unity (since $n_Ba^3=10^{-3}\ll 1$).
The expression of $c_{\bf k}$ can therefore be approximated by:
\begin{align}
c_{\bf k} \simeq 1 + C_d^{-1}(\tilde{k}^2 + \tilde\sigma)
- \sqrt{\big[ 1 + C_d^{-1}(\tilde{k}^2 + \tilde\sigma) \big]^2 - 1}.
\label{Eq:res_ck2}
\end{align}
Notice that this expression of $c_{\bf k}$ reduces to the expression we found before
in the single-mode model, Eq. (\ref{Eq:resultck}), in the limit $\tilde{\sigma}=0$ and $C_d=1$. 
Fig. \ref{Fig:plotck} shows a plot of $c_{\bf k}$ as a function of the dimensionless 
wavevector $\tilde{k}$ for $\tilde\sigma \simeq 0.39$ and $C_d\simeq 0.98$ (lower curve), 
and for $\tilde\sigma = 0$ and $C_d = 1$ (upper curve).
It is seen that the limiting value of $c_{\bf k}$ as $k\to 0$ when $\tilde\sigma\neq 0$
is not equal to unity, but is given by:
\begin{equation}
c_{\bf k\to 0} = 1 + C_d^{-1}\tilde\sigma
- \sqrt{\big( 1  + C_d^{-1}\tilde\sigma \big)^2 - 1} \simeq 0.42.
\label{Eq:res_ck3}
\end{equation}

\medskip

\subsection{Depletion of the condensate in presence of many momentum modes}
\label{Sub:DepletionManyModes}

\begin{figure}[tb]
\centerline{\includegraphics[width=8.89cm, height=5.5cm]{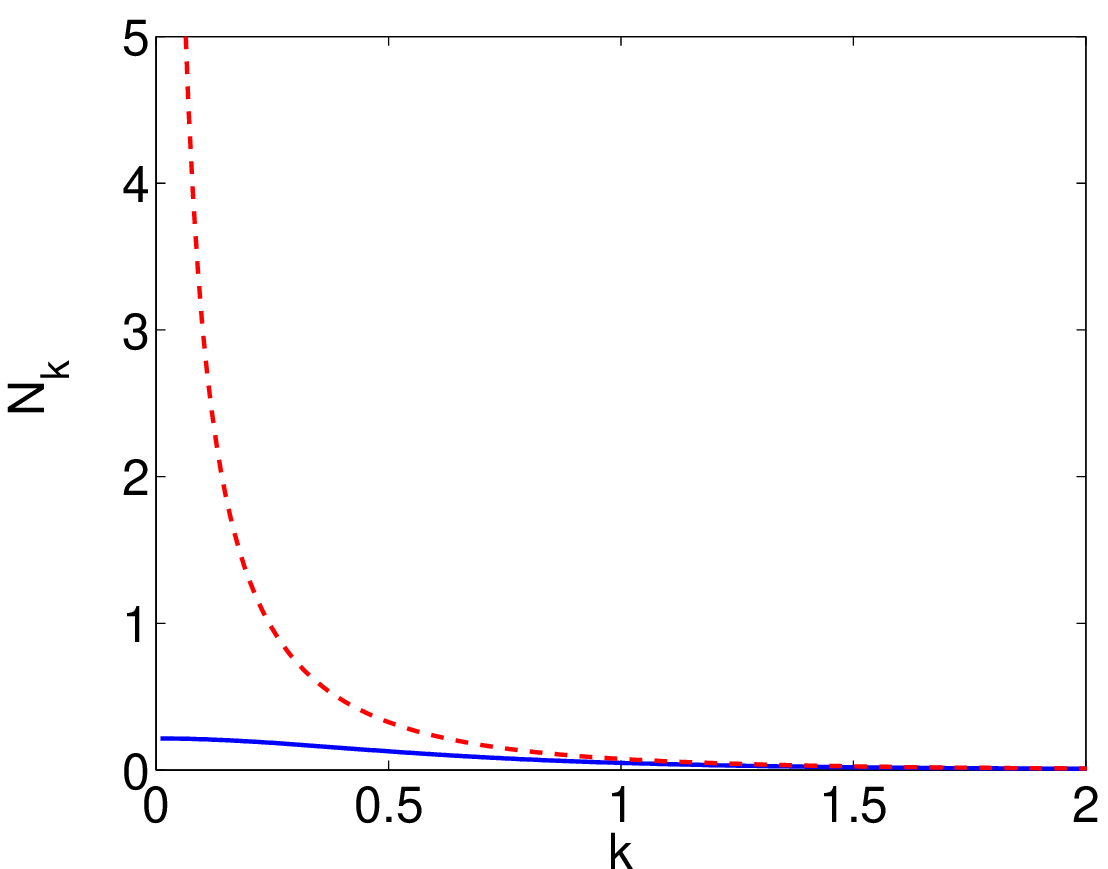}}
\caption[]{Plot of the depletion of the condensate $N_{\bf k}=c_{\bf k}^2/(1-c_{\bf k}^2)$ 
as a function of the dimensionless wavevector
$\tilde{k}$. The dashed line shows the depletion
obtained using the coefficients $c_{\bf k}$ obtained in the single-mode theory, Eq. (\ref{Eq:resultck}).
The solid line, on the other hand, shows the depletion obtained using the coefficients $c_{\bf k}$
obtained from the multi-mode variational treatment of the present section,
Eq. (\ref{Eq:newc_k}), with $\tilde\sigma = 0.39$ and $C_d=0.9762$.
}\label{Fig:plotNk}
\end{figure}

With the knowledge of $c_{\bf k}$, it is easy to find the number of depleted bosons $N_{\bf k}$
in the single momentum state ${\bf k}$, which is given by:
\begin{equation}
N_{\bf k} = \langle\Psi(N)|a_{\bf k}^\dagger a_{\bf k}|\Psi(N)\rangle = \frac{c_{\bf k}^2}{1- c_{\bf k}^2}.
\label{Eq:depltGen}
\end{equation}
In the case of the single mode theory, the coefficient $c_{\bf k}$ as given by
Eq. (\ref{Eq:resultck}) goes to unity as $k\to 0$, leading to a divergence
of the quantity $N_{\bf k}$ at small wavevectors. This happens, as we
have shown in detail in Sec. \ref{Sub:vardepletion}, when we work in the $N\to\infty$ limit 
and extend summations over the coefficients of the wavefunction to infinity; but otherwise
if we insist on working with finite sums, the depletion does not diverge as $k\to 0$. 
Here, by contrast, $c_{\bf k}$ goes to a limit which is smaller
than $1$ as $k\to 0$, and hence $N_{\bf k}$ does not diverge at small values of $k$
{\em even though} the geometric summations (over the coefficients $C_n$ of the wavefunction) 
in the derivation of Eq. (\ref{Eq:depltGen}) are extended to infinity. 
Fig. \ref{Fig:plotNk} shows a plot of the depletion $N_{\bf k}$ as a function 
of the dimensionless wavevector $\tilde{k}$ for the single-mode theory
where $c_{\bf k}$ is given by Eq. (\ref{Eq:resultck}) (upper curve), and 
for the multi-mode approach where $c_{\bf k}$ is given by Eq. (\ref{Eq:res_ck2}),
with $\tilde\sigma = 0.39$ and $C_d=0.98$ (lower curve).

\addvspace{1.5mm}

The total depletion is given by the following expression:
\begin{align}
N_d = \sum_{\bf k\neq 0}\frac{c_{\bf k}^2}{1 - c_{\bf k}^2}.
\label{Eq:expNd}
\end{align}
The summand, after a few manipulations, can be written in the form:
\begin{align}
\frac{c_{\bf k}^2}{1 - c_{\bf k}^2} = \frac{1}{2}
\left(\frac{1 + Q^2}{\sqrt{Q^2(Q^2 +2)}} 
- 1\right).
\end{align}
Transforming the sum in Eq. (\ref{Eq:expNd}) into an integral, we can write the ratio $N_d/N$ 
of the number of depleted bosons $N_d$ to the total number of bosons $N$ in the form:
\begin{align}
\frac{N_d}{N} & = \frac{4\sqrt{2}}{\sqrt\pi}(n_Ba^3)^{\frac{1}{2}}\int_0^\infty d\tilde{k}\;\tilde{k}^2
\left(\frac{1 + Q^2}{\sqrt{Q^2(Q^2 +2)}} 
- 1\right).
\label{Eq:ratioNdN}
\end{align}
In the standard formulation of Bogoliubov's theory, which corresponds to 
setting $\tilde\sigma=0$, $C_d=1$ and $\lambda = 0$ in the above expression, 
the integral in the above equation can be evaluated exactly, and has a value of 
$(\sqrt{2}/3)\simeq 0.4714.$
For non-zero values of $\tilde\sigma$, it is straightforward to evaluate the integral numerically. For example, for
$\tilde\sigma = 0.39$ and $\lambda=0$, the integral evaluates to $0.3276$, 
which translates into a reduction in the number of depleted
bosons by about a third ($0.3276/0.4714 \simeq 0.695$) with respect to the result of the
standard Bogoliubov method.

\medskip

\subsection{Ground state energy in presence of many momentum modes}

\begin{figure}[tb]
\centerline{\includegraphics[width=8.89cm, height=5.5cm]{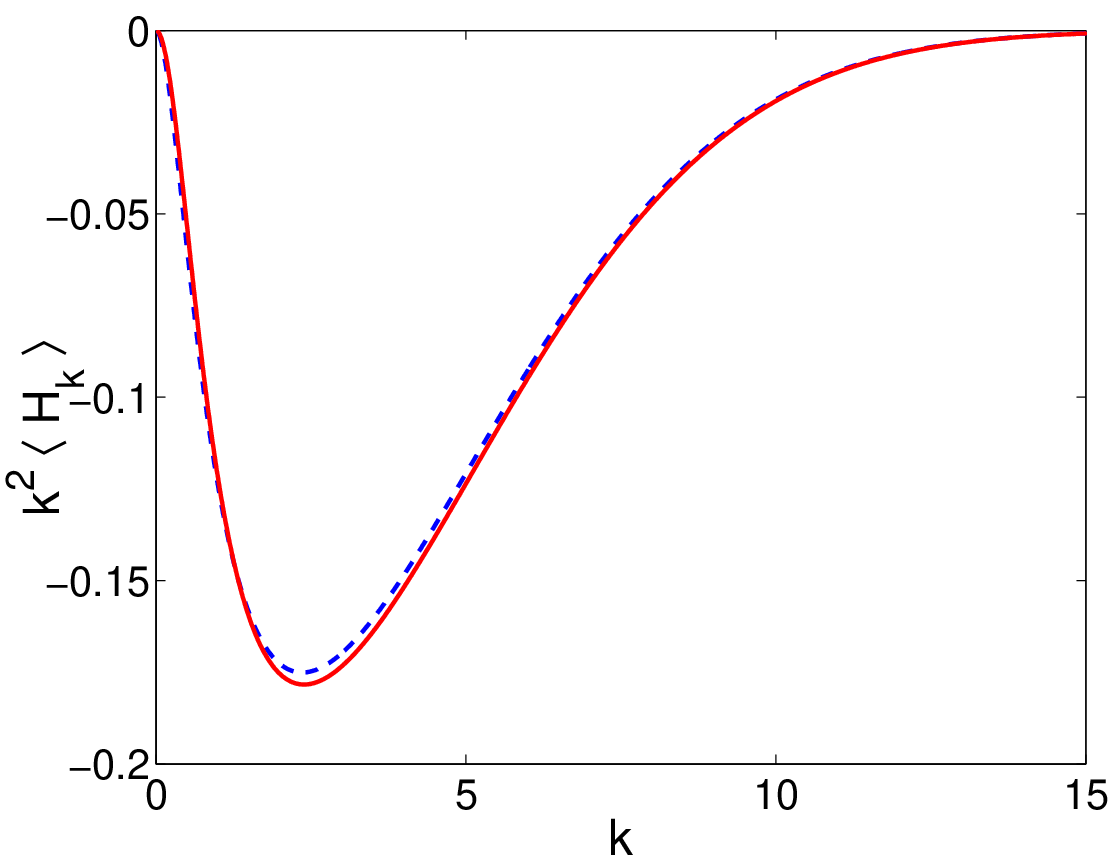}}
\caption[]{Plot of the quantity $\tilde{k}^2\langle\Psi(N)|\hat{H}_{\bf k}|\Psi(N)\rangle$
(where the factor $\tilde{k}^2$ comes from the Jacobian of the integral over wavevectors in three dimensions)
that appears in the calculation of the ground state energy $E=\langle\Psi(N)|\hat{H}|\Psi(N)\rangle$.
The dashed curve corresponds to $\tilde\sigma = 0$ and $C_d = 0.9642$, while the solid curve
corresponds to $\tilde\sigma = 0.39$ and $C_d = 0.9762$.
}\label{Fig:plotHk}
\end{figure}

We are now in a position to find the ground state energy $\langle\hat{H}\rangle=\langle\Psi(N)|\hat{H}|\Psi(N)\rangle$ 
of the system, which is given by:
\begin{equation}
\langle\hat{H}\rangle = \sum_{\bf k\neq 0} \frac{1}{1-c_{\bf k}^2}
\Big\{
\big[\varepsilon_{\bf k} + n_B\bar{v}(k)\big] c_{\bf k}^2 
- n_B \bar{v}(k)c_{\bf k}
\Big\}.
\end{equation}
Using the expression (\ref{Eq:res_ck2}) of the coefficients $c_{\bf k}$ into this last equation leads to an ultraviolet
divergence at large $k$, much like in the standard Bogoliubov approach. To circumvent this difficulty, we here 
shall use the ``regularized" expression (\ref{Eq:res_ck}) of $c_{\bf k}$, which takes into account the fact that 
the interaction potential between bosons falls off at large $k$. Then, if we transform the sum into an integral, we
obtain (in three dimensions):
\begin{align}
\frac{E}{V} & = -\frac{1}{2}gn_B^2\Bigg\{
\frac{8\sqrt{2}}{\sqrt{\pi}}(n_Ba^3)^{\frac{1}{2}}\int_0^\infty d\tilde{k}\;\tilde{k}^2
\Bigg[
C_d e^{-4\pi n_Ba\lambda^2\tilde{k}^2}\Big( 1 + Q^2
-\sqrt{Q^2(Q^2 +2)}\Big)
\nonumber\\
& + \tilde\sigma\Big(
\frac{1 + Q^2}{\sqrt{Q^2(Q^2 + 2)}} - 1
\Big)
\Bigg]
\Bigg\}.
\end{align} 
Numerical evaluation of the above integral for $\lambda = a$ and $n_B a^3 = 0.001$ yields, for $\tilde\sigma = 0$
(the corresponding value of $C_d$, evaluated using Eq. (\ref{Eq:ratioNdN}), is given by $C_d=0.9642$),
\begin{subequations}
\begin{align}
\frac{E}{V}\Big|_{\tilde\sigma = 0} & \simeq \frac{1}{2}gn_B^2\cdot(-13.02)(n_Ba^3)^{\frac{1}{2}},
\\
& \simeq \frac{1}{2} gn_B^2 \cdot (-0.412), 
\end{align}
\end{subequations}
while for $\tilde\sigma = 0.39$ and $C_d= 0.9762$, we obtain:
\begin{subequations}
\begin{align}
\frac{E}{V}\Big|_{\tilde\sigma = 0.39} & \simeq \frac{1}{2}gn_B^2\cdot(-13.20)(n_Ba^3)^{\frac{1}{2}},
\\
& \simeq \frac{1}{2} gn_B^2 \cdot (-0.417). 
\end{align}
\end{subequations}
We hence see that the solution with $\tilde\sigma\neq 0$ for the coefficients $c_{\bf k}$ 
leads a lower overall energy than a solution with $\tilde\sigma=0$ when the depletion factor $C_d$
is used in the calculation (we remind the reader that this factor emerges when
we keep an accurate count of the number of bosons in the condensed state ${\bf k}=0$,
see Eq. (\ref{Eq:defbarv})), 
which constitutes a direct verification of the validity of our variational method.
The fact that the solution with $\tilde\sigma\neq 0$ has a lower energy
than the solution with $\tilde\sigma = 0$ can best be visualized
in Fig. \ref{Fig:plotHk}, where we plot the quantity $\tilde{k}^2\langle\Psi(N)|\hat{H}_{\bf k}|\Psi(N)\rangle$
that appears when we integrate over momentum modes in the calculation of the total energy $E$.
It appears that the area enclosed by the curve with $\tilde\sigma= 0.39$ is greater than
the area enclosed by the curve with $\tilde\sigma=0$, hence showing that the value $\tilde\sigma=0.39$
leads to a lower ground state energy of the system.

\addvspace{1.5mm}

\begin{figure}[t]
\centerline{\includegraphics[width=8.89cm, height=5.5cm]{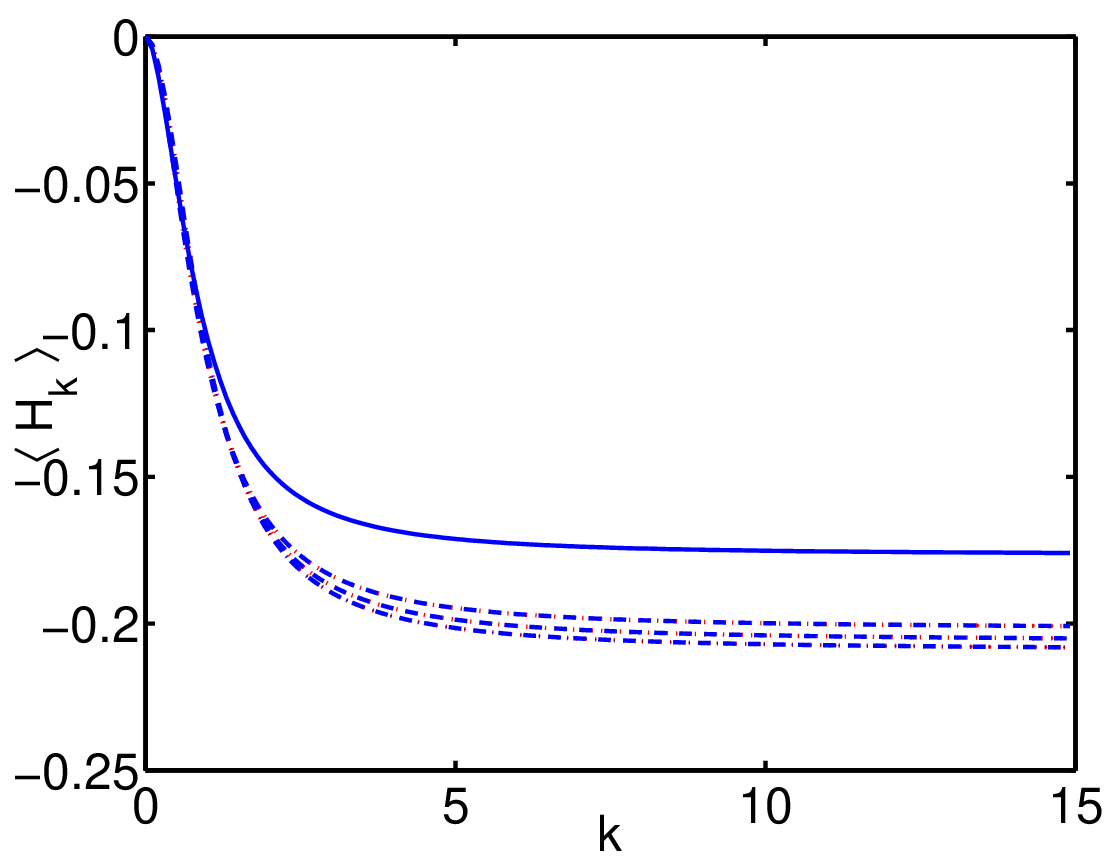}}

\centerline{\includegraphics[width=8.89cm, height=5.5cm]{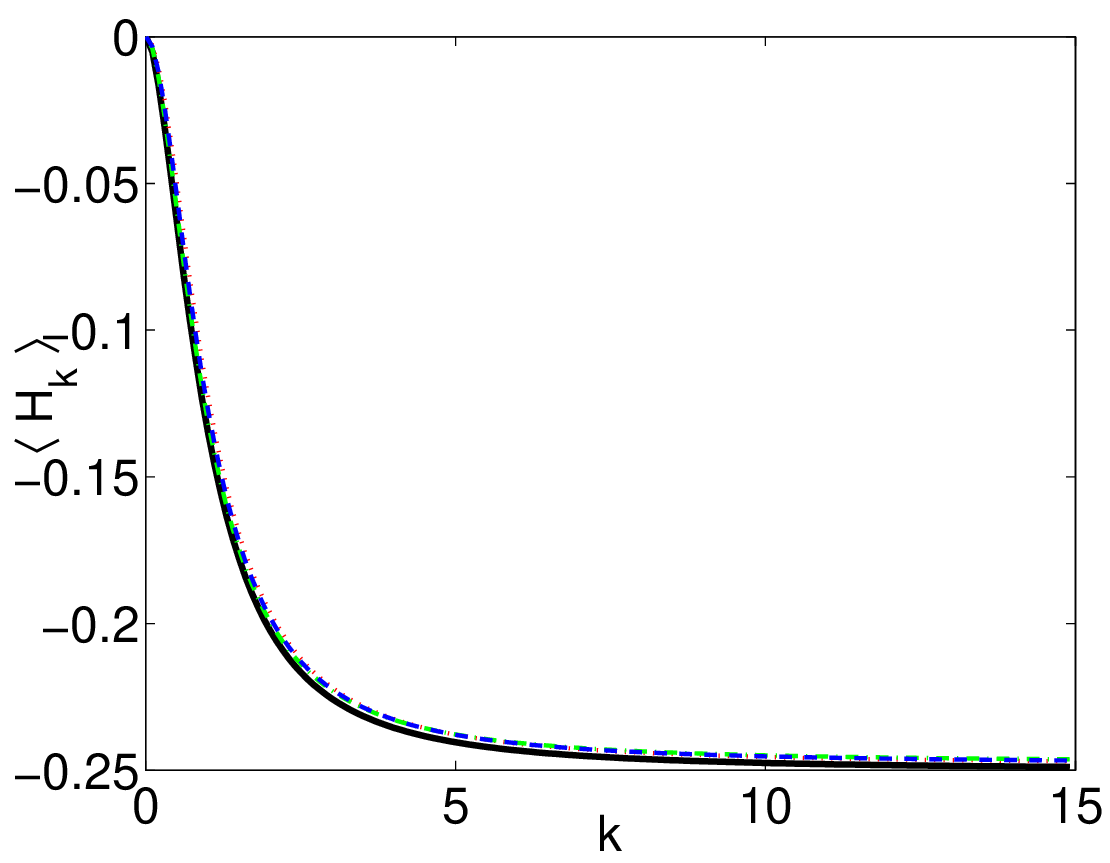}}
\caption[]{Upper panel: Plot of the quantity $\tilde{k}^2\langle\Psi(N)|\hat{H}_{\bf k}|\Psi(N)\rangle$
using Eq. (\ref{Eq:expectation_Hk}) for $\lambda=0$ and $n_Ba^3=0.01$. 
The solid curve corresponds to $\tilde\sigma = 0$ and using the result 
for the constants $c_{\bf k}$ from Bogoliubov's theory, Eq. (\ref{Eq:resultck}). 
The dotted curves are obtained by using Eq. (\ref{Eq:res_ck2}) for $c_{\bf k}$, with
$\tilde\sigma=0.2$, $\tilde\sigma=0.4$ and $\tilde\sigma=0.6$ from top to bottom respectively.
Lower panel: Plot of the quantity $\tilde{k}^2\langle\Psi(N)|\hat{H}_{\bf k}|\Psi(N)\rangle$
using the single-mode result Eq. (\ref{Eq:expectation_h_min}) for $\lambda=0$ and $n_Ba^3=0.01$. 
The solid curve corresponds to $\tilde\sigma = 0$
and $c_{\bf k}$ from Bogoliubov's theory, Eq. (\ref{Eq:resultck}).
There are three other dashed curves which correspond to using Eq. (\ref{Eq:res_ck2}) for $c_{\bf k}$, 
with $\tilde\sigma=0.2$, $\tilde\sigma=0.4$ 
and $\tilde\sigma=0.6$. These three curves are very close to one another, and are all situated
above the solid line corresponding to the result of Bogolibov's theory.
}\label{Fig:plotk2Hk}
\end{figure}

One may wonder at this point if the above results for the ground state energy are not simply 
an artifact of the special choice we made for the interaction potential between bosons, which has a Gaussian 
dependence on $\tilde{k}$ in Fourier space. To clarify this point,
in the upper panel of Fig. \ref{Fig:plotk2Hk} below we plot the ground state expectation value 
of $\hat{H}_{\bf k}$ as a function of the wavevector
$\tilde{k}$ from Eq. (\ref{Eq:expectation_Hk}) using $\lambda=0$ (corresponding to $v({\bf r})=g\delta({\bf r})$)
and with a different value of the parameter $n_Ba^3$, $n_Ba^3=0.01$
(more precisely, we plot the product 
$\tilde{k}^2\langle\hat{H}_{\bf k}\rangle=\tilde{k}^2\langle\Psi(N)|\hat{H}_{\bf k}|\Psi(N)\rangle$ vs. $\tilde{k}$,
the factor $\tilde{k}^2$ representing the Jacobian of the integral over wavevectors in three dimensions
that appears in the calculation of the ground state energy $E=\langle\Psi(N)|\hat{H}|\Psi(N)\rangle$). 
In this last figure, the solid line shows
the result for $\tilde{k}^2\langle\hat{H}_{\bf k}\rangle$
one obtains by plugging the coefficients $c_{\bf k}$ of the single-mode theory, Eq. (\ref{Eq:resultck}),
into Eq. (\ref{Eq:expectation_Hk}).
The dotted lines, on the other hand, show the results for this same quantity one obtains using the coefficients
$c_{\bf k}$ from Eq. (\ref{Eq:res_ck2}) for three nonzero values of $\tilde\sigma$. 
It is again seen that the expectation values $\langle\hat{H}_{\bf k}\rangle$
that are obtained using the coefficients $c_{\bf k}$ from Eq. (\ref{Eq:res_ck2}) 
with $\tilde\sigma\neq 0$ are consistently lower than the one
calculated using the single mode coefficients from Eq. (\ref{Eq:resultck}), which gives further credence to our
minimization procedure in which the number of bosons in the condensate $(N-2\sum_i n_i)$
is kept throughout the calculation. Conversely, in the lower panel of Fig. (\ref{Fig:plotk2Hk}) 
we plot the quantity $\tilde{k}^2\langle\widetilde\psi_{\bf k}|\hat{H}_{\bf k}|\widetilde\psi_{\bf k}\rangle$
from Eq. (\ref{Eq:expectation_h_min}) using the Bogoliubov result (\ref{Eq:resultck}) for $c_{\bf k}$ (solid line),
and using the result (\ref{Eq:res_ck2}) for various nonzero values of $\tilde{\sigma}$. It is seen that 
for the single-mode model the standard result for $c_{\bf k}$ (which corresponds to setting 
$\tilde\sigma=0$ and $C_d=1$ in Eq. (\ref{Eq:res_ck2})) is always smaller
in energy than the result obtained by using a non-zero value of $\tilde\sigma=0$ in the expression of 
$c_{\bf k}$ from Eq. (\ref{Eq:res_ck2}). We therefore conclude that both our minimizations, 
in Sec. \ref{Sub:GSHkBog}
and in the present Section, produce correct functional forms for the constants $c_{\bf k}$ that minimize
the ground state energy of the system, with Eq. (\ref{Eq:resultck}) representing
the correct functional form of $c_{\bf k}$ when we consider the single-mode Hamiltonian $\hat{H}_{\bf k}$
(or, equivalently, when we use the Bogoliubov prescription $a_0\simeq\sqrt{N}$ 
and remove the ${\bf k}=0$ mode from the Hilbert space, so that all the single-mode
Hamiltonians $\hat{H}_{\bf k}$ are decoupled), and Eq. (\ref{Eq:res_ck2}) representing the functional form
that minimizes the ground state energy when we do not use Bogoliubov's prescription and keep the 
correct expression $(N-2\sum_i n_i)$ of the number of bosons in the ${\bf k}=0$ state throughout
the calculation.

\medskip

\subsection{Elementary excitations in presence of many momentum modes}
\label{Sub:ElementaryExcitations}

We now want to examine the elementary excitations of the Bose system in presence of many momentum modes.
By contrast to the situation of Sec. \ref{Sub:ExcOneMode} where we studied the elementary excitations
of the single-mode Hamiltonian $\hat{H}_{\bf k}$, and where the only possible excitations increasing the momentum
of the system by an amount $\hbar {\bf k}$ were $a^\dagger_{\bf k}a_{\bf 0}$ and $a_{\bf 0}^\dagger a_{-\bf k}$,
here since we are dealing with the total Hamiltonian $\hat{H}$, there are an infinite number
of excitations of the form $a^\dagger_{\bf k+q}a_{\bf q}$ which increase the momentum the system
by $\hbar {\bf k}$. The most general excitation of momentum $\hbar {\bf k}$ can be written in the form:
\begin{equation}
\tilde\alpha_{\bf k}^\dagger = \sum_{\bf q}\beta_{\bf q}a^\dagger_{\bf k+q} a_{\bf q},
\label{Eq:alternatealpha}
\end{equation}
with the $\beta_{\bf q}$'s being arbitrary complex numbers. In principle, the possibility that 
the $\beta_{\bf q}$'s may be chosen in such a way that the excitation energies 
of the operators $\tilde\alpha_{\bf k}$ are {\em lower}
than the excitation energies of the usual operators $\alpha_{\bf k}$ cannot be ruled out
(this is because the diagonalization of $\hat{H}$ in terms of the $\alpha_{\bf k}$
is not exact, but only approximate).
However, since we here want to draw a comparison with the standard formulation of Bogoliubov's theory,
we shall proceed to study the diagonalization of $\hat{H}$ in terms of the simple
$\alpha_{\bf k}$ operators, instead of the cumbersome form defined in Eq. (\ref{Eq:alternatealpha}) above,
which is not easy to work with.

\addvspace{1.5mm}

Let us again define the operators $\alpha_{\bf k}$ and $\alpha^\dagger_{\bf k}$ as follows:
\begin{subequations}
\begin{align}
&\alpha_{\bf k}  = \tilde{u}_{\bf k} a_{\bf k} a_{\bf 0}^\dagger + \tilde{v}_{\bf k} a_{\bf 0}a^\dagger_{-\bf k},
\\
&\alpha_{\bf k}^\dagger  = \tilde{u}_{\bf k} a_{\bf k}^\dagger a_{\bf 0} 
+ \tilde{v}_{\bf k} a_{\bf 0}^\dagger a_{-\bf k}.
\end{align}
\label{Eq:againuv_2}
\end{subequations}
It is easy to verify that the action of $\alpha_{\bf k_1}$ on the ket $|\Psi(N)\rangle$ gives the following result:
\begin{align}
&\alpha_{\bf k_1}|\Psi(N)\rangle  = Z \sum_{n_1=1}^{\infty}\sum_{n_2}^\infty\cdots\sum_{n_M}^\infty
C_{n_2}\cdots C_{n_M}
\Bigg[
C_{n_1}\tilde{u}_{\bf k_1}\sqrt{n_1\Big(N+1-2\sum_{i=1}^Mn_i\Big)} 
\nonumber\\
& + C_{n_1-1}\tilde{v}_{{\bf k}_1}\sqrt{n_1\Big(N+2-2\sum_{i=1}^Mn_i\Big)}
\Bigg]
|N+1-2\sum_{i=1}^M n_i;n_1-1, n_1;n_2,n_2;\cdots; n_M,n_M\rangle
\end{align}
Requiring that $\alpha_{\bf k_1}|\Psi(N)\rangle =0 $ gives the condition:
\begin{align}
\frac{\tilde{u}_{\bf k_1}}{\tilde{v}_{\bf k_1}} = -\frac{C_{n_1-1}}{C_{n_1}}\sqrt{\frac{N+2-2\sum_{i=1}^Mn_i}
{N+1-2\sum_{i=1}^Mn_i}}
\end{align}
Approximating the square root with unity, we obtain:
\begin{subequations}
\begin{align}
\frac{\tilde{v}_{\bf k_1}}{\tilde{u}_{\bf k_1}} & \simeq -\frac{C_{n_1}}{C_{n_1-1}} 
\label{Eq:newratiov_ku_ka}\\
& = c_{\bf k_1},
\label{Eq:newratiov_ku_kb}
\end{align}
\label{Eq:newratiov_ku_k}
\end{subequations}
which is the same condition obtained previously for the single-mode case, Eq. (\ref{Eq:ratioCnCn-1}).
Proceeding in the same way as in Sec. \ref{Sub:ExcOneMode}, we can determine 
the constants $\tilde{u}_{\bf k}$ and $\tilde{v}_{\bf k}$ by using Eq. (\ref{Eq:newratiov_ku_k})
and the requirement that the excited state $\alpha_{\bf k}^\dagger|\Psi(N)\rangle$ be normalized
to unity, {\em i.e.} that:
\begin{equation}
\langle\Psi(N)|\alpha_{\bf k}\alpha_{\bf k}^\dagger|\Psi(N)\rangle = 1.
\end{equation}
The above condition leads to the following result for the quantity $\tilde{u}_{\bf k}^2 - \tilde{v}_{\bf k}^2$,
\begin{equation}
\tilde{u}_{\bf k}^2 - \tilde{v}_{\bf k}^2 = \gamma_{\bf k}^2,
\label{Eq:diffukvk2}
\end{equation}
where now the quantity $\gamma_{\bf k}$ is given by (compare with Eq. (\ref{Eq:defgammak})):
\begin{subequations}
\begin{eqnarray}
\gamma_{\bf k}^2 & = & \frac{1}{\langle a_0^\dagger a_0\rangle 
- \langle a_{\bf k}^\dagger a_{\bf k}\rangle}
\label{Eq:defgammaka2}
\\
& \simeq & \frac{1}{\langle a_0^\dagger a_0\rangle},
\label{Eq:defgammakb2}
\end{eqnarray}
\label{Eq:defgammak2}
\end{subequations}
In the above equations, we denote by $\langle\cdots\rangle$ the expectation value
in the ground state $|\Psi(N)\rangle$, {\em i.e.} $\langle\cdots\rangle=\langle\Psi(N)|\cdots|\Psi(N)\rangle$.
(In Sec. \ref{Sub:ExcOneMode}, $\gamma_{\bf k}$ was defined in terms of expectation
values in the normalized ground state $|\widetilde\psi_{\bf k}\rangle$ of the single-mode 
Hamiltonian $\hat{H}_{\bf k}$.) Note also that,
in going from the first to the second line of Eq. (\ref{Eq:defgammak}) we again assumed that 
$\langle a_{\bf k}^\dagger a_{\bf k}\rangle\ll \langle a_0^\dagger a_0\rangle$, 
{\em i.e.} that the depletion of the condensate into {\em any} single-particle state with momentum ${\bf k}$
is small. Using Eqs. (\ref{Eq:newratiov_ku_k}) and (\ref{Eq:diffukvk2}), we now can determine the
values of the constants $\tilde{u}_{\bf k}$ and $\tilde{v}_{\bf k}$. These can be written in the form:
\begin{subequations}
\begin{align}
&\tilde{u}_{\bf k} = \gamma_{\bf k} u_{\bf k}, \quad \tilde{v}_{\bf k} = \gamma_{\bf k} v_{\bf k},
\\
& u_{\bf k} = \frac{1}{\sqrt{1-c_{\bf k}^2}}, \quad v_{\bf k} = \frac{c_{\bf k}}{\sqrt{1-c_{\bf k}^2}}.
\end{align}
\label{Eq:resulttildeukvk2}
\end{subequations}
Using the new result (\ref{Eq:newc_k}) for $c_{\bf k}$ derived in presence of many momentum modes, we finally obtain:
\begin{subequations}
\begin{align}
u_{\bf k}^2 & = \frac{1}{2}\Big(\frac{\varepsilon_{\bf k}+n_Bv({\bf k}) + \sigma}{E_{\bf k}} + 1\Big),
\\
v_{\bf k}^2 & = \frac{1}{2}\Big(\frac{\varepsilon_{\bf k}+n_Bv({\bf k}) + \sigma}{E_{\bf k}} - 1\Big),
\end{align}
\label{Eq:resultukvk2}
\end{subequations}
where now the spectrum $E_{\bf k}$ is given by the expression:
\begin{equation}
E_{\bf k} = n_B v({\bf k})\sqrt{Q^2(Q^2 + 2)}.
\end{equation}

\addvspace{1.5mm}

We now use the general expression (\ref{Eq:App:Hgeneralalphas}) of $\hat{H}_{\bf k}$ in terms of the operators 
$\alpha_{\bf k}$ and $\alpha_{\bf k}^\dagger$
derived in Appendix \ref{AppendixA}, 
which we shall rewrite here for definiteness:
\begin{align}
\hat{H}_{\bf k}  & = \frac{1}{2\gamma_{\bf k}^2}\Big\{
\big[A_{\bf k}(u_{\bf k}^2 + v_{\bf k}^2) - 2B_{\bf k}\big]
(\alpha_{\bf k}^\dagger \alpha_{\bf k} 
+ \alpha_{-\bf k}^\dagger \alpha_{-\bf k})
\nonumber\\
&+ \big[B_{\bf k}(u_{\bf k}^2 + v_{\bf k}^2) - 2A_{\bf k}\big]
(\alpha_{\bf k}^\dagger \alpha_{-\bf k}^\dagger 
+ \alpha_{\bf k} \alpha_{-\bf k})
\nonumber\\
& + \big[A_{\bf k}v_{\bf k}^2 - B_{\bf k}u_{\bf k}v_{\bf k}\big]
\big([\alpha_{\bf k},\alpha_{\bf k}^\dagger]
+ [\alpha_{-\bf k},\alpha_{-\bf k}^\dagger]\big)
\nonumber\\
& + (B_{\bf k}v_{\bf k}^2 - A_{\bf k}u_{\bf k}v_{\bf k}\big)\big([\alpha_{-\bf k}^\dagger,\alpha_{\bf k}^\dagger]
+[\alpha_{-\bf k},\alpha_{\bf k}]\big)
\Big\},
\label{Eq:HAkBk}
\end{align}
where the constants $A_{\bf k}$ and $B_{\bf k}$ are given by
(the constant $\eta_{\bf k}$ is defined in Appendix \ref{AppendixA}):
\begin{equation}
A_{\bf k} = \varepsilon_{\bf k}\eta_{\bf k} + \frac{v({\bf k})}{V}, \quad B_{\bf k} = \frac{v({\bf k})}{V}.
\label{Eq:defAkBk2}
\end{equation}
Note that the coherence factors $u_{\bf k}$ and $v_{\bf k}$ in Eq. (\ref{Eq:HAkBk}) are now given
by the expressions in Eq. (\ref{Eq:resultukvk2}).
With that in mind, we want to choose a value for $\eta_{\bf k}$ that will make the quantity
$[B_{\bf k}(u_{\bf k}^2 + v_{\bf k}^2) - 2A_{\bf k}u_{\bf k}v_{\bf k}]$ vanish. 
Using this constraint and Eq. (\ref{Eq:resultukvk2}), we obtain, after a few manipulations:
\begin{equation}
A_{\bf k} = B_{\bf k}\Big(\frac{\varepsilon_{\bf k} + \sigma}{n_B\bar{v}({\bf k})}+1\Big).
\end{equation}
Using the definitions of $A_{\bf k}$ and $B_{\bf k}$ given in Eq. (\ref{Eq:defAkBk2}), we can find
the expression of the quantity $\eta_{\bf k}$, which is now given by:
\begin{equation}
\eta_{\bf k} = \frac{1}{N}\frac{v({\bf k})}{\bar{v}({\bf k})}\Big(1 + \frac{\sigma}{\varepsilon_{\bf k}}\Big).
\label{Eq:resultetak}
\end{equation}
We now are in a position to calculate the excitation energy associated with Hamiltonian $\hat{H}_{\bf k}$,
which is the coefficient of the quadratic terms $\alpha_{\bf k}^\dagger\alpha_{\bf k}$
and $\alpha_{-\bf k}^\dagger\alpha_{-\bf k}$ in Eq. (\ref{Eq:HAkBk}) 
(it can indeed be shown that the last three terms on the {\em rhs} of this last equation
do not contribute to the excitation energy, see Appendix \ref{AppendixA}), {\em i.e.}:
\begin{subequations}
\begin{align}
\Delta E_{exc}({\bf k}) & = \langle\Psi(N)|\alpha_{\bf k}\hat{H}_{\bf k}\alpha_{\bf k}^\dagger|\Psi(N)\rangle
- \langle\Psi(N)|\hat{H}_{\bf k}|\Psi(N)\rangle,
\\
& = \frac{1}{2\gamma_{\bf k}^2}\big[A_{\bf k}(u_{\bf k}^2 + v_{\bf k}^2) - 2B_{\bf k}u_{\bf k}v_{\bf k}\big].
\end{align}
\end{subequations}
Using the results (\ref{Eq:newc_k}) for the coefficient $c_{\bf k}$ and (\ref{Eq:resultetak}) for 
the quantity $\eta_{\bf k}$, along with the definitions (\ref{Eq:resultukvk2}) 
of the constants $u_{\bf k}$ and $v_{\bf k}$, we find, after a few manipulations:
\begin{align}
\Delta E_{exc}({\bf k}) = \frac{1}{2}E_{\bf k} = \frac{1}{2}n_Bv({\bf k})
\sqrt{Q^2(Q^2+2)},
\end{align}
where we remind the reader that:
\begin{subequations}
\begin{align}
Q^2 & = \frac{\varepsilon_{\bf k} + \sigma}{n_B\bar{v}(k)},
\\
& = C_d^{-1}(\tilde{k}^2 + \tilde\sigma)\exp(4\pi n_B a\lambda^2\tilde{k}^2).
\end{align}
\end{subequations} 
We again would like to emphasize that $\Delta E_{exc}({\bf k})$ is the excitation energy
associated with the Hamiltonian $\hat{H}_{\bf k}$. The full Hamiltonian $\hat{H}=\sum_{\bf k\neq 0}\hat{H}_{\bf k}$
contains an identical contribution $\hat{H}_{-\bf k}$, so that the {\em total} 
energy cost $\Delta E_{exc}^{tot}({\bf k})$ 
to bring the system from its ground state $|\Psi(N)\rangle$ to the excited state
$\alpha_{\bf k}^\dagger|\Psi(N)\rangle$ is given by:
\begin{subequations}
\begin{align}
\Delta E_{exc}^{tot}({\bf k}) & = \Delta E_{exc}({\bf k}) + \Delta E_{exc}(-{\bf k}),
\\
& =  n_Bv({\bf k})\sqrt{Q^2(Q^2+2)}.
\end{align}
\label{Eq:excSpectrum}
\end{subequations}
The above expression of the excitation energy, after a few manipulations, can be written in the form:
\begin{align}
\frac{\Delta E_{exc}^{tot}({\bf k})}{n_Bg} & = \frac{1}{C_d}
\sqrt{(\tilde{k}^2+\tilde\sigma)\Big(\tilde{k}^2+\tilde\sigma 
+ 2C_de^{-4\pi n_Ba\lambda^2\tilde{k}^2}\Big)}.
\label{Eq:spectrumEk0}
\end{align}
Again, given the dilute Bose gas parameters we have considered, the exponential $\exp(-k^2\lambda^2/2)$
is close to unity for the most interesting values of $k$ such that $k\ll \lambda^{-1}$, and hence we can write:
\begin{align}
\frac{\Delta E_{exc}^{tot}({\bf k})}{n_Bg} 
& = \frac{1}{C_d}\sqrt{(\tilde{k}^2+\tilde\sigma)(\tilde{k}^2+\tilde\sigma + 2C_d)}.
\label{Eq:spectrumEk}
\end{align}
Notice that the above expression reduces to the usual Bogoliubov spectrum for the single-mode theory
in the limit $\tilde\sigma = 0$ and $C_d=1$. In the case where $\tilde\sigma\neq 0$, 
however, a major feature of the spectrum
(\ref{Eq:spectrumEk}) is that it displays a {\em finite gap} as $k\to 0$, which is given by:
\begin{align}
\frac{\Delta E_{exc}^{tot}({\bf k=0})}{n_Bg} 
= \sqrt{\frac{\tilde\sigma}{C_d}\Big(\frac{\tilde\sigma}{C_d} + 2\Big)}.
\end{align}
For $\tilde\sigma = 0.39$, the numerical value of the gap is given by:
\begin{align}
\Delta E_{exc}^{tot}({\bf k=0}) \simeq 0.98 \,n_Bg.
\end{align}
This value is comparable to the gap predicted by Hartree-Fock theory, which is given
by $\Delta E_{exc}^{HF}({\bf k=0}) =\,n_Bg$, and to the gap predicted by Girardeau and Arnowitt
a long time ago \cite{Girardeau1959} using a method that is different from the one we employed in the present study.

\begin{figure}[tb]
\centerline{\includegraphics[width=8.89cm, height=5.5cm]{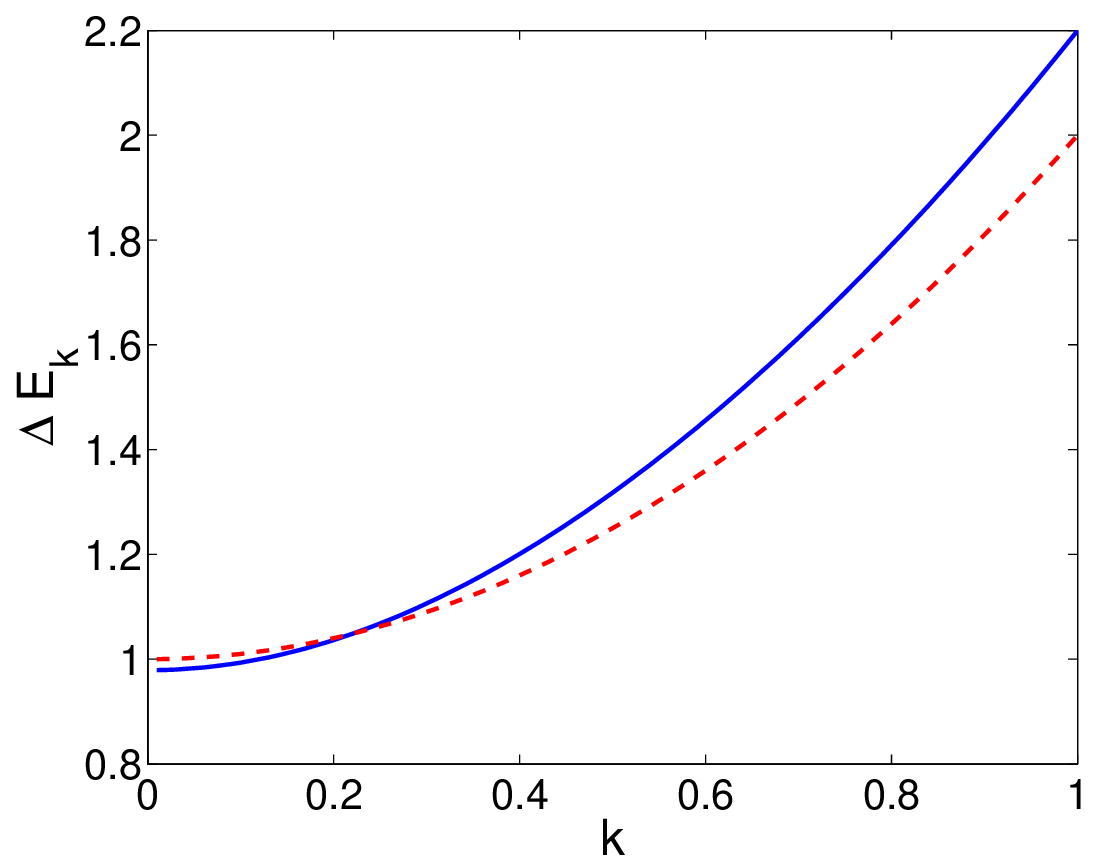}}
\caption[]{Solid line: Plot of the excitation spectrum $\Delta E_{exc}^{tot}({\bf k})$ of Eq. (\ref{Eq:spectrumEk0}).
Dashed line: Hartree-Fock spectrum, $E_{HF} = \varepsilon_{\bf k} + n_Bv({\bf k}) = \varepsilon_{\bf k} + gn_B$.
}\label{Fig:plotExc}
\end{figure}

\addvspace{1.5mm}

At this point, we would like to caution the reader that the above result does not by any means
imply that the actual excitation spectrum of bosons in a real (experimental) system has to be gapped. 
It actually only implies that the truncated Hamiltonian of Bogoliubov's theory has a finite gap in the $k\to 0$ limit 
when this theory is formulated within a number-conserving framework
where an accurate count of the number of bosons in the condensate is kept throughout the calculation.
This author would like to make it clear that he is not by any means advocating
a gapped excitation spectrum for bosons, but merely presenting what Bogoliubov's theory
predicts for this spectrum when the calculation is performed in a tightly number-conserving fashion.
Going back to the Girardeau-Arnowitt number-conserving theory\cite{Girardeau1959} mentioned above, 
which also predicts the existence of a finite gap in the excitation spectrum,
an interesting aspect of this theory is a demonstration that nonpairing ``triplet" 
contributions to the Hamiltonian are of the right order of magnitude to cancel the excitation energy gap.
Takano subsequently showed\cite{Takano1961} that such cancellation 
does indeed occur, and it may well be possible that a similar scenario takes place for
the variational model studied in this paper if additional terms are included in the Bogoliubov Hamiltonian. 
While such a scenario cannot, {\em a priori}, be ruled out, our results are still
important on their own, because they show that, as far as Bogoliubov's Hamiltonian is concerned, 
the conventional excitation spectrum of Bogoliubov's theory becomes gapped 
when the calcualtion is performed in a number-conserving fashion.
More work is needed in order to ascertain under what conditions the 
excitation spectrum of ``real" bosons described by the full Hamiltonian of Eq. (\ref{Eq:HFourierSpace}) 
is indeed gapped or not in the $k\to 0$ limit.

\section{Discussion}
\label{Sec:Discussion}

Having explored the results one obtains for the main physical observables when the full Hamiltonian
$\hat{H}=\sum_{\bf k\neq 0}\hat{H}_{\bf k}$ is diagonalized taking the conservation of boson number into account,
we now want to present a brief summary of our results, and discuss the implications of these results
on the formulation of Bogoliubov's theory of an interacting Bose gas.

\subsection{Summary of our results}

A major goal of this paper has been to clarify the nature 
and meaning of the standard formulation of Bogoliubov's theory. Our investigation
has shown that this theory seeks to diagonalize each single-mode Hamiltonian $\hat{H}_{\bf k}$ independently. 
Comparison with the results of the exact numerical diagonalization of the single-mode Hamiltonian $\hat{H}_{\bf k}$
has shown that Bogoliubov's theory gives an astonishingly accurate description of the ground state energy and excitation
spectrum of $\hat{H}_{\bf k}$. While it is true that
in a number non-conserving framework, where the ${\bf k}=0$ state
is removed from the Hilbert space used to describe the system, 
one can write the ground state wavefunction as a simple product of the ground state
wavefunctions of the $\hat{H}_{\bf k}$'s, and hence
diagonalizing the single-mode Hamiltonians $\hat{H}_{\bf k}$ 
independently from one another makes sense, 
in a number-conserving framework diagonalizing the $\hat{H}_{\bf k}$'s independently is not very helpful,
since the ground state wavefunction of $\hat{H}=\sum_{\bf k\neq 0}\hat{H}_{\bf k}$ {\em cannot} in this case
be written as a simple product of the ground state wavefunctions of the $\hat{H}_{\bf k}$'s
(the reason this is so is because the Hilbert spaces spanned by these single-mode Hamiltonians have
the ${\bf k}=0$ state in common).  
The above observation calls for a more careful diagonalization method where the total Hamiltonian $\hat{H}$
is diagonalized directly. This is what we have attempted to do in this paper by using
the fully number-conserving trial wavefunction of Eq. (\ref{Eq:fullPsi(N)}), and finding the 
variational coefficients $c_{\bf k}$ in exactly the same way as in Ref. \citen{Leggett2001}.
It is actually quite on purpose that we have presented the calculation
of the ground state energy with such a level of detail, both for the single-mode Hamiltonian $\hat{H}_{\bf k}$
thereby reproducing all the results of the conventional Bogoliubov theory, and for the total
Hamiltonian $\hat{H}$. Comparing the two methods underscores the stark differences
between a theory that does enforce the conservation of the total number of bosons $N$ and theories
which do not, and shows explicitly that the results of Bogoliubov's method cannot be obtained when $N$ is conserved
between all the momentum modes, but only in a single-mode approach where $N$ is conserved for a single
momentum mode. In the following, we shall summarize a few salient results of our variational treatment,
and try to discuss how this treatment improves on the standard formulation of Bogoliubov's method.

\medskip

{\em 1. Divergence of the depletion $\langle a_{\bf k}^\dagger a_{\bf k}\rangle$
and of the anomalous average $\langle a_{\bf k} a_{\bf k}\rangle$ in the $k\to 0$ limit ---}
As we have seen in Sec. \ref{Sub:depletionBog}, in the standard, number non-conserving version of 
Bogoliubov's theory, both the depletion $\langle a_{\bf k}^\dagger a_{\bf k}\rangle$
and the anomalous average $\langle a_{\bf k} a_{\bf k}\rangle$ diverge in the $k\to 0$ limit.
The use of a number-conserving approach, as we explicitly have shown in 
Sections \ref{Sub:vardepletion} and \ref{Sub:AnomalousAverages} 
for the single-mode theory and in Sec. \ref{Sub:DepletionManyModes} in presence of many momentum modes,
removes these unphysical divergences and yields results that are always finite for finite $N$.

\medskip

{\em 2. Use of the Hamiltonian $\hat{H}$ instead of $\hat{H}-\mu\hat{N}$ to describe the energy of the system ---}
In addition to removing the unphysical divergences from the quantities
$\langle a_{\bf k}^\dagger a_{\bf k}\rangle$ and $\langle a_{\bf k} a_{\bf k}\rangle$
in the $k\to 0$ limit, the variational method has the advantage of restoring
to the Hamiltonian $\hat{H}$ its usual meaning as the operator representing
the total energy of the system. As we saw in Sec. \ref{Sub:meaningH}, this is not the case
in the conventional formulation of Bogoliubov's theory, where the role of the energy operator
is played by the combination $\hat{H} - \mu \hat{N}$, and where the use of $\hat{H}$ gives rise
to nonsensical results.

\medskip

{\em 3. Commutation relations of the $\alpha_{\bf k}$ operators ---} 
An interesting aspect of our investigation is that we have clarified the origin of the commutation relations
between the excitation operators $\alpha_{\bf k}$ and $\alpha_{\bf k}^\dagger$. Indeed, the number-conserving version 
of these operators, where $\alpha_{\bf k}$ is given by:
\begin{equation}
\alpha_{\bf k} = \gamma_{\bf k}\big(u_{\bf k}a_{\bf k}a_{\bf 0}^\dagger + v_{\bf k} a_{-\bf k}^\dagger a_{\bf 0}\big),
\end{equation}
obeys the following commutation relations: 
\begin{subequations}
\begin{align}
& [\alpha_{-\bf k},\alpha_{\bf k}] = [\alpha_{-\bf k}^\dagger,\alpha_{\bf k}^\dagger] = \gamma_{\bf k}^2 
u_{\bf k}v_{\bf k}\big[a_{\bf k}^\dagger a_{\bf k} - a_{-\bf k}^\dagger a_{-\bf k}\big].
\label{Eq:commutator1}
\\
& [\alpha_{\bf k},\alpha_{\bf k}^\dagger] \simeq \gamma_{\bf k}^2 \big[ a_0^\dagger a_0
- {u}_{\bf k}^2 a_{\bf k}^\dagger a_{\bf k} + {v}_{\bf k}^2 a_{-\bf k}^\dagger a_{-\bf k}\big].
\label{Eq:commutator2}
\end{align}
\end{subequations}
As we have seen in Sec. \ref{Sub:ExcOneMode}, imposing the commutation relation $[\alpha_{\bf k},\alpha_{\bf k}^\dagger]=1$
is nothing more than a convenient way to ensure that the excited state $\alpha_{\bf k}^\dagger|\Psi(N)\rangle$
is normalized to unity. Given that the {\em rhs} of Eq. (\ref{Eq:commutator2})
is an operator and not a c-number, it is not possible to satisfy the commutation relation 
$[\alpha_{\bf k},\alpha_{\bf k}^\dagger]=1$ for all possible states in the Hilbert space, and hence
for weak perturbations one simply requires that this commutation relation be satisfied in an averaged sense
at the ground state, {\em i.e.} $\langle\Psi(N)|[\alpha_{\bf k},\alpha_{\bf k}^\dagger]|\Psi(N)\rangle=1$.
This led us to the following condition on the constant $\gamma_{\bf k}$:
\begin{equation} 
\gamma_{\bf k}^2 = \frac{1}{\langle a_0^\dagger a_0\rangle - \langle a_{\bf k}^\dagger a_{\bf k}\rangle},
\label{Eq:condgamma}
\end{equation} 
with the consequence that $u_{\bf k}$ and $v_{\bf k}$ verify the identity $u_{\bf k}^2-v_{\bf k}^2=1$,
much like in Bogoliubov's approach. It is to be noted that,
even in the single-mode theory of Sec. \ref{Sub:ExcOneMode},
the condition (\ref{Eq:condgamma}) by itself is not sufficient to recover the Bogoliubov
spectrum of excitations, since we also require that the depletion of the ground state into any given momentum mode ${\bf k}$
be small, $\langle a_0^\dagger a_0\rangle \gg \langle a_{\bf k}^\dagger a_{\bf k}\rangle$,
so that we may write $\gamma_{\bf k}^2 \simeq 1/\langle a_0^\dagger a_0\rangle$.
Should a situation arise where the depletion of the ground state is no longer small
(such as for liquid Helium at very low temperatures, 
where the condensate fraction represents only about $10\%$
of the total system\cite{CommentHelium}), 
then the full expression of $\gamma_{\bf k}^2$ given above, Eq. (\ref{Eq:condgamma}), 
has to be used, and there is then no guarantee that the Bogoliubov spectrum will be recovered, 
even in the single-mode theory of Section \ref{Sub:ExcOneMode}.

\medskip

{\em 4. Single-mode theory: number-conserving variational approach for the single-mode
Hamiltonian $\hat{H}_{\bf k}$ ---}
As mentioned in the opening paragraph to this subsection, in this paper we have shown
that Bogoliubov's theory corresponds to a decoupled approach in which
each single-mode Hamiltonian $\hat{H}_{\bf k}$ is diagonalized
separately from the other momentum contributions $\hat{H}_{\bf k'(\neq k)}$.
What is more, we have performed an exact numerical diagonalization of the Hamiltonian $\hat{H}_{\bf k}$.
Comparison of this exact numerical diagonalization and the variational Bogoliubov treatment
shows that Bogoliubov's theory gives spectacularly accurate results for the ground state energy and the excitation
spectra of each of the single-mode Hamiltonians $\hat{H}_{\bf k}$. However, the results of Bogoliubov's method
for the depletion of the condensate are less accurate, as this method, even in its number-conserving incarnation,
overestimates the depletion of the condensate by about one order of magnitude for small values
of the wavevector $k$.

\medskip

{\em 5. Multi-mode theory: variational approach for the full Hamiltonian $\hat{H}$ ---} 
The most important result of this paper, though, has to do with our variational treatment 
of the full Hamiltonian $\hat{H}$, instead of the decoupled kind of treatment done in Bogoliubov's method where
each momentum contribution $\hat{H}_{\bf k}$ is diagonalized separately. In our variational approach,
the ${\bf k}=0$ state is restored to the Hilbert space used to describe the system
(by contrast to Ref. \citen{LeeHuangYang1957} where $a_0$ is replaced by $\sqrt{N}$),
and an accurate count is kept of the number of bosons in the ${\bf k}=0$ state. 
As a result, we have shown that the coefficients 
$c_{\bf k}$ which determine the ground state wavefunction of the system are no longer given by the 
expression (\ref{Eq:resultck}) obtained within the standard Bogoliubov approach. Instead, these
coefficients are now given by the alternate expression (\ref{Eq:newc_k}), which has the profound
consequence of giving rise to a gap in the excitation spectrum of bosons as $k\to 0$, 
by contrast to the excitation spectrum in the standard Bogoliubov theory which is gapless in that limit.
Note that, since the geometrical ansatz $\widetilde{C}_n=\sqrt{1-c_{\bf k}^2}(-c_{\bf k})^n$ for the coefficients 
$\widetilde{C}_n$ of the normalized ground state wavefunction 
$|\widetilde\psi_{\bf k}\rangle=\sum_{n=0}^{N/2}\widetilde{C}_n|n\rangle$
gave such accurate results for the ground state energy of the single-mode Hamiltonian $\hat{H}_{\bf k}$,
we expect the same ansatz to give equally good results when the variational method is applied to
the total Hamiltonian $\hat{H}=\sum_{\bf k\neq 0}\hat{H}_{\bf k}$, as we did in this paper.

\subsection{A comment on the leading correction 
to the Gross-Pitaevskii value of the ground state energy of an interacting Bose gas}

\begin{figure}[tb]
\centerline{\includegraphics[width=8.89cm, height=5.5cm]{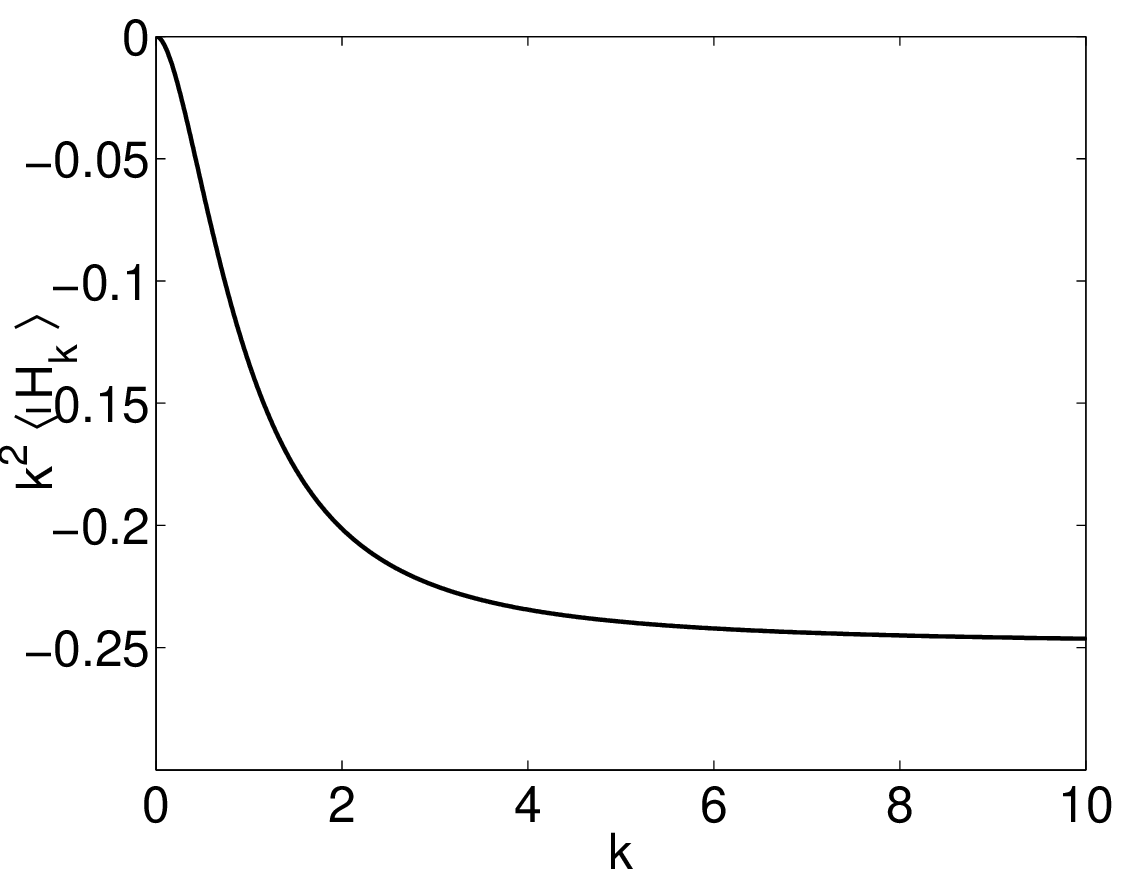}}
\caption[]{Plot of the product 
$k^2\times\langle\widetilde\psi_{\bf k}|\hat{H}_{\bf k}|\widetilde\psi_{\bf k}\rangle$ vs. $k$,
as obtained by exact numerical diagonalization of the single-mode Hamiltonian
$\hat{H}_{\bf k}$ for $N=200$ bosons. Here the wavector $k$ takes the values
$k=0.01,0.02,\ldots,10$. The fact that the product  
$k^2\times\langle\widetilde\psi_{\bf k}|\hat{H}_{\bf k}|\widetilde\psi_{\bf k}\rangle$
in dimensionless units goes to a constant $-0.25$ at large values of $k$ indicates that 
$\langle\widetilde\psi_{\bf k}|\hat{H}_{\bf k}|\widetilde\psi_{\bf k}\rangle$
goes to zero like $-0.25/k^2$ as $k\to\infty$, hence confirming the result of
the analytical Bogoliubov theory for that quantity.
}\label{Fig:ktimesGSEnergyBog}
\end{figure}

Several comments are in order concerning the explicit value of the ground state energy
of an interacting Bose gas. In the standard Bogoliubov approach, it is claimed 
that this quantity is given by the expression (we now include the Gross-Pitaevskii
term $N(N-1)v(0)/2V\simeq N^2v(0)/2V$ in our equations):
\begin{equation}
E_{Bog} = \frac{1}{2}Vn_B^2v({\bf 0})
-\frac{1}{2}\sum_{\bf k\neq 0}\big(\varepsilon_{\bf k} + n_Bv({\bf k}) - E_{\bf k}\big).
\label{Eq:totalE0}
\end{equation}
Of course,  we now know that the above expression can only be derived in a number non-conserving approach,
and is not the correct expression of the ground state energy when the conservation of boson number is taken into
account. However, for the sake
of argument, let us momentarily ignore this conceptual difficulty, and review how the 
final expression of the ground state energy
for the total Hamiltonian $\hat{H}$ is calculated from Eq. (\ref{Eq:totalE0}). 
In the case where the Fourier components of the interaction potential $v({\bf k})$ are all 
given by a single constant, $v({\bf k})=g$, corresponding to 
an interaction potential which is a delta function $v({\bf r})=g\delta({\bf r})$ in real space, 
the summand in the above expression has the asymptotic form
$-n_B^2g^2/(4\varepsilon_{\bf k})=-mn_B^2g^2/(2\hbar^2k^2)$, 
and hence the sum diverges in three dimensions. To take care of this divergence,
a procedure is devised \cite{FW}
whereby the coupling constant $g$ is shifted by the infinite (!) quantity 
$-mg^2\int\frac{d^3\bf k}{(2\pi)^3}(\hbar^2k^2)^{-1}$, and to compensate for this shift a quantity
$+\int\frac{d^3\bf k}{(2\pi)^3}mn_B^2g^2/(2\hbar^2k^2)$ is added to the expression of $E_0$, with the result:
\begin{align}
\frac{E_{Bog}}{V} & = \frac{1}{2}n_B^2\Big(g - \frac{mg^2}{\hbar^2}\int\frac{d^3\bf k}{(2\pi)^3}\,\frac{1}{k^2}\Big) 
\nonumber\\
& + \frac{1}{2}\int\frac{d^3\bf k}{(2\pi)^3}\,\Big(E_{\bf k}-\varepsilon_{\bf k}
- n_Bg +\frac{n_Bg}{2\varepsilon_{\bf k}}\Big).
\label{Eq:E0overV}
\end{align}
It turns out that the two terms between parentheses on the first line of the above equation
represent the first two terms in the expansion of the $s$-wave scattering length $a$ 
in terms of $g$:
\begin{equation}
\frac{4\pi a\hbar^2}{m} \simeq g - \frac{mg^2}{\hbar^2}\int\frac{d^3\bf k}{(2\pi)^3}\,\frac{1}{k^2} + \cdots.
\label{Eq:g_R}
\end{equation}
As it stands, the above expression, literally speaking, predicts a negatively infinite value
of the scattering length ${a}$ (unless, of course, an ultraviolet
cut-off is imposed on the momentum integral in such a way that the second term on the {\em rhs} of 
Eq. (\ref{Eq:g_R}) is smaller than the first, in which
case ${a}$ remains finite and positive; note, however, that if such a cut-off is imposed,
then we are no longer dealing with a delta function potential in real space
$v({\bf r})=g\delta({\bf r})$, but with an approximate form of the latter). 
The result of the above manipulations is that
the integral in the second line of Eq. (\ref{Eq:E0overV}) is now convergent in three dimensions, 
and leads to the celebrated result that is quoted in all standard textbooks on many-particle systems, 
where the final result is expressed not in
terms of the original interaction strength $g$, but in terms of the scattering length ${a}$:
\begin{equation}
\frac{E_{Bog}}{V} = \frac{2\pi {a}\hbar^2n_B^2}{m}\left[ 
1 + \frac{128}{15}\Big(\frac{n_Ba_{0}^3}{\pi}\Big)^{\frac{1}{2}}
\right].
\label{Eq:GSEBog0}
\end{equation}
Strictly speaking, the leading correction on the {\em rhs} of the above equation does not involve the scattering
length ${a}$ defined in Eq. (\ref{Eq:g_R}), but the ``bare" counterpart 
$a_{0}$ such that ${4\pi a_{0}\hbar^2}/{m} = g$.
However, in the limit of a dilute Bose gas where $n_B {a}\ll 1$, one casually
replaces $a_{0}$ by ${a}$ (a very questionable replacement, indeed, given
that the expression of ${a}$ contains a {\em divergent} integral, see Eq. (\ref{Eq:g_R})), leading to:
\begin{equation}
\frac{E_{Bog}}{V} = \frac{2\pi {a}\hbar^2n_B^2}{m}\left[ 
1 + \frac{128}{15}\Big(\frac{n_Ba^3}{\pi}\Big)^{\frac{1}{2}}
\right].
\label{Eq:GSEBog}
\end{equation}

At this point, we want to argue that the manipulations leading
to Eq. (\ref{Eq:GSEBog}) are mathematically untenable, for what we simply did
is shift the divergence from the sum of the expectation values 
$\sum_{\bf k\neq 0}\langle\hat{H}_{\bf k}\rangle_{\bf k}
=\frac{1}{2}\sum_{\bf k\neq 0}(E_{\bf k} - \varepsilon_{\bf k} - n_Bg)$
to the scattering length ${a}$ of Eq. (\ref{Eq:g_R}),
making the latter a divergent quantity, but the original 
divergence of the ground state energy has not been actually removed.
To clarify what we mean by the above statement, let us give
the bare interaction strength $g$ a numerical value, say 
$g=1.0\times 10^{-22}J\cdot \AA^3$ (this is not a value that is experimentally relevant
to any actual system, we just made it up for the sake of argument),
and ask what is the numerical value, in Joules, of the ground state energy per particle.
The presence of the divergent integral on the {\em rhs} of Eq. (\ref{Eq:g_R})
prevents us from giving an answer to this question, and that is what we
mean by the statement that the divergence of the ground state energy
has not been removed by the mathematically questionable manipulations
leading to Eq. (\ref{Eq:GSEBog}).

\addvspace{1.5mm}

In fact, the above manipulations would have been perfectly legitimate
had the integral on the {\em rhs} of Eq. (\ref{Eq:g_R}) been finite
and perturbatively small compared to the bare interaction strength $g$.
The fact that this is not the case, and that the integral we just mentioned
is actually divergent, makes going from $a_{0}$ in the correction term
on the {\em rhs} of Eq. (\ref{Eq:GSEBog0}) to ${a}$ in Eq. (\ref{Eq:GSEBog})
very questionable. Furthermore, and as we mentioned above, including
the divergent integral in the definition of the renormalized scattering length ${a}$,
Eq. (\ref{Eq:g_R}), merely shifts the divergence to this last quantity. In terms 
of the bare scattering length $a_{0}$, or the bare interaction strength $g$,
the ground state energy is still a divergent quantity (since ${a}$ itself is infinite
according to Eq. (\ref{Eq:g_R})), and hence we see that, strictly speaking,  
the manipulations leading to Eq. (\ref{Eq:GSEBog}) did not actually remove
the original divergence of the ground state energy of the system from Eq. (\ref{Eq:totalE0}).

\addvspace{1.5mm}

We now would like to pause for a moment, and plot the product $\tilde{k}^2\langle\hat{H}_{\bf k}\rangle_{\bf k}/n_Bg
=\tilde{k}^2\langle\widetilde\psi_{\bf k}|\hat{H}_{\bf k}|\widetilde\psi_{\bf k}\rangle/n_Bg$
as obtained from the {\em exact} numerical diagonalization of the Hamiltonian $\hat{H}_{\bf k}$
as a function of the wavevector $k$. This plot, given in Fig. \ref{Fig:ktimesGSEnergyBog},
shows that the above product goes to $-0.25$ at large values of $k$,
implying that the exact numerical result for $\langle\hat{H}_{\bf k}\rangle_{\bf k}$
goes to zero like $-n_B^2g^2/(4\varepsilon_{\bf k})$ (in dimensional units) as $k\to\infty$,
in agreement with the result of Bogoliubov's theory.
Unlike divergences that appear in other physical theories (e.g. the divergence
of the perturbative expansion of $\phi^4$-type models) which can be shown to be
artificial divergences and therefore need to be removed, we here are in the opposite situation where
the divergence of the sum $\sum_{\bf k\neq 0}\langle\hat{H}_{\bf k}\rangle_{\bf k}$
when $v({\bf r})=g\delta({\bf r})$ is not artificial at all, but on the contrary is
part and parcel of the {\em exact} diagonalization
of the Hamiltonians $\hat{H}_{\bf k}$. Hence, this is a divergence which {\em should not}
be tampered with and {\em cannot} be removed because it corresponds to the {\em correct}
behavior of the quantity $\sum_{\bf k\neq 0}\langle\hat{H}_{\bf k}\rangle_{\bf k}$
when a delta function interaction between bosons is assumed.

\begin{figure}[tb]
\centerline{\includegraphics[width=8.89cm, height=5.5cm]{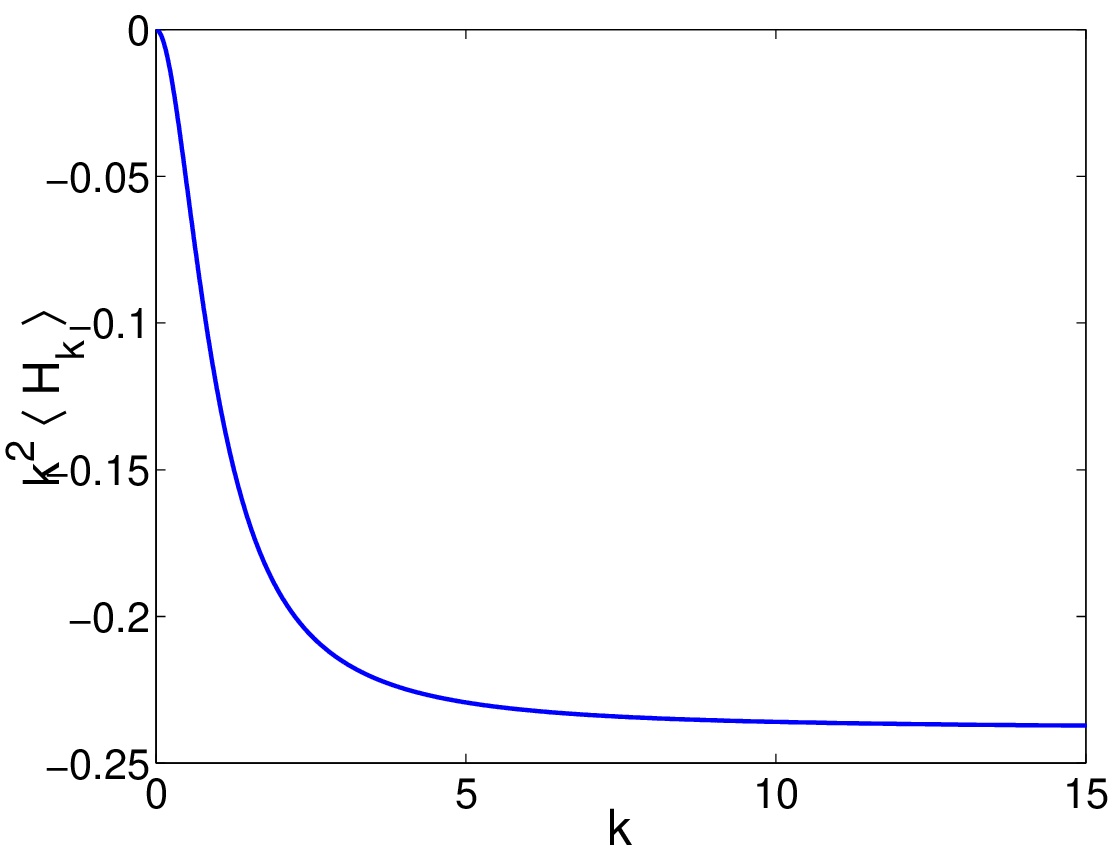}}
\caption[]{Plot of the product 
$\tilde{k}^2\times\langle\Psi|\hat{H}_{\bf k}|\Psi\rangle$ vs. $\tilde{k}$,
for $\tilde\sigma = 0.39$ and $C_d = 0.9762$ (solid line) when we set $\lambda=0$ in the interaction
potential (\ref{Eq:pot_lambda_FS}) between bosons. The fact that the product 
$\tilde{k}^2\times\langle\Psi|\hat{H}_{\bf k}|\Psi\rangle$ 
goes to a constant that is higher than $-0.25$ at large values of $k$ indicates that 
$\langle\Psi|\hat{H}_{\bf k}|\Psi\rangle$ does {\em  not} go to zero like $-0.25/\tilde{k}^2$ 
as $\tilde{k}\gg 1$. 
}\label{Fig:ktimesGSEnergy}
\end{figure}

\addvspace{1.5mm}

With that said, the whole discussion above is somewhat irrelevant, 
because the correct ground state energy of an interacting
Bose gas, when the full Hamiltonian $\hat{H}$ is diagonalized properly, 
is {\em not} given by Eq. (\ref{Eq:totalE0}). Indeed, in the improved variational procedure 
of Sec. \ref{Sec:DiagFullH},
the ground state energy does not diverge like 
$-n_B^2g^2/(4\varepsilon_{\bf k})=-0.25n_B^2g^2/\varepsilon_{\bf k}$,
but rather like $\sim -0.25C_d^2n_B^2g^2/\varepsilon_{\bf k}$ 
(see Fig. \ref{Fig:ktimesGSEnergy}), which makes the whole procedure described above 
(consisting of adding and substracting the quantity 
$-n_B^2g^2/(4\varepsilon_{\bf k})$) of no great help. 
The root cause of the divergence of the ground state energy
being the delta function form of the interaction potential, one should really only consider 
more realistic interactions whose Fourier transform falls off at high momentum, 
hence making the integral giving the ground state energy convergent. 
In this case, the correction to the Gross-Pitaevskii energy
$v({\bf 0})N(N-1)/2V$, expressed in terms of bare coupling constants, 
will be {\em negative}, as it should, and not
confusingly\cite{LeggettNJP} positive as in Eq. (\ref{Eq:GSEBog}).

\subsection{Do the operators $\alpha_{\bf k}^\dagger$ really create collective sound waves ?}
\label{Sec:GoldstoneModes}

We now turn our attention to the $\alpha_{\bf k}^\dagger$ operators, and discuss the validity of interpreting these
operators as creation operators for collective phonon modes in view of the fact that the spectrum of these
excitations, when the Hamiltonian is properly diagonalized in a number-conserving framework, is no longer gapless as 
is the case in the standard formulation of Bogoliubov's method. To gain a little more perspective, let us remind
ourselves of how phonon modes are defined in the case of a periodic monoatomic crystal. 
If we denote by $\hat{\bf d}({\bf x})$ the second-quantized operator describing the displacement 
of atoms from their equilibrium locations, then we can write:
\begin{align}
\hat{\bf d}({\bf x}) \propto \sum_{\bf k}\Big(\frac{\hbar}{2\omega_{\bf k}V}\Big)^\frac{1}{2}
\frac{\bf k}{k}\big[
\beta_{\bf k} e^{i{\bf k}\cdot{\bf x}} - \beta_{\bf k}^\dagger e^{-i{\bf k}\cdot{\bf x}}
\big],
\end{align}
where $\omega_{\bf k}=ck$ ($c$ being the speed of sound in the crystal), and where
$\beta_{\bf k}^\dagger$ and $\beta_{\bf k}$ create and annihilate a phonon mode of wavevector ${\bf k}$,
respectively. We thus see that the $\beta_{\bf k}$'s are related to the {\em displacement} field of the atoms, 
and it would be rather awkward to try to relate these operators to the operators which 
describe adding or removing one atom from the system. Yet, this is exactly 
how the $\alpha_{\bf k}$'s are defined in Bogoliubov's theory,
where these operators are usually interpreted in terms of phonon modes, even though they are not related to an actual
displacement field, {\em i.e.} to a density fluctuation, but merely describe adding or removing a boson from the system.
In the two paragraphs that follow, we want to argue that, 
whether it be from the perspective of the standard, number non-conserving formulation of Bogoliubov's theory,
or from the perspective of a number-conserving approach,
the $\alpha_{\bf k}$'s do not represent collective density oscillations involving {\em all} momentum modes, 
{\em i.e.} phonons, but correspond to single momentum mode excitations instead.

\subsubsection{Nature of the $\alpha_{\bf k}$'s in the standard, number non-conserving Bogoliubov approach}

Let us discuss the nature of the excited modes $\alpha_{\bf k}$ and $\alpha_{\bf k}^\dagger$ that are defined 
in the standard, number non-conserving formulation of Bogoliubov's theory.
For definiteness, let us rewrite the expressions of these operators 
in terms of the boson creation and annihilation operators:
\begin{equation}
\alpha_{\bf k} = u_{\bf k} a_{\bf k} + v_{\bf k} a_{-\bf k}^\dagger, 
\quad
\alpha_{\bf k}^\dagger = u_{\bf k} a_{\bf k}^\dagger + v_{\bf k} a_{-\bf k}.
\end{equation}
From the above equations, we see that $\alpha_{\bf k}^\dagger$ annihilates 
a {single} boson of momentum $-{\bf k}$ with probability $v_{\bf k}$, 
and creates a single boson of momentum ${\bf k}$ with probability $u_{\bf k}$. 
As we already mentioned above, given that $\alpha_{\bf k}^\dagger$ does not act
on the condensed bosons, nor does it act on the depleted bosons with momentum ${\bf k}'\neq {\bf k}$,
it is very surprising that the state $\alpha_{\bf k}^\dagger|\Psi_B\rangle$
has come to be interpreted as a phonon, the latter being a macroscopic excitation involving
{\em all} the bosons in the system. As the expression of $\alpha_{\bf k}$ and $\alpha_{\bf k}^\dagger$
clearly shows, these operators represent physical processes involving at most two bosons, 
and can therefore hardly be described as ``collective excitations", or ``phonons".
Another way to see this is to consider the following excited wavefunctions:
\begin{subequations}
\begin{align}
|\Phi_{\bf k}^{(n)}\rangle & = \frac{(\alpha_{\bf k}^\dagger)^n}{\sqrt{n!}}|\Psi_B\rangle,
\label{Eq:Phi}
\\
|\Psi_{\bf k}^{(n)}\rangle & = C_{\bf k}^{(n)}(a_{\bf k}^\dagger)^n|\Psi_B\rangle.
\label{Eq:Psi}
\end{align}
\end{subequations}
Conventional wisdom teaches us that these are two very different states: $|\Phi_{\bf k}^{(n)}\rangle$ being a state
with $n$ ``collective phonon modes", while the state $|\Psi_{\bf k}^{(n)}\rangle$ of Eq. (\ref{Eq:Psi}) represents 
a state where $n$ bosons have been added to the Bogoliubov ground state. A closer look,
however, reveals that these two states are actually identical, and correspond to the {\em same} quantum state obtained
by adding $n$ bosons of momentum ${\bf k}$ to the Bogoliubov ground state.
Indeed, using the fact that $a^\dagger_{\bf k} = u_{\bf k}\alpha^\dagger_{\bf k} - v_{\bf k} \alpha_{-\bf k}$,
and the fact that the operators $\alpha^\dagger_{\bf k}$ and $\alpha_{-\bf k}$ commute,
we can write:
\begin{subequations}
\begin{align}
& |\Psi_{\bf k}^{(n)}\rangle = C_{\bf k}^{(n)}\big[u_{\bf k}\alpha^\dagger_{\bf k} - v_{\bf k} \alpha_{-\bf k}\big]^n|\Psi_B\rangle,
\\
& = C_{\bf k}^{(n)}\sum_{m=0}^n \frac{n!}{m!(n-m)!}\,(u_{\bf k}\alpha^\dagger_{\bf k})^m(- v_{\bf k} \alpha_{-\bf k})^{n-m}|\Psi_B\rangle.
\end{align}
\end{subequations}
Now, given that $\alpha_{\bf k}|\Psi_B\rangle =0$ for all values of the wavevector ${\bf k}$, we see that all terms in the above 
summation vanish, except for the term $m=n$, which gives:
\begin{align}
|\Psi_{\bf k}^{(n)}\rangle = C_{\bf k}^{(n)}\,(u_{\bf k}\alpha^\dagger_{\bf k})^n|\Psi_B\rangle.
\label{Eq:intermPsi}
\end{align}
Requiring that $|\Psi_{\bf k}^{(n)}\rangle$ be normalized to unity results in the following value of the constant $C_{\bf k}^{(n)}$:
\begin{equation}
C_{\bf k}^{(n)} = \frac{1}{u_{\bf k}^n\sqrt{n!}},
\end{equation}
upon which Eq. (\ref{Eq:intermPsi}) becomes:
\begin{align}
|\Psi_{\bf k}^{(n)}\rangle = \frac{(\alpha_{\bf k}^\dagger)^n}{\sqrt{n!}}|\Psi_B\rangle = |\Phi_{\bf k}^{(n)}\rangle.
\label{Eq:PsiEqualsPhi}
\end{align}
This proves our earlier claim that the states $|\Psi_{\bf k}^{(n)}\rangle$ and 
$|\Phi_{\bf k}^{(n)}\rangle$ correspond to the {\em same} quantum state.
Eq. (\ref{Eq:PsiEqualsPhi}) is an important result, 
which shows that a state of the form 
$|\Phi_{\bf k}^{(1)}\rangle=\alpha_{\bf k}^\dagger |\Psi_B\rangle = \frac{1}{u_{\bf k}}a_{\bf k}^\dagger|\Psi_B\rangle$ 
does {\em not} represent a collective phonon mode at all, and represents merely a state with 
one extra boson of momentum ${\bf k}$ added to the Bogoliubov ground state.

\subsubsection{Nature of the $\alpha_{\bf k}$'s in the variational, number-conserving Bogoliubov formulation}

We now want to discuss the nature of the $\alpha_{\bf k}$ excitations from the point of view of the variational
formulation of Bogoliubov's theory discussed in Section \ref{Sub:GeneralizationFullH}. In this formulation, 
$\alpha_{\bf k}$ is defined by Eq. (\ref{Eq:againuv_2}), and we therefore have for $\alpha_{\bf k}^\dagger$
the following expression:\cite{Leggett2001}
\begin{equation}
\alpha_{\bf k}^\dagger = \tilde{u}_{\bf k} a_{\bf k}^\dagger a_0 + \tilde{v}_{\bf k} a_{-\bf k}a_0^\dagger.
\end{equation}
As we already discussed in Sec. \ref{Sub:ExcOneMode}, the two terms on the {\em rhs} of the above equation
represent the two different ways the system can be excited from the 
ground state $|\Psi(N)\rangle$, which is an eigenstate of the total momentum
operator $\hat{\bf P}$ with eigenvalue $0$, to a state with momentum $+{\bf k}$.
Again here, we can express $a_{\bf k}^\dagger a_0$ in terms of the $\alpha_{\bf k}$'s, with
the result (see Eq. (\ref{Eq:resultbk}) of the Appendix):
\begin{equation}
a_{\bf k}^\dagger a_0 = \frac{1}{\gamma_{\bf k}}\big[{u}_{\bf k}\alpha_{\bf k}^\dagger
-{v}_{\bf k}\alpha_{-\bf k}\big].
\label{Eq:resultakda0}
\end{equation}
Let us apply the operator on the {\em lhs} of the above equation to the ground state $|\Psi(N)\rangle$
of the operator $\hat{H}$. We have:
\begin{align}
a_{\bf k}^\dagger a_0 |\Psi(N)\rangle & = \frac{1}{\gamma_{\bf k}}\big[{u}_{\bf k}\alpha_{\bf k}^\dagger
-{v}_{\bf k}\alpha_{-\bf k}\big]|\Psi(N)\rangle,
\nonumber\\
& = \frac{u_{\bf k}}{\gamma_{\bf k}}\alpha_{\bf k}^\dagger|\Psi(N)\rangle,
\end{align}
where, in going from the first to the second line, we used the fact that $\alpha_{-\bf k}|\Psi(N)\rangle=0$.
As it can be seen from this last equation, the quantum state $\alpha_{\bf k}^\dagger|\Psi(N)\rangle$
is identical to the state $a_{\bf k}^\dagger a_0 |\Psi(N)\rangle$. Since the latter
represents a state where one boson has been removed from the condensate and a boson
has been added to the single-particle state with momentum ${\bf k}$, we conclude here again
that the operator $\alpha_{\bf k}^\dagger$, when applied to the ground state $|\Psi(N)\rangle$,
does not create a collective sound wave
(of the kind that would be created by shaking the walls of the container, 
or varying the parameters of the confining potential in a trapped system ---
that is after all how a sound wave would be created experimentally\cite{Castin1998} --- in which case
all modes of the system would be excited in a collective way, not just one momentum mode as is the case here), 
and merely promotes a single boson from the condensate
with single-particle momentum ${\bf k}={\bf 0}$ to the state with single-particle momentum $+{\bf k}$.

\addvspace{1.5mm}

To summarize this subsection, we conclude that the gapless phonon-like spectrum of the 
$\alpha_{\bf k}^\dagger$ operator in the standard Bogoliubov method,
which in effect creates excited states for the single-mode Hamiltonian $\hat{H}_{\bf k}$,
has no particular physical meaning in terms of travelling sound waves.
Successive application of the operator $\alpha_{\bf k}^\dagger$ to the ground state
$|\psi_{\bf k}\rangle$ of the Hamiltonian $\hat{H}_{\bf k}$ merely adds bosons 
to the single-particle state $e^{i{\bf k}\cdot{\bf r}}/\sqrt{V}$ with momentum $+{\bf k}$,
and hence the corresponding excitations are in no way collective, 
and cannot be identified as sound waves, as they routinely are in the literature.
This conclusion is corroborated by the results we obtained in Sec. \ref{Sub:ElementaryExcitations}
where we have shown that, when the $\bf k=0$ state is restored into
the Hilbert space used to describe the system and the conservation of the number 
of bosons is properly taken into account, the spectrum of the 
$\alpha_{\bf k}$ operators has a gap in the long wavelength limit 
$k\to 0$ and cannot therefore be interpreted as a propagating density disturbance
corresponding to a collective sound wave.
(Incidentally, it is quite instructive to recall that in the BCS theory of superconductivity, 
the elementary excitation operators
are defined in a way that is very similar to how the $\alpha_{\bf k}$'s are 
defined in Bogoliubov's method. Yet, in BCS theory, the $\alpha_{\bf k}$ are {\em not} described as phonons,
but are instead correctly identified as single-particle excitations.)

\subsubsection{Proper description of phonon modes in interacting Bose systems}

Having argued that the $\alpha_{\bf k}$'s do not represent collective phonons, one may wonder how to properly describe density fluctuations in the Bose gas. It turns out that
the operator generating a density excitation of wavevector ${\bf q}$ is nothing but
the $q$-mode fluctuation of the density\cite{Feynman1954} $\hat\rho({\bf r})$: 
\begin{subequations}
\begin{align}
\hat\rho_{\bf q} & = \frac{1}{\sqrt{V}}\int d{\bf r}\; \mbox{e}^{i{\bf q}\cdot{\bf r}}\rho({\bf r}).
\\
& = \frac{1}{\sqrt{V}}\sum_{\bf k} a^\dagger_{\bf q + k} a_{\bf k}.
\label{Eq:rho_q}
\end{align}
\end{subequations}
Hence, the correct phonon mode of the system at wavevector ${\bf q}$ is the one given by the following wavefunction:
\begin{equation}
|\Psi_{\bf q}(N)\rangle = \hat\rho_{\bf q}|\Psi(N)\rangle,
\label{Eq:Psi^G}
\end{equation}
where $|\Psi(N)\rangle$ is the ground state wavefunction of the system.
Now, it is a well-known result that the expectation value of the Hamiltonian in the above state
is given by:\cite{HuangBook}
\begin{align}
E^{(1)}_{\bf q} = E_0 + \frac{\langle\Psi_{\bf q}|\hat{H}|\Psi_{\bf q}\rangle}
{\langle\Psi_{\bf q}|\Psi_{\bf q}\rangle} = E_0 + \frac{\hbar^2q^2}{2mS_{\bf q}}, 
\label{Eq:omega_q}
\end{align}
where $E_0$ is the ground state energy and $S_{\bf q}$ is the static structure factor of the system:
\begin{equation}
S_{\bf q} = \frac{1}{n_B}\langle\Psi(N)|\hat\rho^\dagger_{\bf q}\hat\rho_{\bf q}|\Psi(N)\rangle.
\end{equation}
For a system of bosons, it has been established long ago by Feynman \cite{Feynman1954} on quite general
grounds that the structure factor $S_{\bf q}$ varies linearly with $q$, $S_{\bf q}\sim q$, as $q\to 0$,
which in turns allows us to conclude that the excitation energy $\Delta E_{\bf q}=E^{(1)}_{\bf q}-E_0$ 
of the state $|\Psi_{\bf q}\rangle$
also varies linearly with $q$ in the $q\to 0$ limit:
\begin{equation}
\Delta E_{\bf q} \sim q\quad \mbox{as}\quad q\to 0.
\end{equation}
We therefore conclude that the gapless phonon modes are those given by the wavefunction in 
Eq. (\ref{Eq:Psi^G}). Given the expression (\ref{Eq:rho_q}) of the $q$-mode density fluctuation $\hat\rho_{\bf q}$,
we see that a gapless mode is obtained from the ground state $|\Psi(N)\rangle$ by
using a linear combination of creation and annihilation operators
$\sim\sum_{\bf k} a^\dagger_{\bf q +k} a_{\bf k}$, where the sum extends over {\em all} wavectors ${\bf k}$,
and not simply a truncated sum over a single wavevector ${\bf q}$ as is the case in the definition
of the operators $\alpha_{\bf q}$. This is consistent with our previous conclusion that the $\alpha_{\bf q}$'s
represent single momentum mode excitations whose spectum can have a gap as $q\to 0$, 
much like the analogous operators in BCS theory, while
the more general modes $\hat\rho({\bf q})|\Psi(N)\rangle$ are {\em collective} 
density excitations involving all momentum modes in the system, 
with a gapless excitation spectrum in the $q\to 0$ limit.\cite{HuangBook,Bobrov2010}

\addvspace{1.5mm}

It is interesting at this point to examine whether the standard formulation of Bogoliubov's theory is in agreement with
Feynman's approach reviewed above. To this end, let us calculate the excitation spectrum of Eq. (\ref{Eq:omega_q})
and see if we indeed obtain a linear spectrum in the $q\to 0$ limit. By using the expression of $a_{\bf k}$
in terms of the $\alpha_{\bf k}$'s, Eq. (\ref{Eq:defalphas}), to express
the local density operator $\hat\rho_{\bf q}$ in terms of the $\alpha_{\bf k}$'s,
and using the quadratic expression (\ref{Eq:diagtildeH2}) of the Bogoliubov Hamiltonian 
in terms of these same operators, one easily obtains, after a straightforward calculation:
\begin{subequations}
\begin{align}
E^{(1)}_{\bf q} & = \frac{\langle\Psi(N)|\hat\rho^\dagger_{\bf q}\hat{H}\hat\rho_{\bf q}|\Psi(N)\rangle}
{\langle\Psi(N)|\hat\rho^\dagger_{\bf q}\hat\rho_{\bf q}|\Psi(N)\rangle}  
\\
& = \tilde{E}_2 + \frac{\sum_{\bf k\neq 0}(E_{\bf k} + E_{\bf k+q})\big(v_{\bf k}^2 u_{\bf k+q}^2 
+ u_{\bf k}v_{\bf k}u_{\bf k+q}v_{\bf k+q}\big)}
{\sum_{\bf k\neq 0}\big(v_{\bf k}^2u_{\bf k+q}^2 + u_{\bf k}v_{\bf k}u_{\bf k+q}v_{\bf k+q}\big)}.
\end{align}
\end{subequations}
We thus obtain that the excitation spectrum $\Delta E_{\bf q}= E^{(1)}_{\bf q}-\tilde{E}_2$ 
of the state $|\Psi_{\bf q}\rangle$ in Bogoliubov's theory has a nonzero limiting value as $q\to 0$, given by:
\begin{equation}
\Delta E_{q\to 0} = \frac{2\sum_{\bf k\neq 0}E_{\bf k}u_{\bf k}^2v_{\bf k}^2}{\sum_{\bf k\neq 0}u_{\bf k}^2v_{\bf k}^2}
+ o(q^2). 
\label{Eq:Goldstone_Bogoliubov}
\end{equation}
We therefore see that Bogoliubov's theory does not reproduce the phonon spectrum, in the sense that the spectrum 
of the correct phonon mode $|\Psi_{\bf q}\rangle$ has a finite gap in the $q\to 0$ limit, a result which may
be attributed\cite{Suto2008} to the fact that Bogoliubov's Hamiltonian is a truncated approximation to the full many-body 
Hamiltonian for which Feynman's argument, Eq. (\ref{Eq:omega_q}), holds. Hence, for that reason we expect our number-conserving 
formulation of Bogoliubov's theory to give a result that is similar to the one in Eq. (\ref{Eq:Goldstone_Bogoliubov}),
which is not necessarily gapless as $q\to 0$.

\subsection{Comment on the apparent violation of the Hugenholtz-Pines and Goldstone theorems}

We now want to address the criticism that will inevitably be made about the fact that the variational
theory presented in this paper predicts an excitation spectrum which seems to violate the celebrated Hugenholtz-Pines
\cite{Hugenholtz1959} and Goldstone\cite{Goldstone,Brauner2010} theorems. As it has been observed in Ref. \citen{Suto2008}, 
a finite gap, which appears almost inevitably in number-conserving descriptions of interacting bosons, 
does not in any way indicate an internal inconsistency in such theories. In the previous section we have argued
that it is not correct to identify the $\alpha_{\bf k}$'s as the collective phonon modes 
of the system, and in the present section we shall argue that the presence of a finite gap in the spectrum of the 
$\alpha_{\bf k}$ excitations does not in any way imply the violation of the Goldstone theorem in our approach. But,
before that, we shall start by briefly discussing the apparent violation of the Hugenholtz-Pines theorem.

\subsubsection{Apparent violation of the Hugenholtz-Pines theorem}

In its simplest expression, the HPT states that the
zero wavevector and zero frequency limit of the 
diagonal and off-diagonal self-energies,
$\Sigma_{11}({\bf k},\omega)$ and $\Sigma_{12}({\bf k},\omega)$ respectively,
are related to the chemical potential $\mu$ 
through the equation:\cite{StoofBook}
\begin{equation}
\hbar\Sigma_{11}({\bf 0},0) - \hbar\Sigma_{12}({\bf 0},0) = \mu.
\label{Eq:HPT}
\end{equation}
In the standard field-theoretic formulation of Bogoliubov's theory an immediate consequence of this theorem
is the emergence of an excitation spectrum which does not have a gap as $k\to 0$. In the formalism presented in this
paper, however, a finite gap is found, which can be seen as an indication that our variational theory violates the
HPT.

\addvspace{1.5mm}

As an answer to this potential criticism, we shall note that the HPT has only been established
in the number non-conserving field-theoretic formulation of Bogoliubov's theory. Whether this theorem holds in
number-conserving situations is far from obvious, and is not, by any means, guaranteed.\cite{Misawa1960} 
It is in fact perfectly conceivable that a number-conserving field-theoretic 
formulation of the theory of interacting bosons may not satisfy the {\em same} 
Hugenholtz-Pines theorem satisfied by the number non-conserving version, 
in such a way that the excitation spectrum resulting from the
modified Hugenholtz-Pines theorem for the number-conserving theory is not inconsistent 
with the excitation spectrum derived in the present study. 
To construct a number-conserving field theory for interacting bosons is, however, a 
nontrivial exercise. What is more, to obtain an equivalent to the Hugenholtz-Pines theorem 
in such a theory may not even be possible in the first place, given that a number-conserving theory would most 
likely not require the introduction of the off-diagonal self energy 
$\Sigma_{12}({\bf k},\omega)$ which appears in the HPT. Making any further speculation about this issue is
far beyond the scope of this paper, and we shall therefore content ourselves with the observation made at the
beginning of this paragraph about the fact that the HPT was established within a number non-conserving framework. 
As a result, any claim about our variational theory being in violation of the HPT would be totally unfounded, 
since this theory is a number-conserving one, and we simply have no idea how one can 
write down an appropriate generalization of the HPT that would be valid in number-conserving situations.

\subsubsection{Apparent violation of the Goldstone theorem}

Before we proceed to discuss the apparent violation of the Goldstone theorem in our approach,
we first want to digress on the applicability of this theorem to the standard formulation of Bogoliubov's theory. Using the expression of the operator $a_{\bf k}$ in terms of the field operator $\Psi({\bf r})$:
\begin{equation}
a_{\bf k} = \int d{\bf r}\;\Psi({\bf r})\frac{\mbox{e}^{-i{\bf k}\cdot{\bf r}}}{\sqrt{V}},
\end{equation}
the (truncated) Bogoliubov Hamiltonian 
$\hat{H}=H_0 +\hat{H}_2$ of Eq. (\ref{Eq:defH_i}):
\begin{align}
\hat{H} = \frac{1}{2}V n_0^2v({\bf 0})
+ \sum_{\bf k\neq 0} \Big\{ 
\xi_{\bf k} a_{\bf k}^\dagger a_{\bf k} 
 +  \frac{1}{2}n_0 v({\bf k}) \big[ a_{\bf k}^\dagger a_{-\bf k}^\dagger
+ a_{\bf k} a_{-\bf k}\big]\Big\}
\end{align}
can be rewritten in the form (note that $v({\bf 0})$ still denotes the value of $v({\bf k})$ at ${\bf k}={\bf 0}$):
\begin{align}
\hat{H}  & = \frac{1}{2}V n_0^2v({\bf 0}) + \int d{\bf r}d{\bf r}'\;
\Psi^\dagger({\bf r})\Big[\Big(-\frac{\hbar^2\nabla^2}{2m} + n_0v({\bf 0})\Big)\delta({\bf r}-{\bf r}') + v({\bf r}-{\bf r}')\Big]
\Psi({\bf r}')
\nonumber\\
& + \frac{1}{2} n_0\int d{\bf r}\,d{\bf r}'\;\Big[
\Psi^\dagger({\bf r})v({\bf r}-{\bf r}')\Psi^\dagger({\bf r}') + \Psi({\bf r})v({\bf r}-{\bf r}')\Psi({\bf r}')
\Big]
\nonumber\\
& - \frac{1}{2}n_0v({\bf 0})\int d{\bf r}\,d{\bf r}'\;\Big[\Psi^\dagger({\bf r})\Psi^\dagger({\bf r}') + 
\Psi({\bf r})\Psi({\bf r}') + 4\Psi^\dagger({\bf r})\Psi({\bf r}')\Big].
\end{align}
From this last equation, we see that the two terms on the second line and the first two terms on the third line are {\em not} invariant under the transformation $\Psi({\bf r})\to \Psi({\bf r})e^{i\theta}$. Hence, the central premise of Goldstone's theorem
as applied to Bose systems, requiring that the Hamiltonian be invariant under global gauge transformations, is not satisfied in the standard Bogoliubov model\cite{Suto2008} (this is of course a direct consequence of Bogoliubov's prescription of replacing $a_0$ by an ``inert" $c$-number). One is therefore led to wonder whether Goldstone's theorem is even applicable to the standard Bogoliubov model in the first place, and whether it is legitimate at all to interpret the gapless nature of the spectrum of the $\alpha_{\bf k}$ excitations as having anything to do with the alleged breakdown of the global $U(1)$ gauge symmetry in this system.

\addvspace{1.5mm}

We now turn our attention to the applicability of Goldstone's theorem to our variational model. 
Here, because $a_0$ and $a_0^\dagger$ are not replaced by a $c$-number, one can easily verify that the Bogoliubov Hamiltonian of Eq. (\ref{Eq:HsumHk}) is actually invariant under the transformation $\Psi({\bf r})\to \Psi({\bf r})e^{i\theta}$.
The question now is to know whether the second premise of Goldstone's theorem is satisfied, namely, whether our variational ground state breaks the gauge invariance of the Hamiltonian.
Indeed, unless the ground state of the system breaks a continuous invariance of the Hamiltonian, a gapless mode does not necessarily exist. A case in point is provided by the Hartree-Fock Hamiltonian:
\begin{align}
\hat{H}_{HF} = \frac{v(0)}{2V}\hat{N}(\hat{N}-1)
+\sum_{\bf k\neq 0} \varepsilon_{\bf k}a_{\bf k}^\dagger a_{\bf k}
+ \frac{1}{2V}\sum_{\bf k}\sum_{\bf k'(\neq k)} v({\bf k}-{\bf k}') a_{\bf k}^\dagger a_{\bf k} a_{\bf k'}^\dagger a_{\bf k'}.
\end{align}
It is easy to verify that the kets given by (note that we here no longer explicitly distinguish between ${\bf k}$ and $-{\bf k}$):
\begin{align}
|\Psi_{HF}(\{n_i\})\rangle & = |N-\sum_{i=1}^\infty n_i;n_1;\ldots;n_\infty\rangle,
\nonumber\\
&=\frac{(a_0^\dagger)^{N-\sum_i n_i}}{\sqrt{(N-\sum_i n_i)!}}
\prod_{i=1}^\infty\frac{(a_{{\bf k}_i}^\dagger)^{n_i}}{\sqrt{n_i!}}|0\rangle,
\end{align}
are {\em exact} eigenfunctions of the Hamiltonian $H_{HF}$ with the exact eigenvalues:
\begin{align}
E_{HF}(\{n_i\}) = \frac{v(0)}{2V}N(N-1) + \sum_{i=1}^\infty n_i\varepsilon_{{\bf k}_i} 
+ \frac{1}{2V}\sum_{i=1}^\infty\sum_{j(\neq i)} v({\bf k}_i-{\bf k}_j) n_i n_j.
\end{align}
Assuming for simplicity that the Fourier transform $v({\bf k})$ is positive for all values of the wavevector ${\bf k}$,
one can easily show that the ground state of $\hat{H}_{HF}$ is given by the state where all the bosons are condensed in the ${\bf k} = 0$ single-particle state:
\begin{equation}
|\Psi_{HF}(N)\rangle = |N;0,0;\ldots\rangle.
\end{equation}
This state does not break the $U(1)$ gauge symmetry of the original Hamiltonian, 
in the sense that the action of the generator $\hat{U}(\theta)=\exp(i\theta\hat{N})$
on $|\Psi_{HF}(N)\rangle$ does not create a {\em distinct} degenerate ground state, but merely multiplies $|\Psi_{HF}(N)\rangle$ 
by the trivial overall constant $\exp(i\theta N)$. We therefore find ourselves in a situation where a Goldstone mode does not necessarily exist, and it is indeed found that there is a finite energy gap to any single-particle excitation of the system outside of its ground state. The variational ground state studied in this paper presents us with a very analogous situation, since it too does not break the $U(1)$ gauge symmetry (in the sense explained above), and hence is not bound to verify Goldstone's theorem.

\addvspace{1.5mm}

Having noted the above, we also note that Goldstone bosons do not always emerge in the excitation spectra of many-body systems
even when a continuous symmetry is spontaneously broken. 
A prominent example with a strong analogy to our system is provided by the BCS theory of 
neutral (atomic) Fermi superfluids, where a continuous (gauge) symmetry is spontaneously broken, and yet the spectrum of the
Hamiltonian has a finite gap $\Delta\neq 0$ in the $k\to 0$ limit. In this particular system, a sharp
distinction is made between the spectrum obtained by diagonalizing the Hamiltonian, 
which is identified as one describing {\em single-particle} excitations, 
and the gapless Bogoliubov-Anderson density oscillations, 
which are identified as the correct Goldstone modes of the system.\cite{Giorgini2008} 
A similar line of thought has been pursued for interacting bosons 
in Ref. \citen{Bobrov2010}, where it has been argued that the spectrum 
of interacting bosons should have two branches: a single-particle branch, obtained by diagonalizing the Hamiltonian, which may have a gap, and a collective, gapless branch corresponding to density fluctuations, which emerges as the pole of the density-density correlation function. A similar conclusion has been reached by Kita, who in a series of recent
articles\cite{Kita2009,Kita2010,Kita2011} has examined the nature of Goldstone bosons in an interacting Bose system 
using a careful analysis of the field-theoretic perturbation expansions for the one- and two-particle Green's
functions. A major reason why it is generally accepted that the Bogoliubov spectrum describes the collective Goldstone
mode of an interacting Bose system is that this spectrum determines the poles of both one- 
and two-particle Green's functions, as was claimed by Gavoret and Nozi\`eres,\cite{Gavoret1964}
and reproduced by the dielectric formalism.\cite{Szepfalusy1974,Wong1974}
Kita has argued that, since these theories were based on separate perturbative expansions 
for the one- and two-particle Green's functions, they
may suffer from ambiguities in how to define self-energies and vertices in presence of interactions having only a
single quasiparticle channel, as is the case with the single-component
Bose gas. To address this difficulty, Kita has reinvestigated the perturbative treatment\cite{Kita2009,Kita2010} 
of both Green's functions within a unified framework, whereby the two-particle Green's function
can be obtained from its one-particle counterpart by functional differentiation with respect to an additional
potential. The result that emerged from these investigations was that single-particle excitations as described by
the pole of the one-particle Green's function are subject to severe damping effects, while the Goldstone mode exists as
an isolated pole of the two-particle Green's function.\cite{Kita2011} This conclusion,
along with the conclusion of Ref. \citen{Bobrov2010},
suggests that one may have to look for gapless modes 
in collective density perturbations, just as is the case in BCS theory of neutral superfluids, and not necessarily in the poles of the single-particle Green's function.

\addvspace{1.5mm}

We conclude this subsection by noting that although the phonon modes associated with density perturbations as described by Feynman's
wavefunction and the poles of the two-particle Green's functions of the interacting Bose system are supposed to describe the same physical phenomenon of propagating phonon waves, the deep connection between the two formalisms still needs to be elucidated.   
Such an elucidation would undoubtedly help us understand why and under what conditions gapped modes can exist, and how gapless modes can be meaningfully incorporated in theoretical descriptions of condensed Bose systems.

\subsection{Consequence on field-theoretic formulations of Bogoliubov's theory}

We now would like to make a brief comment on the field-theoretic formulations
of Bogoliubov's method, both at $T=0$ and at finite temperatures. It is important to realize
that the field-theoretic formulations currently in use in the literature 
(see for example Ref. \citen{Shi1998} and references therein)
are based on the Bogoliubov ground state, which is a simple product of the individual
ground states of the single-mode Hamiltonians $\hat{H}_{\bf k}$, as in Eq. (\ref{Eq:ProductPsik}).
Needless to say, the Green's and spectral density functions 
will have a completely different structure when the field theory is formulated using the ground state
discussed in this paper as a starting point. As a result, various other 
quantities, such as density-density correlation functions,\cite{Griffin1993} 
and the damping of quasiparticle excitations,\cite{Shi1998} 
may end up having expressions that are qualitatively very different 
from those derived within the standard, number non-conserving Bogoliubov
approximation. 

\addvspace{1.5mm}

With that said, it would also be interesting to probe whether our variational ground state
can be reproduced using path integration methods. Path integrals have emerged as an elegant
and powerful tool to study many-body systems, and a successful formulation in terms of path integrals
which is able to describe the ground state studied in this paper may help put
the Green's function formalism of interacting bosons at finite temperatures 
on a firmer ground.\cite{CommentFiniteT} However, such a path-integral formulation
may technically prove to be difficult to achieve, since it would most likely require going beyond
standard Gaussian integration.

\subsection{Recent literature on number-conserving formulations of Bogoliubov's theory}

Before we conclude, we want to briefly discuss recent attempts to overcome
the difficulties that arise due to the non-conservation of particle number in Bogoliubov's theory,
as our paper would be rather incomplete without such a discussion.
One notable such attempt is the interesting work of Castin and Dum, Ref. \citen{Castin1998}, 
where the ground state of the interacting Bose system is found by writing the
field operator $\hat\psi({\bf r})$ as a sum of a condensate contribution $\phi_0({\bf r})a_0$
and a contribution $\delta\hat\psi({\bf r})$ from the $\bf k\neq 0$ states:
\begin{equation}
\hat\psi({\bf r}) = \phi_0({\bf r}) a_0 + \delta\hat\psi({\bf r}),
\end{equation}
where $\phi_0({\bf r})$ denotes the wavefunction of the condensed bosons. Then, the quantity $\delta\hat\psi({\bf r})$
is treated as a perturbation with respect to the condensate part $\phi_0({\bf r})a_0$, 
an approximation which is expected to be valid for a dilute Bose gas.
Imposing a condition of the form:
\begin{equation}
\alpha_{\bf k}|\Psi\rangle = 0,
\label{Eq:conditionCastinDum}
\end{equation}
which is reminiscent of Eq. (\ref{Eq:defGS}), and performing a systematic expansion in the small
parameter $\varepsilon=\sqrt{N_d/N_0}$ ($N_d$ being the total number of depleted bosons) 
allows one to recover the results of Bogoliubov's theory for the depletion
of the condensate and the excitation spectrum of the system.
Overall, the approach presented in this work is quite thoughtful, and has the advantage
of providing a clean derivation of the standard Bogoliubov results
that overcomes the need to break the U(1) symmetry, hence showing that this universally accepted
paradigm is not by any means required to describe condensed Bose systems. 
By trying to keep the conservation of the 
total number of bosons intact, it avoids many pitfalls of the conventional
Bogoliubov method. There is a reason, however, why this approach yields Bogoliubov-type results
instead of the results found in the present study. This reason has to do with a number of approximations 
that are made at a few key steps of the calculation, and more specifically the approximation which
consists in replacing the combination $a_0^\dagger a_0$ by the total number of bosons $N$. 
This approximation being, to a certain extent, equivalent to Bogoliubov's prescription, it is not at all
surprising that Bogoliubov's results are recovered. By contrast, in our variational approach, 
no such approximation is used, and great care is exercised in order to 
keep track of the exact number of bosons $(N-2\sum_i n_i)$ in the condensate
(see Appendix \ref{AppendixB}). 
The sharp distinction between using the approximation $a_0^\dagger a_0\simeq N$ on one
hand, and using $a_0^\dagger a_0=(N-2\sum_i n_i)$ on the other hand, 
is what leads to the differing results between the two methods.\cite{commentCastin}

\addvspace{1.5mm}

We now want to comment on the variational approach used by Leggett in Ref. \citen{Leggett2001},
an approach which has been a major inspiration for the present study.
In this approach, the ground state wavefunction of the system 
is given by (see Eq. (8.9) of Ref. \citen{Leggett2001}):
\begin{equation}
\Psi_N = N!^{-1/2}\Big(a_0^\dagger a_0^\dagger - 
\sum_{\bf k>0} c_{\bf k} a_{\bf k}^\dagger a_{-\bf k}^\dagger\Big)^{N/2}|0\rangle.
\label{Eq:PsiLeggett}
\end{equation}
Although the way this wavefunction is written is different from the way we wrote the wavefunction
in Eq. (\ref{Eq:fullPsi(N)}), a closer look reveals that these two wavefunctions are in fact similar to each other
in that both represent an expansion in terms of the basis states
$|N-2\sum_i n_i;n_1,n_1;\ldots;n_i,n_i;\ldots\rangle$. In the course of the calculation, however,
and in order to find the expectation value of the single-mode Hamiltonian $\hat{H}_{\bf k}$ 
in the Bogoliubov ground state, Leggett only retains
the following part of $\Psi_N$ (see Eq. (8.15) of Ref. \citen{Leggett2001}):
\begin{equation}
\Psi_{\bf k} = N!^{-1/2}(1 - c_{\bf k}^2)^{1/2}
\Big(
a_0^\dagger a_0^\dagger - c_{\bf k}a_{\bf k}^\dagger a_{-\bf k}^\dagger
\Big)^{N/2}|0\rangle.
\label{Eq:reducedLeggett}
\end{equation}
This is exactly the ground state wavefunction (\ref{Eq:defpsik}) of the {\em single-mode} Hamiltonian $\hat{H}_{\bf k}$, which
we call $|\widetilde\psi_{\bf k}\rangle$ in our manuscript. 
By contrast to the wavefunction given in Eq. (\ref{Eq:PsiLeggett}) above, the variational
wavefunction in Eq. (\ref{Eq:reducedLeggett}) does not take into account the fact that the number
of bosons in the condensate is given by $(N-2\sum_in_i)$. Hence, this variational 
calculation uses the same kind of approximation we discuss in Sec. \ref{Sec:VariationalMethod}, 
in which the variational constants $c_{\bf k}$ correspond to minimizing the energy 
of each single-mode Hamiltonian $\hat{H}_{\bf k}$ independently from one another.
Again, since the ${\bf k}=0$ state is shared by all the wavefunctions $|\widetilde\psi_{\bf k}\rangle$,
the above procedure to find the $c_{\bf k}$'s is mathematically inaccurate,
even for a dilute Bose gas. This is corroborated by the fact that, 
when the variational calculation is done in a more careful way, where the number $(N-2\sum_i n_i)$ of bosons 
in the ${\bf k}=0$ is kept throughout the calculation, the results obtained for the ground 
state energy and the excitation spectrum are quite different from those of the standard Bogoliubov method, as
we have explicitly shown in Sec. \ref{Sec:DiagFullH} of the present paper.

\addvspace{1.5mm}

Another attempt at formulating a number-conserving theory for the ground
state of interacting bosons was given by Gardiner in Ref. \citen{Gardiner1997}. We will
not discuss this study here, but rather point to a comment by Girardeau, Ref. \citen{Girardeau1998}, where
Gardiner's theory is discussed in great detail, and is shown to be a special case
of the theory developped by Girardeau and Arnowitt in 1959, Ref. \citen{Girardeau1959}.

\section{Conclusions}
\label{Sec:Conclusion}

To summarize, in this paper, we have given a detailed and rather thorough discussion of Bogoliubov's theory 
of an interacting Bose gas. Our main point was to reassert a result which in principle should 
be quite well-known but is often overlooked in the literature, having to do with the fact that Bogoliubov's theory 
is a theory in which each of the single-momentum contributions
$\hat{H}_{\bf k}$ to the total Hamiltonian $\hat{H}=\sum_{\bf k\neq 0}\hat{H}_{\bf k}$ is diagonalized
independently from the other contributions $\hat{H}_{\bf k'(\neq k)}$,
and the ground state wavefunction of the total Hamiltonian $\hat{H}$ is written as a simple
product of the ground state wavefunctions of the $\hat{H}_{\bf k}$'s.
As a way to illustrate this point, we have discussed a variational, number conserving 
formulation of Bogoliubov's method, where the ground state of each single-mode Hamiltonian $\hat{H}_{\bf k}$ is found
independently from the ground states of the other Hamiltonians $\hat{H}_{\bf k'\neq k}$,
and we have derived most of the results of Bogoliubov's standard, number non-conserving formulation within
this variational method, including the gapless excitation spectrum predicted by Bogoliubov's theory. 
Arguing that the above decoupled way of finding the ground state of the total Hamiltonian $\hat{H}$ may
not be accurate, we generalized the above mentioned variational method to the 
total Hamiltonian $\hat{H}$, making sure to keep an accurate count of the number of bosons in the ${\bf k}=0$ state,
which led us to a new ground state which has a {\em lower} overall energy and a much smaller depletion than the standard Bogoliubov ground state.
It also led to an excitation spectrum of bosons which has a finite gap as $k\to 0$, by contrast to Bogoliubov's method
where this gap vanishes. This last feature has allowed us to shed more light on the $\alpha_{\bf k}$
excitations of the standard Bogoliubov theory, which we argued do not represent phonon modes and correspond to single-particle excitations instead. We have argued that the existence of a gap in our number-conserving approach does not imply a violation
of Goldstone's theorem, given that our ground state does not break the $U(1)$ symmetry of the Hamiltonian, which implies that
the Goldstone theorem does not apply in this case. Our results seem to support the conclusions of recent work by Bobrov\cite{Bobrov2010} and Kita,\cite{Kita2011} which have argued that the correct Goldstone modes of condensed Bose systems should be sought in the pole of the two-particle Green's function, altough more work still needs to be done in order to understand the connection between these approaches and the approach developed in the present paper.

Given the above, the author hopes he has made a compelling case 
that the standard formulation of Bogoliubov's theory,
where not much importance is attached to the conservation of boson number,
is far from accurate and hence highly unsatisfactory, to say the least, and is in need of a major revision. 
It is the author's hope that the discussion presented in this work will 
help bring about such a revision, leading to a conceptually more satisfying 
description of dilute Bose systems, both at $T=0$ and at finite temperatures.

\appendix

\section{Diagonalization of the single mode Hamiltonian $\hat{H}_{\bf k}$}
\label{AppendixA}

In this Appendix, we show how one can diagonalize the Hamiltonian $\hat{H}_{\bf k}$ in the number-conserving Bogoliubov
approach. We shall here mostly focus on the case of the single-mode theory of Sec. \ref{Sec:VariationalMethod},
the general case being very similar, apart from a few minor differences discussed at the end of this Appendix.

\subsection{Derivation of the diagonal form of $\hat{H}_{\bf k}$ in terms of the $\alpha_{\bf k}$ operators}

Let us rewrite the expression of $\hat{H}_{\bf k}$:
\begin{align}
\hat{H}_{\bf k} & = \frac{1}{2}\varepsilon_{\bf k}
\big( a_{\bf k}^\dagger a_{\bf k} + a_{-\bf k}^\dagger a_{-\bf k}\big) 
+ \frac{v({\bf k})}{2V}\,\big( a_0^\dagger a_0 a_{\bf k}^\dagger a_{\bf k} 
\nonumber\\
&+a_0^\dagger a_0 a_{-\bf k}^\dagger a_{-\bf k}
+ a_{\bf k}^\dagger a_{-\bf k}^\dagger a_0 a_0
+ a_0^\dagger a_0^\dagger a_{\bf k} a_{-\bf k}\big).
\label{Eq:defHk2}
\end{align}
We now introduce the operator $b_{\bf k}$ such that:
\begin{equation}
b_{\bf k} = a_{\bf k}a_0^\dagger.
\end{equation}
It then follows that the operator $\alpha_{\bf k}=\tilde{u}_{\bf k}a_{\bf k}a_0^\dagger 
+ \tilde{v}_{\bf k}a_{-\bf k}^\dagger a_0$ 
can be written in the form:
\begin{equation}
\alpha_{\bf k}=\tilde{u}_{\bf k}b_{\bf k} + \tilde{v}_{\bf k}b_{-\bf k}^\dagger.
\end{equation}
Taking the adjoint of the above equation, we obtain:
\begin{equation}
\alpha_{\bf k}^\dagger=\tilde{u}_{\bf k}b_{\bf k}^\dagger + \tilde{v}_{\bf k}b_{-\bf k}.
\end{equation}
These last two expressions of $\alpha_{\bf k}$ and $\alpha_{\bf k}^\dagger$
in terms of the $b_{\bf k}$'s can easily be inverted to give the expressions 
of $b_{\bf k}$ in terms of the $\alpha_{\bf k}$'s, with the result: 
\begin{equation}
b_{\bf k} = \frac{1}{\gamma_{\bf k}^2}\big[\tilde{u}_{\bf k}\alpha_{\bf k}
-\tilde{v}_{\bf k}\alpha_{\bf k}^\dagger\big],
\end{equation}
where we remind the reader that $\gamma_{\bf k}^2 = \tilde{u}_{\bf k}^2 - \tilde{v}_{\bf k}^2$. Using the fact that
$\tilde{u}_{\bf k} = \gamma_{\bf k}u_{\bf k}$ and $\tilde{v}_{\bf k} = \gamma_{\
bf k}v_{\bf k}$, we finally can write:
\begin{equation}
b_{\bf k} = \frac{1}{\gamma_{\bf k}}\big[{u}_{\bf k}\alpha_{\bf k}
-{v}_{\bf k}\alpha_{\bf k}^\dagger\big].
\label{Eq:resultbk}
\end{equation}
Now, it is not difficult to see that the pairing terms can be easily written in terms of the $b_{\bf k}$'s:
\begin{subequations}
\begin{align}
&a_{\bf k}^\dagger a_{-\bf k}^\dagger a_0 a_0 = a_{\bf k}^\dagger a_0 a_{-\bf k}^\dagger a_0 
= b_{\bf k}^\dagger b_{-\bf k}^\dagger,
\\
&a_0^\dagger a_0^\dagger a_{\bf k} a_{-\bf k} = a_0^\dagger a_{\bf k} a_0^\dagger a_{-\bf k}
= b_{\bf k} b_{-\bf k}.
\end{align}
\end{subequations}
On the other hand, we can write for the Fock term:
\begin{align}
a_{\bf k}^\dagger a_{\bf k} a_0^\dagger a_0 & = a_{\bf k}^\dagger a_{\bf k}(a_0a_0^\dagger -1),
\nonumber\\
& = a_{\bf k}^\dagger a_0 a_{\bf k} a_0^\dagger - a_{\bf k} a_{\bf k},
\nonumber\\
& = b_{\bf k}^\dagger b_{\bf k} - a_{\bf k}^\dagger a_{\bf k}.
\end{align}
Hence, we can rewrite for $\hat{H}_{\bf k}$ the following expression:
\begin{align}
\hat{H}_{\bf k} & = \frac{1}{2}\Big(\varepsilon_{\bf k} - \frac{v({\bf k})}{V}\Big)
\big( a_{\bf k}^\dagger a_{\bf k} + a_{-\bf k}^\dagger a_{-\bf k}\big) 
\nonumber\\
& + \frac{v({\bf k})}{2V}\,\big( b_{\bf k}^\dagger b_{\bf k} + b_{-\bf k}^\dagger b_{-\bf k}
+ b_{\bf k}^\dagger b_{-\bf k}^\dagger 
+ b_{\bf k} b_{-\bf k}\big).
\label{Eq:defHk3}
\end{align}
To the above Hamiltonian, we add and substract the quantity 
$\frac{1}{2}\eta_{\bf k}\varepsilon_{\bf k}
(b_{\bf k}^\dagger b_{\bf k} + b_{-\bf k}^\dagger b_{-\bf k})$,
and rewrite the result in the form:
\begin{align}
\hat{H}_{\bf k} = \hat{H}_{1\bf k} 
+ \frac{1}{2}\Big(\varepsilon_{\bf k} \eta_{\bf k} + \frac{v({\bf k})}{V}\Big)
\big( b_{\bf k}^\dagger b_{\bf k} + b_{-\bf k}^\dagger b_{-\bf k}\big) 
+ \frac{v({\bf k})}{2V}\,\big(b_{\bf k}^\dagger b_{-\bf k}^\dagger 
+ b_{\bf k} b_{-\bf k}\big),
\label{Eq:defHk4}
\end{align}
where the {\em excess} Hamiltonian $\hat{H}_{1\bf k}$ is given by:
\begin{align}
\hat{H}_{1\bf k} & = \frac{1}{2}\varepsilon_{\bf k}
(a_{\bf k}^\dagger a_{\bf k} + a_{-\bf k}^\dagger a_{-\bf k})
\nonumber\\
& - \frac{1}{2}\varepsilon_{\bf k}\eta_{\bf k}(b_{\bf k}^\dagger b_{\bf k} + b_{-\bf k}^\dagger b_{-\bf k})
\nonumber\\
& - \frac{v({\bf k})}{2V}
(a_{\bf k}^\dagger a_{\bf k} + a_{-\bf k}^\dagger a_{-\bf k}).
\label{Eq:excessH}
\end{align}
For weak perturbation, this Hamiltonian will have a very small expectation value, owing to the fact that the two
terms on the first and second lines cancel each other (we remind the reader that $b_{\bf k}=a_{\bf k}a_0^\dagger$, 
and hence $b_{\bf k}^\dagger b_{\bf k}=a_{\bf k}^\dagger a_{\bf k} a_0^\dagger a_0$; 
on the other hand we will see below that $\eta_{\bf k}=\gamma_{\bf k}^2$, 
and since $\gamma_{\bf k}^2\sim 1/N_0$, we see that 
$\gamma_{\bf k}^2b_{\bf k}^\dagger b_{\bf k}\sim a_{\bf k}^\dagger a_{\bf k}$), 
and will henceforth be neglected altogether. (Note that,
because it only involves a single momentum mode,
the term on the last line on Eq. (\ref{Eq:excessH})
can be neglected in the thermodynamic limit $V\to \infty$). 

\addvspace{1.5mm}

We now can use Eq. (\ref{Eq:resultbk}) to rewrite the part of $\hat{H}_{\bf k}$ that is quadratic in the $b_{\bf k}$'s
in terms of the $\alpha_{\bf k}$'s. A somewhat tedious but straightforward calculation gives:
\begin{align}
&\hat{H}_{\bf k}  = \frac{1}{2\gamma_{\bf k}^2}\Big\{
\big[A_{\bf k}(u_{\bf k}^2 + v_{\bf k}^2) - 2B_{\bf k}\big]
(\alpha_{\bf k}^\dagger \alpha_{\bf k} 
+ \alpha_{-\bf k}^\dagger \alpha_{-\bf k})
\nonumber\\
& + \big[B_{\bf k}(u_{\bf k}^2 + v_{\bf k}^2) - 2A_{\bf k}\big]
(\alpha_{\bf k}^\dagger \alpha_{-\bf k}^\dagger 
+ \alpha_{\bf k} \alpha_{-\bf k})
\nonumber\\
& + \big[A_{\bf k}v_{\bf k}^2 - B_{\bf k}u_{\bf k}v_{\bf k}\big]
\big([\alpha_{\bf k},\alpha_{\bf k}^\dagger]
+ [\alpha_{-\bf k},\alpha_{-\bf k}^\dagger]\big)
\nonumber\\
& + (B_{\bf k}v_{\bf k}^2 - A_{\bf k}u_{\bf k}v_{\bf k}\big)\big([\alpha_{-\bf k}^\dagger,\alpha_{\bf k}^\dagger]
+[\alpha_{-\bf k},\alpha_{\bf k}]\big)
\Big\},
\label{Eq:App:Hgeneralalphas}
\end{align}
where we defined the constants $A_{\bf k}$ and $B_{\bf k}$ such that:
\begin{equation}
A_{\bf k} = \varepsilon_{\bf k}\eta_{\bf k} + \frac{v({\bf k})}{V}, \quad B_{\bf k} = \frac{v({\bf k})}{V}.
\label{Eq:defAkBk}
\end{equation}
We now want to fix the value of the constant $\eta_{\bf k}$ in such a way as to make the coefficient
of the term $(\alpha_{\bf k}^\dagger \alpha_{-\bf k}^\dagger + \alpha_{\bf k} \alpha_{-\bf k})$ vanish,
{\em i.e.} by imposing the condition:
\begin{equation}
\big[B_{\bf k}(u_{\bf k}^2 + v_{\bf k}^2) - 2A_{\bf k}\big] = 0.
\label{Eq:conditionetak}
\end{equation}
Using Eqs. (\ref{Eq:defAkBk}) and the expressions (\ref{Eq:resulttildeukvk})-(\ref{Eq:resultukvk}) 
of the coherence factors
$\tilde{u}_{\bf k}$ and $\tilde{v}_{\bf k}$, we obtain, after a few steps:
\begin{equation}
\eta_{\bf k} = \gamma_{\bf k}^2.
\label{Eq:App:resultetak}
\end{equation}

Using the definition of $\alpha_{\bf k}$ in terms of the $a_{\bf k}$ operators, one can easily evaluate the commutator
$[\alpha_{-\bf k},\alpha_{\bf k}]$, with the result:
\begin{align}
[\alpha_{-\bf k},\alpha_{\bf k}] = [\alpha_{-\bf k}^\dagger,\alpha_{\bf k}^\dagger] = \gamma_{\bf k}^2 u_{\bf k}v_{\bf k}\big[
a_{\bf k}^\dagger a_{\bf k} - a_{-\bf k}^\dagger a_{-\bf k}
\big].
\label{Eq:App:commutator1}
\end{align}
For all practical purposes, the last term on the {\em rhs} of the above equation can be
neglected, since for weak perturbation $\gamma_{\bf k}^2\sim 1/N_0\sim 1/N$ and the expectation values of 
$a_{\bf k}^\dagger a_{\bf k}$ and $a_{-\bf k}^\dagger a_{-\bf k}$ should be almost equal,
hence canceling each other. We therefore can approximate:
\begin{align}
[\alpha_{-\bf k},\alpha_{\bf k}] = [\alpha_{-\bf k}^\dagger,\alpha_{\bf k}^\dagger] \simeq 0.
\label{Eq:App:commutator1b}
\end{align}
On the other hand, we have already established in Eq. (\ref{Eq:commutatoralphas1}) of
the text that the commutator $[\alpha_{\bf k},\alpha_{\bf k}^\dagger]$
is given by:
\begin{equation}
[\alpha_{\bf k},\alpha_{\bf k}^\dagger] \simeq \gamma_{\bf k}^2 a_0^\dagger a_0
- \tilde{u}_{\bf k}^2 a_{\bf k}^\dagger a_{\bf k} + \tilde{v}_{\bf k}^2 a_{-\bf k}^\dagger a_{-\bf k}.
\label{Eq:App:commutator2}
\end{equation}
Given that $\tilde{u}_{\bf k}=\gamma_{\bf k}u_{\bf k}=u_{\bf k}/\sqrt{N_0}$, and 
$\tilde{v}_{\bf k}=\gamma_{\bf k}v_{\bf k}=v_{\bf k}/\sqrt{N_0}$, assuming weak
depletion the second and last terms on the {\em rhs} of the above equation can be neglected.
This implies, since $\gamma_{\bf k}^2\sim 1/N_0$ that we can approximate:
\begin{equation}
[\alpha_{\bf k},\alpha_{\bf k}^\dagger] \approx \gamma_{\bf k}^2 a_0^\dagger a_0 \simeq 1.
\label{Eq:App:commutator3}
\end{equation}
Under these circumstances, if we use the approximation $\gamma_{\bf k}^2\simeq 1/N_0\simeq 1/N$, so that
$V\gamma_{\bf k}^2\simeq 1/n_B$, and then use the expressions of $u_{\bf k}$ and $v_{\bf k}$ 
in terms of $c_{\bf k}$, Eqs. (\ref{Eq:resulttildeukvk}), one can show that:
\begin{equation}
n_Bv({\bf k})\big(u_{\bf k}^2 + v_{\bf k}^2\big) = 2\big(\varepsilon_{\bf k} + n_Bv({\bf k})\big)u_{\bf k}v_{\bf k},
\end{equation}
then the expression of $\hat{H}_{\bf k}$ reduces to:
\begin{align}
\hat{H}_{\bf k} & = \Big\{
\big[\varepsilon_{\bf k} + n_Bv({\bf k})\big]v_{\bf k}^2 - n_Bv({\bf k}) u_{\bf k}v_{\bf k}
\Big\}
\nonumber\\
& +\frac{1}{2}\Big\{
\big[\varepsilon_{\bf k} + n_Bv({\bf k})\big](u_{\bf k}^2 + v_{\bf k}^2) 
-2n_Bv({\bf k})u_{\bf k}v_{\bf k}
\Big\}
(\alpha_{\bf k}^\dagger \alpha_{\bf k} + \alpha_{-\bf k}^\dagger \alpha_{-\bf k}).
\label{Eq:intermHk}
\end{align}
If we again use the expression of $u_{\bf k}$ and $v_{\bf k}$ in terms of $\varepsilon_{\bf k}$, 
$n_B$ and $v({\bf k})$, Eqs. (\ref{Eq:resultukvk}) to evaluate the terms
between curly braces, then we obtain after a few manipulations:
\begin{align}
\hat{H}_{\bf k} \simeq -\frac{1}{2}\big[\varepsilon_{\bf k} + n_Bv({\bf k}) - E_{\bf k}\big] +
\frac{1}{2}E_{\bf k}\big(\alpha_{\bf k}^\dagger\alpha_{\bf k} + \alpha^\dagger_{-\bf k}\alpha_{-\bf k}\big),
\label{Eq:Hcanonical2}
\end{align}
which is nothing but Eq. (\ref{Eq:Hcanonical}) of the text.

\subsection{Proof that the excess Hamiltonian $\hat{H}_{1\bf k}$ gives a negligible contribution to
the excitation energy}

In this Subsection, we want to make sure that the excess Hamiltonian $\hat{H}_{1\bf k}$ does not contribute to the
excitation spectrum. We are interested in evaluating the following quantity:
\begin{subequations}
\begin{align}
\Delta\hat{H}_{1\bf k} & = \langle\Psi(N)|\alpha_{\bf k}\hat{H}_{1\bf k}\alpha_{\bf k}^\dagger|\Psi(N)\rangle 
- \langle\Psi(N)|\hat{H}_{1\bf k}|\Psi(N)\rangle,
\\
& = \langle\Psi(N)|\big[\alpha_{\bf k},\hat{H}_{1\bf k}\big]\alpha_{\bf k}^\dagger|\Psi(N)\rangle,
\end{align}
\end{subequations}
where, in going from the first to the second line, use has been made of the commutation relation
$[\alpha_{\bf k},\alpha_{\bf k}^\dagger]\simeq 1$. In what follows, we shall find it convenient to decompose
$\hat{H}_{1\bf k}$ into the follwing sum:
\begin{equation}
\hat{H}_{1\bf k} = \hat{h}_{1\bf k} + \hat{h}_{2\bf k},
\end{equation}
with:
\begin{subequations}
\begin{align}
\hat{h}_{1\bf k} & = \frac{1}{2}\Big(\varepsilon_{\bf k} - \frac{v({\bf k})}{V}\Big)
\big(a_{\bf k}^\dagger a_{\bf k} + a_{-\bf k}^\dagger a_{-\bf k}\big),
\\
\hat{h}_{2\bf k} & = -\frac{1}{2}\varepsilon_{\bf k}\eta_{\bf k}\big(b_{\bf k}^\dagger b_{\bf k} +
b_{-\bf k}^\dagger b_{-\bf k}\big)
\end{align}
\end{subequations}
Calculating the commutator $[\alpha_{\bf k},\hat{h}_{1\bf k}]$, we find:
\begin{align}
[\alpha_{\bf k},\hat{h}_{1\bf k}] = \frac{1}{2\gamma_{\bf k}^2}\Big(\varepsilon_{\bf k} - \frac{v({\bf k})}{V}\Big)
\big[
(\tilde{u}_{\bf k}^2 + \tilde{v}_{\bf k}^2)\alpha_{\bf k} - 2\tilde{u}_{\bf k}\tilde{v}_{\bf k}\alpha_{-\bf k}^\dagger 
\big].
\end{align}
We therefore can write:
\begin{subequations}
\begin{align}
&\langle\Psi(N)|[\alpha_{\bf k},\hat{h}_{1\bf k}]\alpha_{\bf k}^\dagger|\Psi(N)\rangle =
\frac{1}{2\gamma_{\bf k}^2}\Big(\varepsilon_{\bf k} - \frac{v({\bf k})}{V}\Big)
\nonumber\\
&\times\langle\Psi(N)|(\tilde{u}_{\bf k}^2 + \tilde{v}_{\bf k}^2)\alpha_{\bf k}\alpha_{\bf k}^\dagger|\Psi(N)\rangle,
\\
& = \frac{\tilde{u}_{\bf k}^2 + \tilde{v}_{\bf k}^2}{2\gamma_{\bf k}^2}
\Big(\varepsilon_{\bf k} - \frac{v({\bf k})}{V}\Big)
\end{align}
\label{Eq:expech1}
\end{subequations}
Similarly, calculating the commutator $[\alpha_{\bf k},\hat{h}_{2\bf k}]$,
we find:
\begin{align}
[\alpha_{\bf k},\hat{h}_{2\bf k}] & = -\frac{1}{2\gamma_{\bf k}^4}\varepsilon_{\bf k}\eta_{\bf k}
\big[
(\tilde{u}_{\bf k}^2 + \tilde{v}_{\bf k}^2)\alpha_{\bf k} - 2\tilde{u}_{\bf k}\tilde{v}_{\bf k}\alpha_{-\bf k}^\dagger
\big],
\end{align}
and hence:
\begin{align}
\langle\Psi(N)|[\alpha_{\bf k},\hat{h}_{2\bf k}]\alpha_{\bf k}^\dagger|\Psi(N)\rangle =
-\varepsilon_{\bf k}\eta_{\bf k}\frac{\tilde{u}_{\bf k}^2 + \tilde{v}_{\bf k}}{\gamma_{\bf k}^2}.
\label{Eq:expech2}
\end{align}
Adding the two contributions from Eqs. (\ref{Eq:expech1}) and (\ref{Eq:expech2}) together, we find:
\begin{align}
&\langle\Psi(N)|[\alpha_{\bf k},\hat{H}_{1\bf k}]\alpha_{\bf k}^\dagger|\Psi(N)\rangle =
\frac{\tilde{u}_{\bf k}^2 + \tilde{v}_{\bf k}^2}{2\gamma_{\bf k}^2}
\notag\\
&\times\Big[
\varepsilon_{\bf k}\Big(1 - \frac{\eta_{\bf k}}{\gamma_{\bf k}^2}\Big) - \frac{v({\bf k})}{V}
\Big].
\end{align}
Given that $\eta_{\bf k} = \gamma_{\bf k}^2$, and using the fact that $\tilde{u}_{\bf k}=\gamma_{\bf k}u_{\bf k}$
and $\tilde{v}_{\bf k} = \gamma_{\bf k} v_{\bf k}$, we finally obtain:
\begin{align}
&\langle\Psi(N)|[\alpha_{\bf k},\hat{H}_{1\bf k}]\alpha_{\bf k}^\dagger|\Psi(N)\rangle = -
\frac{v({\bf k})}{2V}(u_{\bf k}^2 + v_{\bf k}^2),
\end{align}
which indeed gives a negligible contribution to the excitation energy in the thermodynamic limit $V\to\infty$.

\addvspace{1.5mm}

At the end of this Appendix, we briefly discuss the diagonalization of $\hat{H}_{\bf k}$
when a multi-mode model governed by the
full Hamiltonian $\hat{H}=\sum_{\bf k\neq 0}\hat{H}_{\bf k}$ is considered. In this case,
the coherence factors are given by Eq. (\ref{Eq:resultukvk2}), and hence the requirement (\ref{Eq:conditionetak}) 
above gives a different result for $\eta_{\bf k}$ than Eq. (\ref{Eq:App:resultetak}),
namely Eq. (\ref{Eq:resultetak}). This, in turn, leads to the boson excitation
spectrum given in Eq. (\ref{Eq:excSpectrum}) of the text.

\addvspace{1.5mm}

\section{Expectation value of the Hamiltonian when we keep an accurate tally of the number of condensed bosons}
\label{AppendixB}

In this Appendix, we give the salient features of how we calculate the expectation value of the Hamiltonian
\begin{align}
\hat{H}_{{\bf k}_j} & = \frac{1}{2} \varepsilon_{{\bf k}_j}\big(a_{{\bf k}_j}^\dagger a_{{\bf k}_j}
+ a_{-{\bf k}_j}^\dagger a_{-{\bf k}_j}\big) 
+ \frac{v({\bf k}_j}{2V}\big(a_0^\dagger a_0 a_{{\bf k}_j}^\dagger a_{{\bf k}_j}
\notag\\
& + a_0^\dagger a_0 a_{{\bf k}_j}^\dagger a_{{\bf k}_j}  
+ a_0 a_0 a_{{\bf k}_j}^\dagger a_{-{\bf k}_j}^\dagger
+ a_0^\dagger a_0^\dagger a_{{\bf k}_j} a_{-{\bf k}_j}\big)  
\end{align}
in the variational ground state $|\Psi(N)\rangle$ of Eq. (\ref{Eq:fullPsi(N)}), namely:
\begin{align}
|\Psi(N)\rangle = Z\sum_{n_1=0}^\infty\cdots\sum_{n_M=0}^\infty C_{n_1}\cdots C_{n_M} 
|N-2\sum_{i=1}^M n_i;n_1,n_1;\ldots;n_M,n_M\rangle.
\end{align}
We have:
\begin{align}
a_{{\bf k}_j}^\dagger a_{{\bf k}_j}|\Psi(N)\rangle  
= Z\sum_{n_1=0}^\infty\cdots\sum_{n_M=0}^\infty C_{n_1}\cdots C_{n_M} n_j 
|N-2\sum_{i=1}^M n_i;n_1,n_1;\ldots;n_M,n_M\rangle,
\end{align}
and hence:
\begin{subequations}
\begin{align}
\langle\Psi(N)|a_{{\bf k}_j}^\dagger a_{{\bf k}_j}|\Psi(N)\rangle & = \frac{\sum_{n_j=0}^\infty n_jC_{n_j}^2}
{\sum_{n_j=0}^\infty C_{n_j}^2},
\\
& = \frac{c_{{\bf k}_j}^2}{1 - c_{{\bf k}_j}^2},
\label{Eq:App:kinetictermfinal}
\end{align}
\end{subequations}
where, in going from the first to the second line, we made use of the fact that $C_{n_j}=(-c_{{\bf k}_j})^{n_j}$,
and of the summation formulae (valid for $|x|<1$):
\begin{subequations}
\begin{align}
\sum_{n=0}^\infty x^n & = \frac{1}{1 - x},
\\
\sum_{n=0}^\infty nx^n & = \frac{x}{1 - x}.
\end{align}
\label{Eq:App:geometricsums1}
\end{subequations}
An identical result is obtained for
the expectation value of the operator $a_{-{\bf k}_j}^\dagger a_{-{\bf k}_j}$.

\addvspace{1.5mm}

Let us now turn our attention to the Fock terms. We have:
\begin{align}
a_0^\dagger a_0 a_{{\bf k}_j}^\dagger a_{{\bf k}_j}|\Psi(N)\rangle  
&= Z\sum_{n_1=0}^\infty\cdots\sum_{n_M=0}^\infty C_{n_1}\cdots C_{n_M} 
\notag\\
&\times n_j\Big(N - 2\sum_{i=1}^M n_i\Big) 
|N-2\sum_{i=1}^M n_i;n_1,n_1;\ldots;n_M,n_M\rangle.
\end{align}
We shall rewrite the above result in the form:
\begin{align}
a_0^\dagger a_0 a_{{\bf k}_j}^\dagger a_{{\bf k}_j}|\Psi(N)\rangle  
& = Z\sum_{n_1=0}^\infty\cdots\sum_{n_M=0}^\infty C_{n_1}\cdots C_{n_M} 
n_j\big(N-2n_j\big) 
\nonumber\\
&\times|N-2\sum_{i=1}^M n_i;n_1,n_1;\ldots;n_M,n_M\rangle
\notag\\
& - 2Z \sum_{n_1=0}^\infty\cdots\sum_{n_M=0}^\infty C_{n_1}\cdots C_{n_M}
n_j\Big(\sum_{i=1(\neq j)}^M n_i\Big) 
\nonumber\\
&\times|N-2\sum_{i=1}^M n_i;n_1,n_1;\ldots;n_M,n_M\rangle.
\end{align}
Hence:
\begin{subequations}
\begin{align}
&\langle\Psi(N)|a_0^\dagger a_0 a_{{\bf k}_j}^\dagger a_{{\bf k}_j}|\Psi(N)\rangle  
= Z^2\sum_{n_1=0}^\infty C_{n_1}^2\cdots\sum_{n_j=0}^\infty C_{n_j}^2 n_j\big(N-2n_j\big)
\cdots\sum_{n_M=0}^\infty C_{n_M}^2 
\notag\\
& - 2Z^2 \sum_{i=1(\neq j)}^\infty\Big(
\sum_{n_1=0}^\infty C_{n_1}^2\times\cdots\times \sum_{n_i=0}^\infty n_iC_{n_i}^2
\times\cdots\times \sum_{n_j=0}^\infty n_j C_{n_j}^2 \times\cdots\times\sum_{n_M=0}^\infty C_{n_M}^2
\Big),
\\
{}\nonumber
\\
& = \frac{\sum_{n_j=0}^\infty C_{n_j}^2 n_j(N-2n_j)}{\sum_{n_j=0}^\infty C_{n_j}^2} 
- 2 \sum_{i=1(\neq j)}^\infty\frac{\sum_{n_i=0}^\infty C_{n_i}^2 n_i \times \sum_{n_j=0}^\infty C_{n_j}^2 n_j}
{\sum_{n_i=0}^\infty C_{n_i}^2 \times \sum_{n_j=0}^\infty C_{n_j}^2 }.
\end{align}
\end{subequations}
Using the summation formulae of Eqs. (\ref{Eq:App:geometricsums1}), and the additional formula:
\begin{equation}
\sum_{n=0}^\infty n^2x^n = \frac{x}{(1 - x)^2} + \frac{2x^2}{(1-x)^3},
\label{Eq:App:geometricsums2}
\end{equation}
we obtain, after a few manipulations:
\begin{align}
\langle\Psi(N)|a_0^\dagger a_0 a_{{\bf k}_j}^\dagger a_{{\bf k}_j}|\Psi(N)\rangle  
\simeq  N\frac{c_{{\bf k}_j}^2}{1 - c_{{\bf k}_j}^2}\Big[
 1 - \frac{2}{N} \sum_{i=1(\neq j)}^\infty \frac{c_{{\bf k}_i}^2}{1 - c_{{\bf k}_i}^2}
\Big].
\label{Eq:App:Focktermfinal}
\end{align}
Note that, in arriving to this result, a term of the form $-2\Big[\frac{c_{{\bf k}_j^2}}{1- c_{{\bf k}_j^2}}
+ \frac{2c_{{\bf k}_j^4}}{(1 - c_{{\bf k}_j^2)^3}}\Big]$ has been negelected (this is legitimate given the
fact that the terms kept have an overall factor of $N$ at the front, which makes the term
neglected very small by comparison in the thermodynamic $N\to\infty$ limit). Note also that
an identical result is obtained for the expectation value of the other Fock term,
$a_0^\dagger a_0 a_{-{\bf k}_j}^\dagger a_{-{\bf k}_j}$.

\addvspace{1.5mm}

We now want to find the expectation value of the pairing term $a_0 a_0 a_{{\bf k}_j}^\dagger a_{-{\bf k}_j}^\dagger$.
We have:
\begin{subequations}
\begin{align}
a_0a_0 & a_{{\bf k}_j}^\dagger a_{-{\bf k}_j}^\dagger|\Psi(N)\rangle  
 = Z\sum_{n_1=0}^\infty \cdots \sum_{n_M=0}^\infty C_{n_1}\cdots C_{n_M}
\sqrt{(n_j+1)^2}
\nonumber\\
&\times\sqrt{\Big(N-2\sum_{i=1}^M n_i\Big)\Big(N -1 -2\sum_{i=1}^M n_i\Big)}
\nonumber\\
& \times |N-2-2\sum_{i=1}^M n_i; n_1,n_1;\ldots;n_j+1,n_j+1;\ldots;n_M,n_M\rangle,
\\
& \simeq Z\sum_{n_1=0}^\infty \cdots \sum_{n_M=0}^\infty C_{n_1}\cdots C_{n_M}
(n_j+1)\Big(N - 2n_j - \frac{1}{2} - 2\sum_{i=1(\neq j)}^M n_i\Big)
\nonumber\\
& \times |N-2(n_j+1)-2\sum_{i=1(\neq j)}^M n_i; n_1,n_1;\ldots;n_j+1,n_j+1;\ldots;n_M,n_M\rangle,
\end{align}
\end{subequations}
where, in going from the first to the second line, we used the approximation in Eq. (\ref{Eq:ApproxSqrt}) to write:
\begin{equation}
\sqrt{\Big(N-2\sum_{i=1}^M n_i\Big)\Big(N -1 -2\sum_{i=1}^M n_i\Big)} \simeq
N - 2n_j - \frac{1}{2} - 2\sum_{i=1(\neq j)}^M n_i.
\end{equation}
If we now change the index of summation for the momentum ${\bf k}_j$ from $n_j$ to $m_j=n_j+1$, we can write:
\begin{align}
a_0a_0 a_{{\bf k}_j}^\dagger a_{-{\bf k}_j}^\dagger|\Psi(N)\rangle  
& = Z\sum_{n_1=0}^\infty \cdots \sum_{m_j=1}^\infty\cdots \sum_{n_M=0}^\infty 
C_{n_1}\cdots C_{m_j-1}\cdots C_{n_M} m_j\Big( N - 2m_j + \frac{3}{2}\Big)
\nonumber\\
& \times |N-2m_j-2\sum_{i=1(\neq j)}^M n_i; n_1,n_1;\ldots;m_j,m_j;\ldots;n_M,n_M\rangle
\nonumber\\
& - 2 Z \sum_{i=1(\neq j)}^M\Bigg(
\sum_{n_1=0}^\infty \cdots \sum_{m_j=1}^\infty\cdots \sum_{n_M=0}^\infty 
C_{n_1}C_{n_2}\cdots C_{m_j-1}\cdots C_{n_M} m_jn_i
\Bigg)
\nonumber\\
& \times |N-2m_j-2\sum_{i=1(\neq j)}^M n_i; n_1,n_1;\ldots;m_j,m_j;\ldots;n_M,n_M\rangle.
\end{align}
Hence, we obtain for the expectation value:
\begin{align}
\langle\Psi(N)|a_0a_0 a_{{\bf k}_j}^\dagger a_{-{\bf k}_j}^\dagger|\Psi(N)\rangle & =
\frac{\sum_{m_j=1}^\infty m_j C_{m_j}C_{m_j-1}\Big[N-2m_j + \frac{3}{2}\Big]}{\sum_{m_j=0}^\infty C_{m_j}^2} 
\nonumber\\
& - 2\sum_{i=1 (\neq j)}^M \Bigg(
\frac{\sum_{m_j=1}^\infty m_j C_{m_j}C_{m_j-1}}{\sum_{m_j=0}^\infty C_{m_j}^2}
\frac{\sum_{n_i=1}^\infty n_i C_{n_i}^2}{\sum_{n_i=0}^\infty C_{n_i}^2}
\Bigg).
\end{align}
It can be verified that the only term that one has to keep in the first summation is the term proportional to $N$
(note that this is also how the calculation is conducted 
in the single-mode variational theory of Sec. \ref{Sec:VariationalMethod}). Then,
if we use the summation formulas (\ref{Eq:App:geometricsums1}), we obtain, after a few manipulations:
\begin{align}
\langle\Psi(N)|a_0a_0 a_{{\bf k}_j}^\dagger a_{-{\bf k}_j}^\dagger|\Psi(N)\rangle =
- N \frac{c_{{\bf k}_j}}{1 - c_{{\bf k}_j}^2} \Big[
1 - \frac{2}{N}\sum_{i=1(\neq j)}^M\frac{c_{{\bf k}_i^2}}{1 - c_{{\bf k}_i}^2}
\Big].
\label{Eq:App:pairingtermfinal}
\end{align}
A similar contribution is also obtained for the expectation value of the operator 
$a_0^\dagger a_0^\dagger a_{{\bf k}_j}^\dagger a_{-{\bf k}_j}^\dagger$.

\medskip

Collecting all terms, Eqs. (\ref{Eq:App:kinetictermfinal}), (\ref{Eq:App:Focktermfinal}) 
and (\ref{Eq:App:pairingtermfinal}), we obtain:
\begin{align}
\langle\Psi(N)|\hat{H}_{{\bf k}_j}|\Psi(N)\rangle & = 
\varepsilon_{{\bf k}_j}\frac{c_{{\bf k}_j}^2}{1 - c_{{\bf k}_j}^2}
+ n_B v({\bf k}_j)\frac{c_{{\bf k}_j}^2}{1 - c_{{\bf k}_j}^2}
\Big[
1 - \frac{2}{N}\sum_{i=1(\neq j)}^M\frac{c_{{\bf k}_i^2}}{1 - c_{{\bf k}_i}^2}
\Big]
\nonumber\\
& - n_B v({\bf k}_j)\frac{c_{{\bf k}_j}}{1 - c_{{\bf k}_j}^2}
\Big[
1 - \frac{2}{N}\sum_{i=1(\neq j)}^M\frac{c_{{\bf k}_i^2}}{1 - c_{{\bf k}_i}^2}
\Big].
\end{align}
For the total Hamiltonian $\hat{H} = \sum_{j=1}^M\hat{H}_{{\bf k}_j}$, we obtain:
\begin{align}
\langle\Psi(N)|\hat{H}|\Psi(N)\rangle & = \sum_{j=1}^M\Bigg\{
\Big[ 
\varepsilon_{{\bf k}_j} + n_B v({\bf k}_j)
\Big(
1 - \frac{2}{N}\sum_{i=1(\neq j)}^M\frac{c_{{\bf k}_i}^2}{1 - c_{{\bf k}_i}^2}
\Big) \Big]\frac{c_{{\bf k}_j}^2}{1 - c_{{\bf k}_j}^2}
\nonumber\\
& - n_B v({\bf k}_j)
\Big(
1 - \frac{2}{N}\sum_{i=1(\neq j)}^M\frac{c_{{\bf k}_i}^2}{1 - c_{{\bf k}_i}^2}
\Big)\frac{c_{{\bf k}_j}}{1 - c_{{\bf k}_j}^2}
\Bigg\},
\end{align}
which is nothing but Eq. (\ref{Eq:expvaluefullH}) of the text.


\section{Minimization of the expectation value of the total Hamiltonian}
\label{AppendixC}

In this Appendix, we show how we minimize the expectation value of the Hamiltonian given in 
Eq. (\ref{Eq:expvaluefullH}) over the coefficients $c_{{\bf k}_j}$. To this end, we shall rewrite
this expectation value in the form:
\begin{align}
\langle\Psi(N)|\hat{H}|\Psi(N)\rangle  & = 
\Big[ 
\varepsilon_{{\bf k}_j} + n_B v({\bf k}_j)
\Big(
1 - \frac{2}{N}\sum_{i=1(\neq j)}^M\frac{c_{{\bf k}_i}^2}{1 - c_{{\bf k}_i}^2}
\Big) \Big]\frac{c_{{\bf k}_j}^2}{1 - c_{{\bf k}_j}^2}
\nonumber\\
&- n_B v({\bf k}_j)
\Big(
1 - \frac{2}{N}\sum_{i=1(\neq j)}^M\frac{c_{{\bf k}_i}^2}{1 - c_{{\bf k}_i}^2}
\Big)\frac{c_{{\bf k}_j}}{1 - c_{{\bf k}_j}^2}
\nonumber\\
&+ \sum_{\ell=1(\neq j)}^M\Bigg\{
\Big[ 
\varepsilon_{{\bf k}_\ell} + n_B v({\bf k}_\ell)
\Big(
1 - \frac{2}{N}\sum_{i=1(\neq \ell)}^M\frac{c_{{\bf k}_i}^2}{1 - c_{{\bf k}_i}^2}
\Big) \Big]\frac{c_{{\bf k}_\ell}^2}{1 - c_{{\bf k}_\ell}^2}
\nonumber\\
&- n_B v({\bf k}_\ell)
\Big(
1 - \frac{2}{N}\sum_{i=1(\neq\ell)}^M\frac{c_{{\bf k}_i}^2}{1 - c_{{\bf k}_i}^2}
\Big)\frac{c_{{\bf k}_\ell}}{1 - c_{{\bf k}_\ell}^2}
\Bigg\}.
\end{align}
Taking the partial derivative with respect to $c_{{\bf k}_j}$, we obtain:
\begin{align}
\frac{\partial\langle\hat{H}\rangle}{\partial c_{{\bf k}_j}} & = 
\Big[ 
\varepsilon_{{\bf k}_j} + n_B v({\bf k}_j)
\Big(
1 - \frac{2}{N}\sum_{i=1(\neq j)}^M\frac{c_{{\bf k}_i}^2}{1 - c_{{\bf k}_i}^2}
\Big) 
\Big]\frac{2c_{{\bf k}_j}}{\big(1 - c_{{\bf k}_j}^2\big)^2}
\nonumber\\
&- n_B v({\bf k}_j)
\Big(
1 - \frac{2}{N}\sum_{i=1(\neq j)}^M\frac{c_{{\bf k}_i}^2}{1 - c_{{\bf k}_i}^2}
\Big)\frac{1 + c_{{\bf k}_j}^2}{\big(1 - c_{{\bf k}_j}^2\big)^2}
\nonumber\\
&+ \sum_{\ell=1(\neq j)}^M\Big\{
n_B v({\bf k}_\ell)
\Big(- \frac{2}{N}\frac{2c_{{\bf k}_j}}{\big(1 - c_{{\bf k}_j}^2\big)^2}\Big) 
\frac{c_{{\bf k}_\ell}^2}{1 - c_{{\bf k}_\ell}^2}
\nonumber\\
&- n_B v({\bf k}_\ell)
\Big(-\frac{2}{N}\frac{2c_{{\bf k}_j}}{\big(1 - c_{{\bf k}_i}^2\big)^2}
\Big)\frac{c_{{\bf k}_\ell}}{1 - c_{{\bf k}_\ell}^2}
\Big\}.
\end{align}
Rearranging the terms on the {\em rhs} of the above equation, and using the notation 
\begin{equation}
\bar{v}({\bf k}_j) = v({\bf k}_j)
\Big(1 - \frac{2}{N}\sum_{i=1(\neq j)}^M\frac{c_{{\bf k}_i}^2}{1 - c_{{\bf k}_i}^2}\Big), 
\end{equation}
we can write:
\begin{align}
\frac{\partial\langle\hat{H}\rangle}{\partial c_{{\bf k}_j}} & = \Big[
\varepsilon_{{\bf k}_j} + n_B \bar{v}({\bf k}_j)
-\frac{2}{N}\sum_{\ell=1(\neq j)}^M n_Bv({\bf k}_\ell)
\frac{c_{{\bf k}_\ell}(c_{{\bf k}_\ell} -1)}{1 - c_{{\bf k}_\ell}^2}
\Big]\frac{2c_{{\bf k}_j}}{\big(1 - c_{{\bf k}_j}^2\big)^2}
\nonumber\\
& - n_B\bar{v}({\bf k}_j)\frac{1 + c_{{\bf k}_j}^2}{\big(1 - c_{{\bf k}_j}^2\big)^2}
\end{align}


We hence see that the equation $(\partial\langle\hat{H}\rangle/\partial c_{{\bf k}_j})=0$ 
can be written in the form:
\begin{equation}
\frac{1}{\big(1 - c_{{\bf k}_j}^2\big)^2}\Big[
2\tilde{\cal E}_{{\bf k}_j}c_{{\bf k}_j} - n_B\bar{v}({\bf k}_j)\big(1+c_{{\bf k}_j}^2\big)
\Big] = 0,
\label{Eq:App:minHck1}
\end{equation}
with
\begin{align}
\tilde{\cal E}_{{\bf k}_j} = \varepsilon_{{\bf k}_j} + n_B \bar{v}({\bf k}_j)
+\frac{2}{N}\sum_{\ell=1(\neq j)}^M n_Bv({\bf k}_\ell)
\frac{c_{{\bf k}_\ell}}{1 + c_{{\bf k}_\ell}}.
\end{align}
Equation (\ref{Eq:App:minHck1}) can in turn be rewritten in the form:
\begin{equation}
c_{{\bf k}_j}^2 - 2\Big(\frac{\tilde{\cal E}_{{\bf k}_j}}{n_B\bar{v}({\bf k}_j)}\Big)c_{{\bf k}_j} + 1 =0,
\end{equation}
which is nothing but Eq. (\ref{Eq:Eqckfull}) of the text.


\begin{thebibliography}{99}


\bibitem{Anderson1995} M.H. Anderson, J.R. Ensher, M.R. Matthews, C.E. Wieman and E.A.Cornell,
Science {\bf 269}, 198 (1995).

\bibitem{Davis1995} K.B. Davis, M.-O. Mewes, M.R. Andrews, N.J. van Druten, D.S. Durfee,
D.M. Kurn and W. Ketterle, Phys. Rev. Lett. {\bf 75}, 3969 (1995).

\bibitem{Bogoliubov1947} N.N. Bogoliubov, J. Phys. U.S.S.R. {\bf 5}, 71 (1947).

\bibitem{Lee1957} T.D. Lee and C.N. Yang, Phys. Rev. {\bf 105}, 1119 (1957).

\bibitem{Bruckner1957} K.A. Bruckner and K. Sawada, Phys. Rev. {\bf 106}, 1117 (1957).

\bibitem{Beliaev1958} S.T. Beliaev, Soviet Physics JETP {\bf 7}, 104 (1958); {\bf 7}, 289 (1958).

\bibitem{LeeHuangYang1957} T.D. Lee, K. Huang and C.N. Yang, Phys. Rev. {\bf 106}, 1135 (1957).

\bibitem{Hugenholtz1959} N.M. Hugenholtz and D. Pines, Phys. Rev. {\bf 116}, 489 (1959).

\bibitem{Sawada1959} K. Sawada, Phys. Rev. {\bf 116}, 1344 (1959).

\bibitem{Gavoret1964} J. Gavoret and P. Nozi\`eres, Ann. Phys. (New York) {\bf 28}, 349 (1964).

\bibitem{Hohenberg1965} P.C. Hohenberg and P.C. Martin, Ann. Phys. (New York) {\bf 34}, 291 (1965).

\bibitem{Cheung1971} T.H. Cheung and A. Griffin, Phys. Rev. A {\bf 4}, 237 (1971).

\bibitem{Wong1974} V.K. Wong and H. Gould, Ann. Phys. {\bf 83}, 252 (1974).

\bibitem{Popov1965} V.N. Popov, Soviet Physics JETP {\bf 20}, 1185 (1965).

\bibitem{Singh1967} K.K. Singh, Physica {\bf 34}, 285 (1967).

\bibitem{Szepfalusy1974} P. Sz\'epfalusy and I. Kondor, Ann. Phys. (New York) {\bf 82}, 1 (1974).

\bibitem{Griffin1993} A. Griffin, {\em Excitations in a Bose-Condensed Liquid}, 
Cambridge University Press, New York, 1993.

\bibitem{Griffin1996} A. Griffin, Phys. Rev. B {\bf 53}, 9341 (1996).

\bibitem{Bijlsma1997} M. Bijlsma and H.T.C. Stoof, Phys. Rev. A {\bf 55}, 498 (1997).

\bibitem{Shi1998} H. Shi and A. Griffin, Phys. Rep. {\bf 304}, 1 (1998).

\bibitem{Andersen2004} J.O. Andersen, Rev. Mod. Phys. {\bf 76}, 599 (2004).

\bibitem{Zagrebnov2001} V.A. Zagrebnov, J.-B. Bru, Physics Reports {\bf 350}, 291 (2001).

\bibitem{Suto2008} A. S\"ut\H o and P\'eter Sz\'epfalusy, Phys. Rev. A {\bf 77}, 23606 (2008).

\bibitem{Yukalov2006} V.I. Yukalov and H. Kleinert, Phys. Rev. A {\bf 73}, 063612 (2006).

\bibitem{Girardeau1959} M. Girardeau and R. Arnowitt, Phys. Rev. {\bf 113}, 755 (1959).

\bibitem{Gardiner1997} C.W. Gardiner, Phys. Rev. A {\bf 56}, 1414 (1997); 

\bibitem{Girardeau1998} M.D. Girardeau, Phys. Rev. A {\bf 58}, 775 (1998).

\bibitem{Castin1998} Y. Castin and R. Dum, Phys. Rev. A {\bf 57}, 3008 (1998).
See also Y. Castin, arXiv:cond-mat/0105058.

\bibitem{Ginibre1968} J. Ginibre, Comm. Math. Phys. {\bf 8}, 26 (1968).

\bibitem{Lieb2005} E.H. Lieb,R. Seiringer and J. Yngvason, Phys. Rev. Lett. {\bf 94}, 80401 (2005).

\bibitem{FW} A.L. Fetter and J.D. Walecka, {\em Quantum Theory of Many-Particle Systems}, 
Dover Publications, USA, 2003.

\bibitem{deGennes} P.G. de Gennes, {\em Superconductivity of Metals and Alloys}, 
Reading, Massachusetts, 1966; Section 4.3.

\bibitem{CommentMeaningH} This is actually equivalent to writing for the average energy $E$
of a given system at temperature $T$ in the grand-canonical ensemble an expression of the form
$E = \mbox{Tr}\big[(\hat{H}-\mu\hat{N})e^{-\beta(\hat{H}-\mu\hat{N})}\big]/Z_{gr}$
(with $\beta=(k_BT)^{-1}$ and $Z_{gr}$ the grand-canonical partition funtion), which does not
of course make much sense.
In the grand-canonical ensemble, the energy is {\em always} given by the thermal average 
of the Hamiltonian $\hat{H}$ itself,
$E = \mbox{Tr}\big[\hat{H}e^{-\beta(\hat{H}-\mu\hat{N})}\big]/Z_{gr}$.
Likewise, in the standard formulation of Bogoliubov's method, 
the ground state energy should be given by the quantum average $\langle\Psi|\hat{H}|\Psi\rangle$,
and not by the quantity $\langle\Psi|\hat{H}-\mu\hat{N}|\Psi\rangle$, which represents 
the grand potential of the system at $T=0$.

\bibitem{Leggett2001} A.J. Leggett, Rev. Mod. Phys. {\bf 73}, 307 (2001).



\bibitem{FieldTheoreticFormulation} It should be mentioned here that Eq.(\ref{Eq:condalphak}) plays a crucial
role in the field-theoretic formulation of Bogoliubov's theory, since it ensures that the action of any normal-ordered
product of the excitation operators $\alpha_{\bf k}^\dagger$ and $\alpha_{\bf k}$ on the Bogoliubov ground state
is automatically zero.

\bibitem{MahanBook} G.D. Mahan, {\em Many-Particle Physics}, Kluwer Academic, New York, USA, 2000.


\bibitem{Takano1961} F. Takano, Phys. Rev. {\bf 123}, 699 (1961).

\bibitem{CommentHelium} It is quite remarkable that, despite repeated claims that
Bogoliubov's theory is only valid in the dilute limit $n_B{a}^3\ll 1$, this theory has routinely
been used to analyze experimental data for liquid Helium.
Unfortunately, the fact that the depletion in Helium is quite substantial makes any use
of the canonical commutation relations between the excitation operators $\alpha_{\bf k}$
and any theoretical results obtained from these commutation relations of a rather dubious character.

\bibitem{LeggettNJP} Indeed, the fact that the first
order correction to the ground state energy in Eq. (\ref{Eq:GSEBog}) is positive is quite opposite of what we expect,
namely that the more involved Bogoliubov ground state has {\em lower} energy than the 
Gross-Pitaevskii ground state $|\Psi_{GP}(N)\rangle = |N;0,\ldots\rangle$,
the energy of which is generally written in the form $E_{GP} = 2\pi {a} \hbar^2 n_B^2/m$.
For a discussion of this ``paradox", see A.J. Leggett, New J. Phys. {\bf 5}, 103 (2003).



\bibitem{Feynman1954} R.P. Feynman, Phys. Rev. {\bf 94}, 262 (1954).



\bibitem{HuangBook} K. Huang, {\em Statistical mechanics}, Second Edition, Wiley, New York, 1987.


\bibitem{Goldstone} J. Goldstone, Il Nuovo Cimento {\bf 19}, 154 (1961).

\bibitem{Brauner2010} For a recent review, see for example T. Brauner, Symmetry {\bf 2}, 609 (2010), and
references therein.

\bibitem{StoofBook} For a modern derivation of the Hugenholtz-Pines theorem, see 
H.T.C. Stoof, K.B. Gubbels and D.B. M. Dickerscheid, {\em Ultracold Quantum Fields},
Springer, The Netherlands, 2009.

\bibitem{Misawa1960} There is indeed evidence from previously published work that the HPT does not hold,
at least in its currently used form, when the conservation of particle number is taken into account; see
S. Misawa, Prog. Theor. Phys. {\bf 24}, 1224 (1960).

\bibitem{Giorgini2008} S. Giorgini, L.P. Pitaevskii and S. Stringari, Rev. Mod. Phys. {\bf 80}, 1215 (2008), 
and references therein.


\bibitem{Bobrov2010} V.B. Bobrov, S.A. Trigger and I.M. Yurin, Physics Letters A {\bf 374},1938 (2010).


\bibitem{Kita2009} T. Kita, Phys. Rev. B {\bf 80}, 214502 (2009).


\bibitem{Kita2010} T. Kita, Phys. Rev. B {\bf 81}, 214513 (2010).


\bibitem{Kita2011} T. Kita, J. Phys. Soc. Jap. {\bf 80}, 084606 (2011). 



\bibitem{CommentFiniteT} Note, however, that in order to be able to describe the statistical mechanics
of the interacting Bose gas all the way to the transition temperature into the normal state $T_c$, 
one has to take into account the Fock interactions between depleted
bosons, {\em i.e.} the terms we neglected in going from Eq. (\ref{Eq:V_F'}) 
to Eq. (\ref{Eq:V_F'0}). Note also that, taking these neglected Fock terms into
account in the variational approach for the Bogoliubov Hamiltonian at $T=0$ 
may modify the ground state energy and the 
total depletion of the interacting gas with respect to the results
of Sec. \ref{Sub:DepletionManyModes}.

\bibitem{commentCastin} It is worth mentioning at this point that it is not even clear whether 
the perturbative approach of Ref. \citen{Castin1998} can be reformulated to produce a finite result
for the number $N_{\bf k}$ of depleted bosons in the state of wavevector ${\bf k}$ in the $k\to 0$ limit, 
by analogy with the variational result (\ref{Eq:fullNkvariational})
obtained within the single-mode approach.
The approach developed in Refs. \citen{LeeHuangYang1957} and \citen{Leggett2001}
and in the present paper has the advantage of allowing us to obtain such a finite result for $N_{\bf k}$,
both in the single-mode theory of Sec. \ref{Sub:vardepletion}
and in the multi-mode approach of Sec. \ref{Sub:DepletionManyModes}.


\end{thebibliography}
\end{document}